\documentclass[twocolumn,floatfix,amsmath,amssymb,superscriptaddress,nofootinbib,longbibliography,prx]{revtex4-2}

\usepackage{lipsum}
\usepackage{verbatim}
\usepackage[colorlinks=true,linkcolor=blue,citecolor=blue]{hyperref}
\usepackage[utf8]{inputenc}
\usepackage[T1]{fontenc}
\usepackage{xspace}
\usepackage{enumitem}

\usepackage{bm}
\usepackage{array}
\usepackage{mathtools}
\usepackage[mathscr]{euscript}

\usepackage{physics}
\usepackage{siunitx}

\usepackage[usenames,dvipsnames]{xcolor}
\usepackage{graphicx}
\usepackage{tikz}
\usepackage{pgfplots}

\usepackage{tabularx}
\usepackage{booktabs}
\usepackage{multirow}

\usepackage{tcolorbox}

\DeclarePairedDelimiter\ceil{\lceil}{\rceil}
\DeclarePairedDelimiter\floor{\lfloor}{\rfloor}

\DeclareMathOperator*{\sgn}{sgn}

\DeclareMathOperator*{\wt}{wt}
\newcommand{\trans}[1]{#1^{\mathsf{T}}}

\newtheorem{result}{Result}

\usepackage[caption=false]{subfig}
\newcommand{\phantomsubfloat}[1]{
    {%
        \captionsetup[subfigure]{labelformat=empty}
        \subfloat[][]{#1}
    }%
}

\newcolumntype{x}[1]{>{\centering\arraybackslash\hspace{0pt}}p{#1}}

\appdef \turnpage {%
  \AddToHookNext{shipout/after}{%
    \global\pdfpageattr\expandafter{\the\pdfpageattr/Rotate 90}%
    \AddToHookNext{shipout/after}{%
      \global\pdfpageattr\expandafter{\the\pdfpageattr/Rotate 0}%
    }%
  }%
}

\usepackage{cleveref}
\crefname{equation}{Eq.}{Eqs.}
\crefname{section}{Sec.}{Secs.}
\crefname{subsection}{Sec.}{Secs.}
\crefname{appendix}{Appendix}{Appendices}
\crefname{figure}{Figure}{Figures}
\crefname{table}{Table}{Tables}
\crefname{result}{Result}{Results}
\crefname{algorithm}{Algorithm}{Algorithms}

\newcommand{\methodsabove}[1]{\hyperref[#1]{above}\xspace}
\newcommand{\methodsbelow}[1]{\hyperref[#1]{below}\xspace}

\newcommand{\ie}{\textit{i}.\textit{e}.}
\newcommand{\eg}{\textit{e}.\textit{g}.}

\definecolor{daxcolor}{rgb}{0.8, 0.3, 0.4}

\usepackage{orcidlink}

\usepackage{xpatch}
\xpatchcmd\bibsection{19}{10}{}{}
\xpatchcmd\bibsection{\begingroup}{\vskip-0pt\begingroup}{}{}

\newcounter{algorithm}

\begin{document}

\preprint{}
\title{Readout Error Mitigation for Mid-Circuit Measurements and Feedforward}

\author{Jin Ming Koh\,\orcidlink{0000-0002-6130-5591}}
\affiliation{Department of Physics, Harvard University, Cambridge, Massachusetts 02138, USA}
\affiliation{A*STAR Quantum Innovation Centre (Q.InC), Institute of High Performance Computing (IHPC), \\ Agency for Science, Technology and Research (A*STAR), 1 Fusionopolis Way, \#16-16 Connexis, Singapore 138632, Republic of Singapore\looseness=-1}

\author{Dax Enshan Koh\,\orcidlink{0000-0002-8968-591X}}
\affiliation{A*STAR Quantum Innovation Centre (Q.InC), Institute of High Performance Computing (IHPC), \\ Agency for Science, Technology and Research (A*STAR), 1 Fusionopolis Way, \#16-16 Connexis, Singapore 138632, Republic of Singapore\looseness=-1}

\author{Jayne Thompson\,\orcidlink{0000-0002-3746-244X}}
\affiliation{A*STAR Quantum Innovation Centre (Q.InC), Institute of High Performance Computing (IHPC), \\ Agency for Science, Technology and Research (A*STAR), 1 Fusionopolis Way, \#16-16 Connexis, Singapore 138632, Republic of Singapore\looseness=-1}

\begin{abstract}
Current quantum computing platforms suffer from readout errors, where faulty measurement outcomes are reported by the device. These errors are particularly harmful in quantum programs that rely on branch statements, where operations in the second half of the program are dynamically determined by mid-circuit measurements. We propose a general protocol for mitigating mid-circuit measurement errors, offering an efficient solution that works for any number of feedforward layers without increasing circuit depth or two-qubit gate counts, making it highly suitable for noisy intermediate-scale quantum (NISQ) devices. Our method demonstrates up to a ${\sim} 60\%$ reduction in error on superconducting quantum processors across several practically relevant feedforward circuits, including dynamic qubit resets, shallow-depth GHZ state preparation, and multi-stage quantum teleportation. This work paves the way for more resilient adaptive quantum circuits, crucial for both current and future quantum computing applications.
\end{abstract}

\maketitle
\date{\today}

In quantum computation, mid-circuit measurements and feedforward represent the analogue of classical \texttt{if}-\texttt{else} branching statements in code. They allow us to make measurements on certain qubits during computation and decide which subsequent quantum operations to apply depending on the measurement outcomes. This capability is of immediate relevance in the present noisy intermediate-scale quantum (NISQ) era---enabling ancillary qubits to be reset mid-circuit so they can be reused during computation~\cite{decross2023qubit} and driving resource-saving protocols such as circuit knitting and stitching~\cite{piveteau2024circuit, vazquez2024scaling}. It also underpins quantum error correction~\cite{gottesman2010introduction, google2023suppressing, sivak2023real, sundaresan2023demonstrating}, which functions by way of repeated syndrome measurements and conditioned recovery operations. Indeed, such operations underlie many iconic quantum primitives and algorithms across diverse domains (\eg~resource distillation and injection~\cite{knill2005quantum, bartolucci2023fusion, gupta2024encoding}, state and gate teleportation~\cite{pirandola2015advances, wan2019quantum, hu2023progress}, quantum simulation~\cite{rost2021demonstrating, han2021experimental, del2024robust}, optimization algorithms~\cite{decross2023qubit}, and efficient state preparation~\cite{baumer2023efficient, zhu2023nishimori, tantivasadakarn2023shortest, fossfeig2023experimental, smith2023deterministic, iqbal2024non}).

However, this technology has an Achilles heel: it is highly vulnerable to readout errors, where a quantum device reports incorrect measurement outcomes~\cite{cai2023quantum, lin2021independent}. In a program with mid-circuit measurements and feedforward, such errors in the measurement record induce the application of incorrect feedforward operations (\cref{fig:readout-errors/illustration}), thus initiating a cascade wherein the intermediary state of the unmeasured qubits is subsequently fed into an incorrect branch of the program. Existing readout error-mitigation techniques assume all measurements happen at the end of the circuit and are only concerned with retrieving correct conglomerate measurement statistics through classical post-processing~\cite{kandala2017hardware, kandala2019error, jurcevic2021demonstration}---they are not designed to handle this type of branching error occurring in the middle of computation.

In this work, we present \textit{probabilistic readout error mitigation} (\texttt{PROM}), a general readout error mitigation method for quantum circuits containing mid-circuit measurements and feedforward. Our protocol works by modifying the feedforward operations---we probabilistically sample from an engineered ensemble of feedforward trajectories, and then perform an averaging process in post-processing. The method accommodates multiple layers of mid-circuit measurements and feedforward. It incurs no increase in circuit depth nor additional two-qubit gate count cost. We provide analyses on resource requirements in terms of the number of extra shots of the circuit required, sensitivity to miscalibration of the device, and we experimentally demonstrate the effectiveness of the protocol on superconducting quantum processors in several settings of practical interest, namely qubit resets, shallow-depth GHZ state preparation, and multi-stage quantum state teleportation.

Our method integrates directly with probabilistic error cancellation~\cite{van2023probabilistic, gupta2023probabilistic}, which neglects readout errors, for complementary mitigation of quantum errors on dynamic circuits. More broadly, any quantum error suppression or mitigation technique (\eg~see \cite{li2017efficient, temme2017error, giurgica2020digital, czarnik2021error, liao2023machine, wallman2016noise, hashim2021randomized, sun2024quantum}) can be used in conjunction for a comprehensive targeting of hardware noise.

\section{Framework}
\label{sec:framework}

Our set-up begins with the fundamental rationale of quantum error mitigation---that in most quantum computational settings, what one ultimately cares about is the measurement outcome statistics produced by the program. That is, many quantum algorithms involve (i) enacting a quantum process $\mathcal{P}$ on $n$ qubits to produce some state $\rho_{\mathrm{out}}$, (ii) measuring a collection of observables $\vb{O} = \mqty[O_1 & O_2 & \ldots & O_B]$ of bounded spectra on $\rho_{\mathrm{out}}$, and (iii) repeating this process over $N$ trials to obtain estimates of the expectation values $\smash{\expval{\overline{\vb{O}}}}$ for some sufficiently large $N$. In such settings, the computation is considered successful provided one can estimate $\langle \overline{O}_b\rangle$ for every $b \in [B]$ to a designated accuracy $\epsilon$ with some tolerable error probability $\delta$. Quantum error mitigation describes a family of protocols to successfully compute in NISQ settings where our capacity to precisely implement $\mathcal{P}$ is compromised, at the expense of requiring a larger number $N' > N$ of trials. 

In our context, we focus on $\mathcal{P}$ enacted through \emph{mid-circuit measurements} and \emph{feedforward}. For simplicity, we begin with the setting of a single layer of feedforward. Such a process involves
\begin{enumerate}
    \item Application of a unitary quantum circuit resulting in an intermediary quantum state $\rho$.
    \item Measurement of a subset of $m \leq n$ qubits in the computational basis\footnote{Mid-circuit measurements in other bases can be absorbed into the state preparation of $\rho$ and the feed-forward operations; thus this computational basis assumption can be made without loss of generality.} resulting in a measurement outcome (bitstring) $s \in \mathcal{S} = \{0, 1\}^{\otimes m}$.
    \label{item:framework/process/measurement}
    \item Application of an $s$-dependent unitary operation $V_s$ as dictated by a feedforward function that designates one unitary to apply for each $s$. 
    \label{item:framework/process/feedforward}
\end{enumerate}

We refer to step \ref{item:framework/process/measurement} as the mid-circuit measurements and step \ref{item:framework/process/feedforward} as the feedforward operation. 

\begin{figure}[!t]
    \centering
    \includegraphics[width=\linewidth]{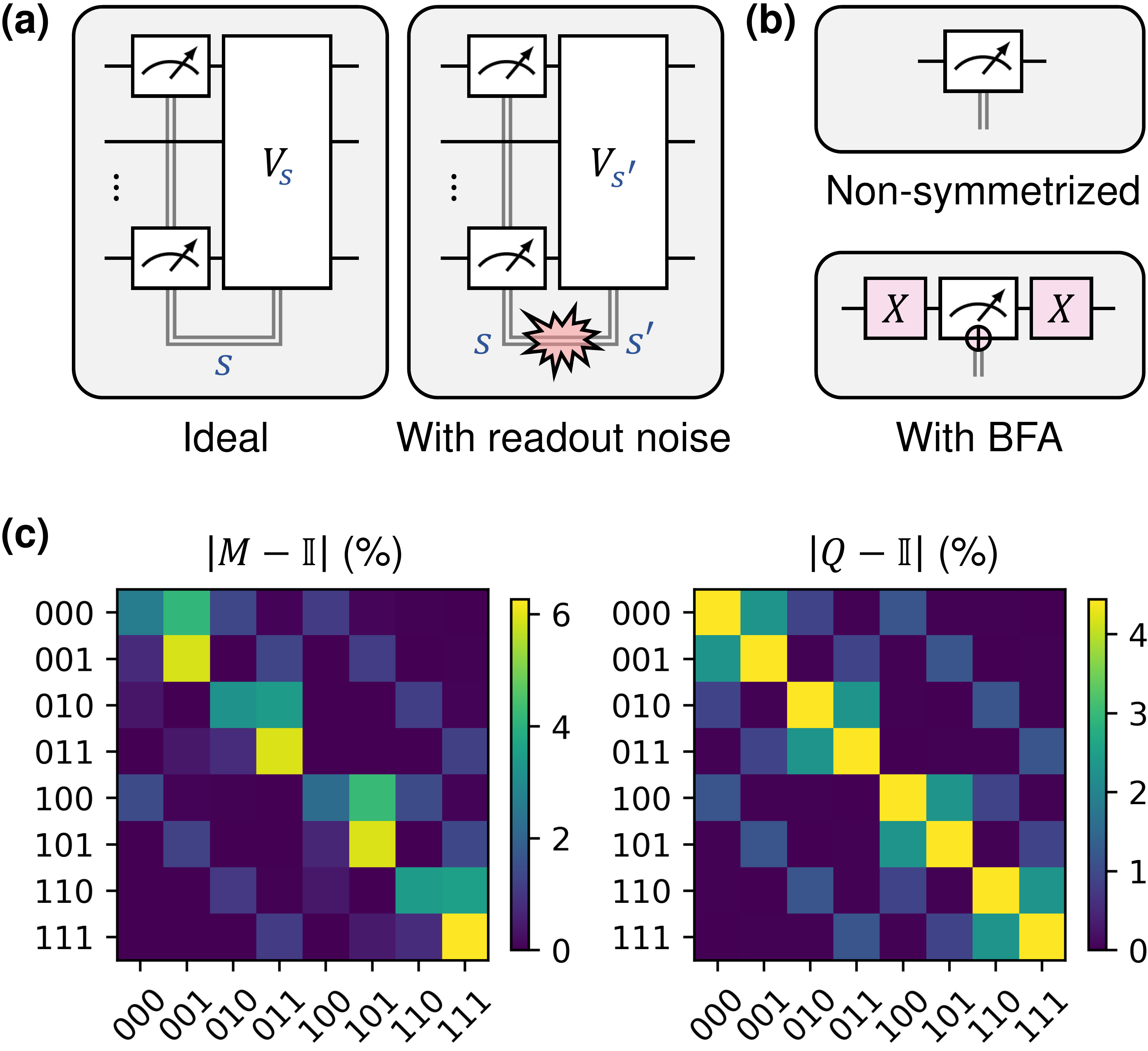}
    \phantomsubfloat{\label{fig:readout-errors/illustration}}
    \phantomsubfloat{\label{fig:readout-errors/bfa}}
    \phantomsubfloat{\label{fig:readout-errors/matrices}}
    \vspace{-0.5\baselineskip}
    \caption{\textbf{Readout errors.} \textbf{(a)} In an ideal noiseless setting, a layer of mid-circuit measurements produce an outcome (bitstring) $s$, which determines a subsequent feedforward quantum operation $V_s$. But in practice, readout errors can cause a measurement outcome $s'$ to be reported instead, resulting in an incorrect operation $V_{s'}$ applied. Such errors occur on a per-shot basis and their effect cannot be corrected in post-processing. \textbf{(b)} Bit-flip averaging (BFA) involves randomly sandwiching each measurement between a pair of inserted $X$ gates (shaded pink) and applying a classical bit-flip to the outcome of the measurement. This symmetrizes the readout error channel~\cite{smith2021qubit, hicks2021readout}. \textbf{(c)} Without symmetrization, readout error channels are characterized by a confusion matrix $M$ recording the probabilities of mis-reporting measurement outcomes (see \cref{sec:framework}). Under BFA, the confusion matrix $M$ is symmetrized into $Q$, which is completely defined by its first column $\vb{q}$ with entries describing readout error syndrome probabilities (see \cref{sec:theory-single/main-results}). The matrices displayed here are calibrated from superconducting quantum device \textit{ibmq\_kolkata}.}
    \label{fig:readout-errors}
\end{figure}

In this work, we are concerned with \emph{readout errors} occurring during mid-circuit measurements (see \cref{fig:readout-errors/illustration}), wherein the measurement apparatus reports a different outcome $s'\in \mathcal{S}$ when the true outcome is $s \in \mathcal{S}$ with some confusion probability $M_{s's}$. This mis-reporting of measurement outcome causes two correlated sources of error: \begin{itemize}
    \item While the true measurement outcome is $s$ and the quantum state after measurement is collapsed to $\rho_s \propto \Pi_s \rho \Pi_s$
    for $\Pi_s$ the projector associated with outcome $s$, the measurement outcome reported by the device is $s'$. 
    \item The incorrect feedforward operation $V_{s'}$ is applied in place of $V_s$. 
\end{itemize} 

Such errors break the deterministic causal relation between the measurement outcome $s$ and the feedforward operation $V_{s}$, resulting in an impossible combination of operations: $V_{s'}$ acting on states $\rho_s$. The quantum state of the qubits is fed into an incorrect conditional branch of the circuit. Ultimately, this mangles the measured expectation values $\langle \vb{O} \rangle$ such that they differ from the ideal values $\langle \overline{\vb{O}} \rangle$. We aim to develop techniques that mitigate the effects of these mid-circuit readout errors, so that the measured $\langle \vb{O} \rangle$ accurately estimates $\langle \overline{\vb{O}} \rangle$.

Existing readout error mitigation (REM) techniques are concerned only with mitigating readout errors for terminal measurements occurring at the end of the circuit~\cite{kandala2017hardware, kandala2019error, jurcevic2021demonstration}. Formally, suppose $\vb{c}$ is a vector of counts whose element $c_s$ corresponds to the number of times we observe outcome $s$ over $N$ shots of the circuit, and let $\vb{c}'$ be the hypothetical ideal vector of counts were there no readout errors. Then for $N \gg 1$, we have the relation $\vb{c} \approx M \vb{c}'$ where $M$ is referred to as the confusion matrix~\cite{chow2010detecting} (see \cref{fig:readout-errors/matrices}). Various REM techniques aim to estimate $\vb{c}'$ when given $\vb{c}$, assuming $M$ has been calibrated through device benchmarking\footnote{Though certain recent protocols consider cases where our knowledge of $M$ is partial, or indirectly constructs the data in $M$ to reduce computational costs~\cite{bravyi2021mitigating, barron2020measurement, nation2021scalable}.}.

Applying terminal REM schemes to our context of mid-circuit measurements with feedforward encounters immediate problems. Approximating the noiseless counts of the terminal measurements and thereby $\langle \vb{O} \rangle$ in the fashion above, without attention to the erroneous branching of the program, cannot lead to a correct estimation of the true $\langle \overline{\vb{O}} \rangle$ regardless of the number of shots we take. At the same time, correcting the erroneous combinations of $V_{s'}$ applied on $\rho_s$ naïvely requires knowledge of the true measurement outcomes $s$ at the single-shot level, which is inaccessible information. Fundamentally, \emph{standard readout error mitigation do not work for mid-circuit measurement errors}. 

We must therefore go beyond terminal readout error mitigation to develop a paradigm that functions for mid-circuit measurements and feedforward. Naturally, such a paradigm subsumes terminal measurements as a special case. In what follows, we present such a class of methods. \Cref{sec:theory-single} begins with a general solution for the context introduced above involving a single layer of mid-circuit measurements and feedforward. We then extend our solution to arbitrarily many layers in \cref{sec:theory-multiple}.

\section{Core Protocol}
\label{sec:theory-single}

\subsection{Main results}
\label{sec:theory-single/main-results}

\begin{figure*}[!t]
    \centering
    \includegraphics[width=0.75\linewidth]{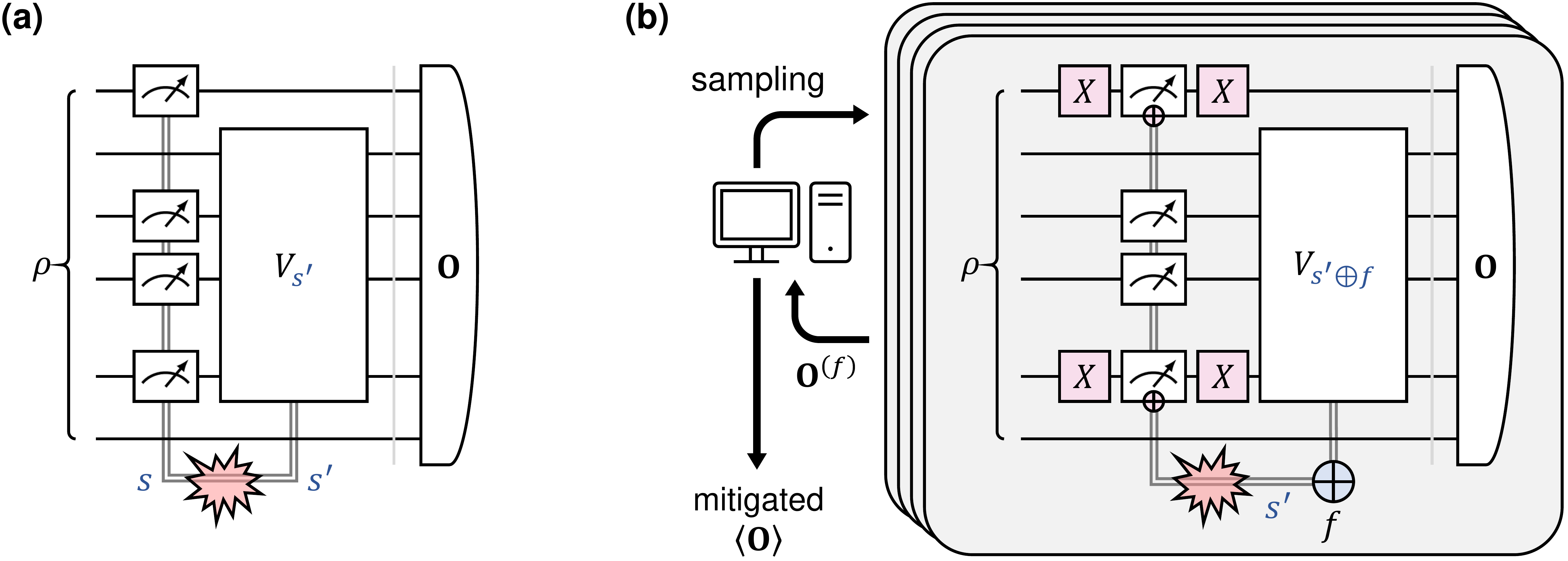}
    \phantomsubfloat{\label{fig:theory-single/input-circuit}}
    \phantomsubfloat{\label{fig:theory-single/schematic}}
    \vspace{-0.5\baselineskip}
    \caption{\textbf{Probabilistic readout error mitigation for a single feedforward layer.} \textbf{(a)} Structure of the quantum circuit considered, comprising state preparation, a layer of mid-circuit measurements and feedforward, and terminal measurements of observables $\vb{O}$. From shot to shot, the correct measurement outcome (bitstring) $s$ is corrupted into $s'$ by readout errors, and the reported outcome $s'$ conditions the feedforward unitary $V_{s'}$ that is applied. \textbf{(b)} Schematic of our readout error mitigation protocol. Bit-flip averaging is applied by randomly twirling measurements with quantum and classical $X$ gates (shaded pink), and the XOR between the reported measurement outcome and a bitmask $f$ is used for feedforward. By sampling $f$ over a suitable probability distribution and with some classical post-processing, mitigated expectation values of $\vb{O}$ are produced.}
    \label{fig:theory-single}
\end{figure*}

We first begin by leveraging bit-flip averaging~\cite{smith2021qubit, hicks2021readout}. This involves sandwiching each measurement between a pair of Pauli-$X$ gates with probability $1 / 2$, while simultaneously flipping the outcome (see \cref{fig:readout-errors/bfa}). This symmetrizes readout errors in our system, such that the probability of any error occurring is rendered independent of the input state. The readout error channels are then completely characterized by a $2^m$-element probability vector $\vb{q}$ in place of the more cumbersome $2^m \times 2^m$ confusion matrix $M$ (see \cref{fig:readout-errors/matrices}). Each element $q_s$ represents the probability that error syndrome $s$ occurs---\ie~if the true measurement outcome is $a \in \mathcal{S}$, then $q_s$ is the probability of reporting $(a \oplus s)$, where $\oplus$ denotes the modulo-$2$ addition, or XOR, of two bitstrings. Note that, other than enabling a more succinct description of the readout error channels, this process does not affect the operational behavior of the mid-circuit measurements in any way.

Our key idea is then to intervene in the feedforward operations on a shot-by-shot basis. We facilitate this by introducing a bitmask $f \in \mathcal{S}$ in each shot to be applied to the outcome of the $m$ qubits we measure mid-circuit. When the mid-circuit measurements report an outcome $s' \in \mathcal{S}$, we invoke the feedforward operation $V_{s' \oplus f}$ instead of $V_{s'}$. Our first key result is that there exists a systematic procedure to choose a distribution for $f$ such that the effect of readout errors on the measured $\langle \vb{O} \rangle$ is mitigated.

Specifically, to determine the bitmask to pick in each shot of $\mathcal{P}$, we introduce a vector of weights $\smash{\vb*{\alpha} = \trans{\smash{\mqty[\alpha_{\vb{0}} & \ldots & \alpha_{\vb{1}}]}}}$ with norm $\xi = \smash{\norm*{\vb*{\alpha}}_1} = \smash{\sum_{f \in \mathcal{S}}} \abs{\alpha_f}$. These weights are not necessarily non-negative, and the normalized vector $\vb*{\alpha} / \xi$ can be regarded as a quasiprobability distribution. The weights $\vb*{\alpha}$ can be systematically computed as a function of our readout error model (see below). We are now equipped to introduce our probabilistic readout error mitigation (\texttt{PROM}) protocol. 

\begin{tcolorbox}[boxsep=8pt,
    left=0pt,right=0pt,top=0pt,bottom=0pt]
    \refstepcounter{algorithm}
    \label{alg:prom-single-layer}
    
    \textbf{Algorithm \thealgorithm}: \textit{Probabilistic readout error mitigation (\texttt{PROM}) for a single feedforward layer}. \\[-0.5\baselineskip]
    
    \textbf{Inputs:} Circuit $\mathcal{P}$ containing a layer of mid-circuit measurements and feedforward (see \cref{fig:theory-single/input-circuit}). Probability vector $\vb{q}$ characterizing readout error channels affecting mid-circuit measurements. Arbitrary measured observables $\vb{O}$. Number of shots $N'$ to be executed. \\[-0.5\baselineskip]

    \textbf{Output:} Estimate $\smash{\expval*{\widehat{\vb{O}}}}$ of expectation values. \\[-0.5\baselineskip]

    We construct a bit-flip-averaged (see \cref{fig:theory-single/schematic}) version of $\mathcal{P}$ and an appropriate weight vector $\vb*{\alpha}$ from $\vb{q}$ (see \cref{result:single/correctness}). For shot $k = 1$ to $N'$, we 
    \begin{enumerate}
        \item Sample a bitmask $f \in \mathcal{S}$ with probability $\abs*{\alpha_f} / \xi$.
        \item Execute the mid-circuit measurements with bitmask $f$ applied to the feedforward. 
        \item Measure $\vb{O}$ to obtain outcome $\vb{m}^{(k)}$. Accordingly each entry $m^{(k)}_b$ is an eigenvalue of $O_b$.
        \item Record $\smash{\vb{o}^{(k)} = \vb{m}^{(k)} \sgn(\alpha_f)}$.
    \end{enumerate}

    Output the estimate $\smash{\expval*{\widehat{\vb{O}}}} = \xi \cdot \frac{1}{N'} \sum_{k = 1}^{N'} \vb{o}^{(k)}$.
\end{tcolorbox}

Our key observation is that $\smash{\vb*{\alpha}}$ can be systematically chosen such that the above protocol produces an unbiased estimator $\smash{\widehat{\vb{O}}}$ for the observables $\vb{O}$.

\begin{result}[Correctness] 
    \label{result:single/correctness}
    Consider \cref{alg:prom-single-layer} with readout error channels characterized by $\vb{q}$. If we choose
    \begin{equation}\begin{split}
        \vb*{\alpha} 
        = \frac{1}{2^m} \mathcal{W} \left(
            \frac{\vb{1}}{\mathcal{W}(\vb{q})} \right),
    \end{split}\end{equation}
    where $\mathcal{W}(\cdot)$ denotes the unnormalized Walsh-Hadamard transform, then the $\smash{\expval*{\widehat{\vb{O}}}}$ obtained from the protocol is an unbiased estimate of the ideal noiseless observable expectation values $\smash{\expval*{\overline{\vb{O}}}}$. That is, $\smash{\expval*{\widehat{\vb{O}}} = \expval*{\overline{\vb{O}}}}$ in the limit of a large number of shots.
\end{result}

We give the intuition behind this result in \cref{sec:theory-single/general} and proof in \cref{app-sec:theory/single/solution-general}. In practice, standard numerical algorithms~\cite{ahmed1975walsh, beer1981walsh} can be used to efficiently compute the Walsh-Hadamard transforms required above.

As \cref{alg:prom-single-layer} modifies the input quantum circuit only through introducing the feedforward bitmasks $f$, no two-qubit gate count nor circuit depth costs are incurred. In addition to being unbiased, statistical analysis shows that this error mitigation protocol carries a sampling overhead of $\xi^2$ compared to a hypothetical noiseless experiment. This is made precise by the following result.

\begin{result}[Convergence]
    \label{result:single/convergence}
    Using \cref{alg:prom-single-layer} with $\vb*{\alpha}$ chosen for an unbiased estimator (as in \cref{result:single/correctness}), to obtain the same precision of measured $\expval{\vb{O}}$ at the same confidence as $N$ hypothetical noiseless shots on the input circuit $\mathcal{P}$, a number of noisy shots scaling as $N' = \xi^2 N$ is sufficient.
\end{result}

In fact, under general conditions, for a total readout error probability $\eta = 1 - q_{\vb{0}} < 1/2$, the overhead factor can be shown to be bounded by $\xi < 1 / (1 - 2 \eta)$. In the noiseless limit $\alpha_{\vb{0}} = \delta_{f \vb{0}}$ and $\xi = 1$. We refer readers to \cref{app-sec:theory/sampling-overhead} for the analysis underlying these results.

Lastly we remark on the computational resource costs of the error mitigation protocol. In this general setting there are ${\sim} 2^m$ independent entries of $\vb*{\alpha}$ to be computed, which demand classical resources scaling exponentially in $m$. But, thereafter, using standard algorithms~\cite{marsaglia2004fast, vose1991linear, walker1977efficient} to sample from the computed probability distribution, the drawing of bitmasks $f \in \mathcal{S}$ can be performed in constant time per shot. These complexities are summarized in \cref{tab:summary}, with further details elaborated in \cref{app-sec:theory/single/solution-general}. By taking advantage of known structure in the readout error channels, the exponential classical cost in $m$ can be removed, as we describe in \cref{sec:theory-single/simplified}.

\begin{table*}[!t]
    \centering
    \begin{tabular}{@{} p{1.5cm} p{5.5cm} x{2cm} x{2cm} x{2.5cm} x{3.5cm} @{}}
        \toprule 
        Layers
            & Assumptions
            & \multicolumn{3}{c}{Classical Resources}
            & Overhead $\xi^2$ \\
        \cmidrule(lr){3-5}
        {} 
            & 
            & Init.~Time 
            & Init.~Space 
            & Sampling Time \\
        \midrule 
        \multirow{4}{*}{$1$} 
            & -
            & $\order{m \cdot 2^m}$ 
            & $\order{2^m}$ 
            & $\order{1}$ 
            & $\norm{\vb*{\alpha}}_1^2$ \\
        {}
            & Independent errors
            & $\order{m}$ 
            & $\order{m}$ 
            & $\order{m}$ 
            & $\prod_{\ell = 1}^m (1 - 2 r_\ell)^{-2}$ \\
        {}
            & Uniform error
            & $\order{1}$ 
            & $\order{1}$ 
            & $\order{m}$ 
            & $(1 - 2 r)^{-2m}$ \\
        \midrule 
        \multirow{5}{*}{$L$} 
            & -
            & $\order{m \cdot 2^m}$ 
            & $\order{2^m}$ 
            & $\order{1}$
            & $\norm{\vb*{\alpha}}_1^2$ \\
        {}
            & Independent errors between layers
            & $\order{m \cdot 2^{\overline{m}}}$ 
            & $\order{L \cdot 2^{\overline{m}}}$ 
            & $\order{L}$
            & $\prod_{l = 1}^L (\xi^{[l]})^2$ \\
        {}
            & Fully independent errors
            & $\order{m}$ 
            & $\order{m}$ 
            & $\order{m}$
            & $\prod_{\ell = 1}^m (1 - 2 r_\ell)^{-2}$ \\
        {}
            & Uniform error
            & $\order{1}$ 
            & $\order{1}$ 
            & $\order{m}$
            & $(1 - 2 r)^{-2m}$ \\
        \bottomrule
    \end{tabular}
    \caption{Summary of characteristics of the probabilistic readout error mitigation protocol (\texttt{PROM}) studied in this paper. Above $m$ is the total number of mid-circuit measurements and $\overline{m}$ is the maximum number of measurements in a feedforward layer. Under independence assumptions, $r_\ell \in [0, 1/2)$ are characteristic readout error probabilities for measurement $\ell \in [m]$, and under uniformity assumptions $r_\ell = r$ identically. The `Init.~Time' and `Init.~Space' columns refer to the initial classical time and memory costs of computing coefficients for the mitigated estimator, `Sampling Time' refers to the classical sampling time to generate a bitmask per experiment shot (see \cref{sec:theory-single}), and `Overhead' refers to a multiplicative factor in the number of experiment shots required for the mitigated estimator to converge to the same precision at the same confidence as hypothetical noiseless circuit execution. The single-layer variants are special cases of the multi-layer variants.}
    \label{tab:summary}
\end{table*}

\subsection{Underlying principles}
\label{sec:theory-single/general}

We now outline the underlying mechanism behind our protocol. We start from the quantum state after the mid-circuit measurements and feedforward in the absence of readout errors, which can be written as
\begin{equation}\begin{split}
    \overline{\sigma} = \sum_{s \in \mathcal{S}} p_s V_s \rho_s V_s^\dag,
    \label{eq:theory-single/general/state-ideal}
\end{split}\end{equation}
where the measurement outcome $s$ defines the quantum trajectory that the system has taken, and results in a normalized quantum state $\smash{\rho_s = (\Pi_s \rho \Pi_s) / p_s}$ when $p_s \neq 0$. Here $\Pi_s$ is the measurement projector associated with the outcome $s$, and the probability of the trajectory occurring is $\smash{p_s = \tr(\Pi_s \rho)}$. We define the tensor
\begin{equation}\begin{split}
    T_{b s s'}
    = p_s \tr \left( O_b V_{s'} \rho_s V_{s'}^\dag \right),
    \label{eq:theory-single/general/T-definition}
\end{split}\end{equation}
such that the entry $T_{b s s'}$ is the expectation value of observable $O_b$ measured on quantum trajectories that had measurement outcome $s$ but feedforward unitary $\smash{V_{s'}}$ applied, weighted by the probability $p_s$ of those trajectories occurring. Thus, the diagonal and off-diagonal entries of the matrix $T_b$ describe trajectories in which feedforward is correctly and incorrectly performed, respectively. Then
\begin{equation}\begin{split}
    \expval{\overline{O}_b} 
    = \tr \left(O_b \overline{\sigma}\right)
    = \tr T_b.
    \label{eq:theory-single/general/expval-O-ideal}
\end{split}\end{equation}

Now we consider the presence of readout errors. On a per-shot basis, a true mid-circuit measurement outcome $s$ is corrupted into $s'$, and in addition we impose a bitmask $f \in \mathcal{S}$ such that feedforward unitary $V_{s' \oplus f}$ is applied, in accordance to the \texttt{PROM} protocol. The quantum state after the mid-circuit measurements and feedforward is
\begin{equation}\begin{split}
    \sigma^{(f)}
    = \sum_{s \in \mathcal{S}} \sum_{s' \in \mathcal{S}} 
        q_{s \oplus s'} p_s 
        V_{s' \oplus f} \rho_s V_{s' \oplus f}^\dag,
    \label{eq:theory-single/general/state-noisy}
\end{split}\end{equation}
where $q_{s \oplus s'}$ is the probability of the readout error occurring (as described in \cref{sec:theory-single/main-results}). Accordingly
\begin{equation}\begin{split}
    \expval{O_b^{(f)}}
    = \tr \left( O_b \sigma^{(f)} \right)
    &= \sum_{s \in \mathcal{S}} \sum_{s' \in \mathcal{S}} 
        q_{s \oplus s'} T_{b s (s' \oplus f)}.
    \label{eq:theory-single/general/expval-O-f}
\end{split}\end{equation}

Desiring an unbiased estimator $\smash{\expval*{\widehat{\vb{O}}} = \expval*{\overline{\vb{O}}}}$, we substitute \cref{eq:theory-single/general/expval-O-ideal} and \cref{eq:theory-single/general/expval-O-f} into
\begin{equation}\begin{split}
   \langle \widehat{\vb{O}} \rangle
    \coloneqq \sum_{f \in \mathcal{S}} \alpha_f \langle \vb{O}^{(f)}\rangle
    = \xi \sum_{f \in \mathcal{S}} 
        \frac{\abs{\alpha_f}}{\xi} \sgn(\alpha_f) \langle \vb{O}^{(f)}\rangle.
    \label{eq:theory-single/protocol/estimator-definition}
\end{split}\end{equation}

This results in an equality enforced for every $b \in [B]$. We do not assume further knowledge of the problem at hand; thus, the tensor $T$ is unknown and the mitigation protocol must work for arbitrary $T$. Then the equality necessitates that the coefficients of every $T_{bss'}$ match on both sides. This yields a linear system encoding a unique $\vb*{\alpha}$, with closed-form solution
\begin{equation}\begin{split}
    \alpha_f
    = \frac{1}{2^m} \sum_{k \in \mathcal{S}} 
        (-1)^{f \cdot k} \left(\frac{1}{\lambda_k}\right),
    \label{eq:theory-single/general/alpha-solution}
\end{split}\end{equation}
where $\vb*{\lambda}$ are the eigenvalues of the symmetrized confusion matrix $Q$ defined by $Q_{sf} = q_{s \oplus f}$ for $s, f \in \mathcal{S}$ (see \cref{fig:readout-errors/matrices}). Note that $Q$ and every of its recursively halved quadrants are symmetric about their diagonals and anti-diagonals. Thus $Q \in \mathrm{span}(\{\mathbb{I}, X\}^{\otimes m})$ and has closed-form eigenvalues
\begin{equation}\begin{split}
    \lambda_k 
    = \sum_{s \in \mathcal{S}} (-1)^{k \cdot s} q_s
    = 1 - 2 \sum_{\substack{s \in \mathcal{S} \\ k \cdot s \neq 0}} q_s,
    \label{eq:readout-errors/lambda-solution}
\end{split}\end{equation}
for each $k \in \mathcal{S}$---see \cref{app-sec:preliminaries/Q-diagonalization} for technical elaboration. Equivalently we can write $\vb*{\lambda} = \mathcal{W}(\vb{q})$ where $\mathcal{W}$ is the unnormalized Walsh-Hadamard transform~\cite{ahmed1975walsh, beer1981walsh} and similarly $\vb*{\alpha} = \mathcal{W}(\vb{1} / \vb*{\lambda}) / 2^m$, as presented in \cref{result:single/correctness}.

A full derivation of the algorithm, and alternate forms for $\vb*{\alpha}$, are given in \cref{app-sec:theory/single/solution-general}.

\subsection{Robustness to calibration errors}
\label{sec:theory-single/robustness}

A pertinent further question concerns the sensitivity or well-conditionedness of the error mitigation protocol. That is, given differing $\vb{q}$ and $\vb{q}'$ inputs characterizing the readout errors present, how large are the differences in the mitigated expectation values $\smash{\expval*{\widehat{\vb{O}}'}}$ and $\smash{\expval*{\widehat{\vb{O}}}}$, and secondarily the overhead factors $\xi$ and $\xi'$? This question is relevant as in reality, calibration of $\vb{q}$ on a quantum device inevitably carries small errors up to a finite experimental precision; moreover, it may be desirable to employ approximations on $\vb{q}$, for example as described next in \cref{sec:theory-single/simplified}. We show that small errors in inputs propagate mildly through the protocol. 

\begin{result}[Robustness to calibration errors]
    \label{result:single/robustness}
    Given two different characterizations of readout errors $\vb{q}$ and $\vb{q}'$ distinguished by total variation distance $d$, the respective mitigated expectation values $\smash{\expval*{\widehat{\vb{O}}}}$ and $\smash{\expval*{\widehat{\vb{O}}'}}$ produced by \cref{alg:prom-single-layer} and the overhead factors $\xi$ and $\xi'$ satisfy
    \begin{subequations}\begin{align}
        \abs{ \expval{\widehat{O}'_b} - \expval{\widehat{O}_b} }
        &\leq \frac{2 \xi^2 d}{1 - 2 \xi d}
            \cdot \norm{O_b}_2,
        \label{eq:theory-single/general/sensitivity-bounds-expval}
        \\
        \abs{ \xi' - \xi }
        &\leq \frac{2 \xi^2 d}{1 - 2 \xi d},
        \label{eq:theory-single/general/sensitivity-bounds-xi}
    \end{align}\end{subequations}
    for $2 \xi d < 1$ and every $b \in [B]$, where $\norm{\cdot}_2$ denotes the spectral norm.
\end{result}

In particular, invoking \cref{result:single/robustness} and taking $\vb{q}'$ and $\vb{q}$ as the (unknown) exact characterization and the characterization used in the error mitigation protocol respectively, the error in mitigated observables scale linearly as ${\sim} 2 \xi^2 d$ to leading order where $d$ is small given sufficiently good calibration of readout error probabilities on the device.

Secondarily, suppose that the characterized readout errors $\vb{q}$ on the quantum device contain small correlations, but the errors are otherwise close to being independent. Then \cref{eq:theory-single/general/sensitivity-bounds-xi} implies that the overhead factor of the mitigation protocol is close to that assuming independent errors, which is easy to evaluate, as analyzed next in \cref{sec:theory-single/simplified}. We refer readers to \cref{app-sec:theory/sensitivity-analysis} for derivations of the sensitivity bounds above.

\begin{figure*}[!t]
    \centering
    \includegraphics[width=\linewidth]{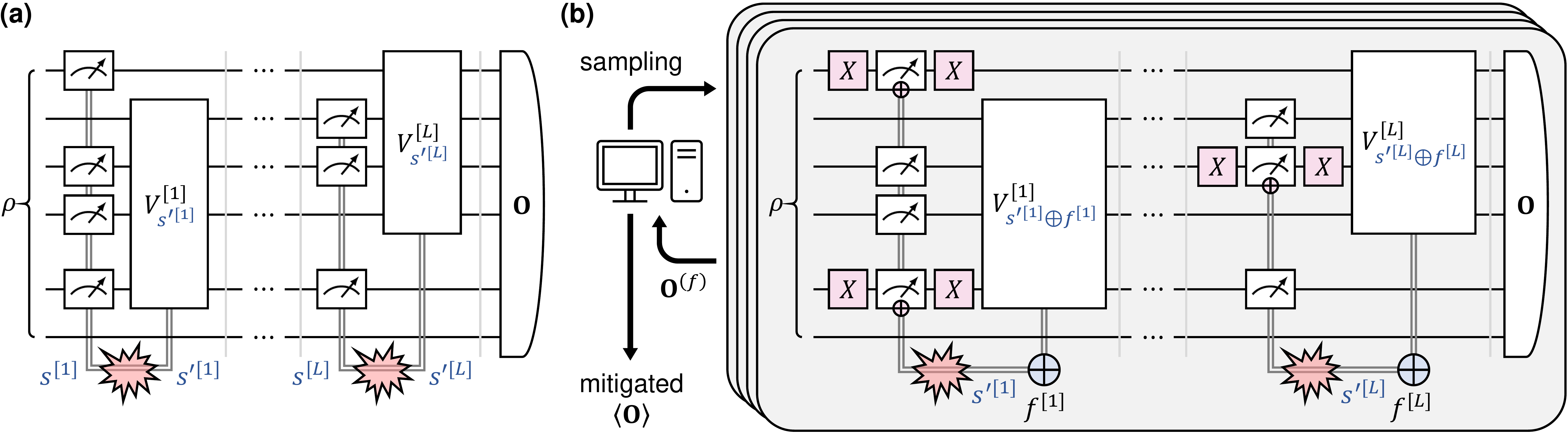}
    \phantomsubfloat{\label{fig:theory-multiple/input-circuit}}
    \phantomsubfloat{\label{fig:theory-multiple/schematic}}
    \vspace{-0.5\baselineskip}
    \caption{\textbf{Probabilistic readout error mitigation for multiple feedforward layers.} \textbf{(a)} Structure of the quantum circuit considered, comprising state preparation, $L$ layers of mid-circuit measurements and feedforward, and terminal measurements of observables $\vb{O}$. From shot to shot, in layer $l$, the correct measurement outcome (bitstring) $\smash{s^{[l]}}$ is corrupted into $\smash{s'^{[l]}}$ by readout errors, and the reported outcome $\smash{s'^{[l]}}$ determines a feedforward unitary $\smash{V^{[l]}_{s'^{[l]}}}$ that is applied. \textbf{(b)} Schematic of our readout error mitigation protocol. Bit-flip averaging is applied by randomly twirling measurements with quantum and classical $X$ gates (shaded pink), and in layer $l$ the XOR between the reported measurement outcome and a bitmask $\smash{f^{[l]}}$ is used for feedforward. By sampling $\smash{f = f^{[1]} f^{[2]} \ldots f^{[L]}}$ over a suitable probability distribution and with some classical post-processing, mitigated expectation values of $\vb{O}$ are produced.}
    \label{fig:theory-multiple}
\end{figure*}

\subsection{Simplified solutions for uncorrelated channels}
\label{sec:theory-single/simplified}

In a setting where readout errors affecting the $m$ mid-circuit measurements are assumed to be independent, the solution for $\vb*{\alpha}$ factorizes,
\begin{equation}\begin{split}
    \vb*{\alpha} 
    = \bigotimes_{\ell = 1}^m \vb*{\alpha}^{(\ell)},
    \qquad
    \vb*{\alpha}^{(\ell)}
    = \frac{1}{1 - 2 r_\ell} \mqty[1 - r_\ell \\ -r_\ell],
    \label{eq:theory-single/simplified/tensored-alpha-solution}
\end{split}\end{equation}
where $r_\ell \in [0, 1 / 2)$ is the readout error probability for the $\ell^\text{th}$ measurement. Writing $\smash{\xi^{(\ell)} = \norm*{\vb*{\alpha}^{(\ell)}}_1}$, the overhead factor may be expressed simply as
\begin{equation}\begin{split}
    \xi 
    = \prod_{\ell = 1}^m \xi^{(\ell)}
    = \prod_{\ell = 1}^m \frac{1}{1 - 2 r_\ell}.
    \label{eq:theory-single/simplified/tensored-overhead-factor}
\end{split}\end{equation}

The factorized structure of $\vb*{\alpha}$ implies that $\abs*{\vb*{\alpha}} / \xi$ is a product over individual $\abs*{\vb*{\alpha}^{(\ell)}} / \xi^{(\ell)}$ probability distributions. This enables a reduction of classical resources required to initialize the mitigation protocol to $\order{m}$ time and space, with $\order{m}$ sampling time for each shot. In an even simpler setting of uniform readout channels ($r_\ell = r$), all $\smash{\vb*{\alpha}^{(\ell)}}$ are identical and classical initialization costs are trivially reduced to $\order{1}$ time and space. These complexities are reflected in \cref{tab:summary}, and further technical details are provided in \cref{app-sec:theory/single/solution-tensored}.

In practice, readout errors on a quantum device expectedly exhibit at least a small amount of correlations or non-uniformity. How large would the consequences of assuming independence or uniformity be? Our prior discussion surrounding \cref{eq:theory-single/general/sensitivity-bounds-expval} has established a mild scaling of errors in the mitigated observables versus ideal values, with respect to the (small) total variation distance between the true $\vb{q}$ and that used in the protocol. For small distances, errors made by an approximate mitigation assuming independence or uniformity are small, and importantly, the approximation enables significant resource complexity savings against the general setting. The utility of such simplifications thus depends on the quantum device and the size of the quantum circuit. We discuss this further in our experimental study in \cref{sec:expts}.

\section{Generalization to Multiple Feedforward Layers}
\label{sec:theory-multiple}

Many operational scenarios involve multiple layers of mid-circuit measurements and feedforward, thus extending beyond the single-layer setting introduced in \cref{sec:framework} and analyzed in \cref{sec:theory-single}. This multi-layer setting is illustrated in \cref{fig:theory-multiple/input-circuit}, and is described by
\begin{enumerate}
    \item Application of a unitary quantum circuit resulting in an intermediary quantum state $\rho$.
    \item For layers $l = 1, \ldots, L$:
    \begin{enumerate}
        \item Measurement of $\smash{m^{[l]}} \leq n$ qubits in the computational basis\footnote{Mid-circuit measurements in other bases can be absorbed into the state preparation of $\rho$ and the feed-forward operations of the previous and current layers; thus this computational basis assumption is without loss of generality.} resulting in a measurement outcome (bitstring) $\smash{s^{[l]} \in \mathcal{S}^{[l]}} = \smash{\{0, 1\}^{m^{[l]}}}$.
        \item Application of an $s^{[l]}$-dependent unitary operation $\smash{V^{[l]}_{s^{[l]}}}$ as dictated by a feedforward function that designates one unitary to apply in this layer for each $s^{[l]}$.
    \end{enumerate}
\end{enumerate}

In this generalized setting readout errors can occur in every mid-circuit measurement and feedforward layer---in each layer $l$ readout errors can cause a measurement outcome $\smash{s'^{[l]}}$ to be reported while the true outcome is $\smash{s^{[l]}}$, resulting in the application of an incorrect feedforward unitary $\smash{V^{[l]}_{s'^{[l]}}}$ in that layer (see \cref{fig:theory-multiple/input-circuit}). As before, we denote the total number of measurements as $m$.

\subsection{Multi-layer error mitigation protocol}
\label{sec:theory-multiple/protocol}

While multi-layer feedforward incurs readout errors in each layer, which can cause a sequence of incorrect feedforward unitaries to be applied, we show that our error mitigation scheme continues to work with slight modifications. Here, operationally, symmetrized readout error channels are still characterized by a vector $\vb{q}$, where each element $q_s$ represents the probability that the multi-layer error syndrome $s \in \mathcal{S} = \{0, 1\}^m$ occurs. That is, if the true measurement outcome is $a = \smash{a^{[1]} a^{[2]} \ldots a^{[L]}} \in \mathcal{S}$ concatenated across the layers, then $q_s$ is the probability that the quantum device reports $(a \oplus s)$, where $s = \smash{s^{[1]} s^{[2]} \ldots s^{[L]}}$ is the syndrome across layers; here $\smash{a^{[l]}} \in \smash{\mathcal{S}^{[l]}}$ and $\smash{s^{[l]}} \in \smash{\mathcal{S}^{[l]}}$ are the true measurement outcome and syndrome in layer $l$ respectively. 

The idea remains to sample a bitmask $f \in \mathcal{S}$ in each experiment shot. We then partition $\smash{f = f^{[1]} f^{[2]} \ldots f^{[L]}} \in \mathcal{S}$ into layer-wise bitmasks $\smash{f^{[l]} \in \mathcal{S}^{[l]}}$, and we apply $\smash{f^{[l]}}$ to the feedforward in layer $l$---that is, when the mid-circuit measurements reports an outcome $s'^{[l]} \in \mathcal{S}^{[l]}$, we apply the feedforward unitary
$\smash{V^{[l]}_{s'^{[l]} \oplus f^{[l]}}}$ instead of $\smash{V^{[l]}_{s'^{[l]}}}$. We detail this protocol below.

\begin{tcolorbox}[boxsep=8pt,
    left=0pt,right=0pt,top=0pt,bottom=0pt]
    \refstepcounter{algorithm}
    \label{alg:prom-multiple-layer}
    
    \textbf{Algorithm \thealgorithm}: \textit{Probabilistic readout error mitigation (\texttt{PROM}) for multiple feedforward layers}. \\[-0.5\baselineskip]
    
    \textbf{Inputs:} Circuit $\mathcal{P}$ containing $L$ layers of mid-circuit measurements and feedforward (see \cref{fig:theory-multiple/input-circuit}). Probability vector $\vb{q}$ characterizing mid-circuit measurements. Arbitrary measured observables $\vb{O}$. Number of shots $N'$ to be executed. \\[-0.5\baselineskip]

    \textbf{Output:} Estimate $\smash{\expval*{\widehat{\vb{O}}}}$ of expectation values. \\[-0.5\baselineskip]

    We construct a bit-flip-averaged (see \cref{fig:theory-multiple/schematic}) version of $\mathcal{P}$ and an appropriate weight vector $\vb*{\alpha}$ from $\vb{q}$ (see \cref{result:multi/correctness}). For shot $k = 1$ to $N'$, we 
    \begin{enumerate}
        \item Sample a bitmask $\smash{f = f^{[1]} f^{[2]} \ldots f^{[L]}} \in \mathcal{S}$ with probability $\abs*{\alpha_f} / \xi$.
        \item For layers $l = 1, \ldots, L$, execute the mid-circuit measurements in layer $l$ with bitmask $\smash{f^{[l]}}$ applied to the feedforward.
        \item Measure $\vb{O}$ to obtain outcome $\vb{m}^{(k)}$. 
        % (Each entry $m^{(k)}_b$ is an eigenvalue of $O_b$).
        \item Record $\smash{\vb{o}^{(k)} = \vb{m}^{(k)} \sgn(\alpha_f)}$.
    \end{enumerate}

    Output the estimate $\smash{\expval*{\widehat{\vb{O}}}} = \xi \cdot \frac{1}{N'} \sum_{k = 1}^{N'} \vb{o}^{(k)}$.
\end{tcolorbox}

Our key result in this generalized setting is that the weights $\vb*{\alpha}$ can be systematically computed with a similar approach as in the single-layer setting, to produce an unbiased estimator $\widehat{\vb{O}}$ for the observables $\vb{O}$. Moreover, the same characterization of the sampling overhead factor $\xi^2 = \smash{\norm*{\vb*{\alpha}}_1^2}$ of the protocol applies.

\begin{result}[Multi-layer correctness, convergence]
    \label{result:multi/correctness}
    Consider \cref{alg:prom-multiple-layer} with readout error channels characterized by $\vb{q}$. Then \cref{result:single/correctness,result:single/convergence} originally stated for \cref{alg:prom-single-layer} continue to hold for the multi-layer case under the condition that $\vb{q}$ is the multi-layer probability vector.
\end{result}

As before, when the total readout error probability $\eta < 1/2$, the general bound $\xi < 1 / (1 - 2 \eta)$ applies. The same robustness to calibration errors (\cref{result:single/robustness}) also holds with $\vb{q}$ and $\vb{q}'$ being multi-layer probability vectors. 

The single-layer context discussed in \cref{sec:theory-single} is thus subsumed as a special case with $L = 1$. We give the intuition for this extension of the error mitigation scheme to multiple feedforward layers in \cref{sec:theory-multiple/general} below and proof in \cref{app-sec:theory/multiple}. Similarly to the single-layer setting, we continue to enjoy no increase in two-qubit gate count or depth in the circuits executed in the protocol relative to the input circuit. We summarize classical processing and sampling costs for multi-layer scenarios in \cref{tab:summary}.

\subsection{Principles underlying multi-layer extension}
\label{sec:theory-multiple/general}

The main ideas leading to the solution $\vb*{\alpha}$ follows closely the structure laid out in the single-layer context in \cref{sec:theory-single/general}. Here, with $\smash{s = s^{[1]} s^{[2]} \ldots s^{[L]} \in \mathcal{S}}$ the concatenation of measurement outcomes across all layers and $\smash{\Pi^{[l]}_{s^{[l]}}}$ the measurement projector in layer $l$ associated with the outcome $\smash{s^{[l]}}$, we write the sequence $W$ of mid-circuit measurements and feedforward unitaries, and likewise define the tensor $T$,
\begin{equation}\begin{split}
    W_{ss'} = 
    \prod_{l = L}^1 V^{[l]}_{s'^{[l]}} \Pi^{[l]}_{s^{[l]}},
    \quad
    T_{b s s'}
    = \tr \left[ O_b 
        W_{s s'} \rho W_{s s'}^\dag
    \right].
    \label{eq:theory-multiple/general/T-definition}
\end{split}\end{equation}

Directly analogous to \cref{eq:theory-single/general/T-definition}, the entry $T_{b s s'}$ is the expectation value of $O_b$ measured on quantum trajectories that had true measurement outcomes $s$ but feedforward unitaries applied according to $s'$, weighted by the probability of those trajectories occurring. 

Then the ideal expectation value of $O_b$ in a hypothetical noiseless setting, $\smash{\expval*{\overline{O}_b}}$, is of exactly the same form as in \cref{eq:theory-single/general/expval-O-ideal} but with $T$ defined above. Moreover, the expectation value of $O_b$ with readout errors and with bitmask $f \in \mathcal{S}$ applied in the feedforward, $\smash{\expval*{O_b^{(f)}}}$, is of the same form as in \cref{eq:theory-single/general/expval-O-f}. Just as in \cref{sec:theory-single/general}, we demand $\smash{\expval*{\widehat{\vb{O}}} = \expval*{\overline{\vb{O}}}}$ and thus equate $\smash{\expval*{\overline{O}_b}}$ and $\smash{\expval*{O_b^{(f)}}}$ through the definition of estimator $\smash{\widehat{\vb{O}}}$, leading to a requirement that the coefficient of every $T_{b s s'}$ match on both sides. 

As $\smash{\expval*{\overline{O}_b}}$ and $\smash{\expval*{O_b^{(f)}}}$ have remained identical in form, we have the same solution for $\vb*{\alpha}$---namely, that presented in \cref{result:single/correctness}. We provide an explicit derivation and further technical details in \cref{app-sec:theory/multiple/solution-general}. The classical resource costs are likewise unchanged (listed in \cref{tab:summary}).

\subsection{Simplified solutions for uncorrelated channels}
\label{sec:theory-multiple/simplified}

As the measurements in different layers of the circuit are separated in time, a plausible assumption is that readout errors affecting them are independent, consistent with standard Markovian assumptions on noise channels~\cite{suter2016colloquium, clerk2010intro}. Measurements within the same layer, which may occur simultaneously or in overlapping time intervals, may still exhibit correlations in errors. In such a setting, the confusion vector $\vb{q}$ factorizes and so do the solutions for $\vb*{\lambda}$ and $\vb*{\alpha}$,
\begin{equation}\begin{split}
    \vb{q} = \bigotimes_{l = 1}^L \vb{q}^{[l]},
    \qquad
    \vb*{\lambda} = \bigotimes_{l = 1}^L \vb*{\lambda}^{[l]},
    \qquad 
    \vb*{\alpha} = \bigotimes_{l = 1}^L \vb*{\alpha}^{[l]},
    \label{eq:theory-multiple/simplified/alpha-factorization}
\end{split}\end{equation}
where $\smash{k = k^{[1]} k^{[2]} \ldots k^{[L]}} \in \mathcal{S}$ and $\smash{f = f^{[1]} f^{[2]} \ldots f^{[L]}} \in \mathcal{S}$ partition the bitstrings over the layers, and $\smash{\vb{q}^{[l]}}$, $\smash{\vb*{\lambda}^{[l]}}$ and $\smash{\vb*{\alpha}^{[l]}}$ are the confusion vector, eigenvalues and mitigator coefficients for layer $l$ respectively. Explicitly, the latter two are given by \cref{eq:readout-errors/lambda-solution,eq:theory-single/general/alpha-solution} but with $(\mathcal{S}, m, \vb{q}, k, f)$ replaced by their layer counterparts $\smash{(\mathcal{S}^{[l]}, m^{[l]}, \vb{q}^{[l]}, k^{[l]}, f^{[l]})}$. Writing layer-wise overheads $\smash{\xi^{[l]} = \norm*{\vb*{\alpha}^{[l]}}_1}$, the overhead factor likewise factorizes,
\begin{equation}\begin{split}
    \xi = \prod_{l = 1}^m \xi^{[l]}.
    \label{eq:theory-multiple/simplified/xi-factorization}
\end{split}\end{equation}

Then $\abs{\vb*{\alpha}} / \xi$ is a product over the layer-wise $\smash{\abs*{\vb*{\alpha}^{[l]}} / \xi^{[l]}}$ probability distributions. This allows a reduction of the classical resources required to $\smash{\order*{m \cdot 2^{\overline{m}}}}$ time and $\smash{\order*{L \cdot 2^{\overline{m}}}}$ space for solution computation, where $\smash{\overline{m} = \max_{l \in [L]} m^{[l]}}$ is the maximum number of measurements in a layer, and $\order{L}$ classical sampling time per shot, as summarized in \cref{tab:summary}. Further analysis details are provided in \cref{app-sec:theory/multiple/solution-layer-tensored}.

For yet further simplification, a stronger assumption is to impose independence on readout errors across all measurements regardless of their location in layers, and even stronger still, to assume that all readout error channels are uniform. These settings enforce additional structure in the solutions and land us in the context of \cref{sec:theory-single/simplified}. The results for $\vb*{\alpha}$ and classical resource costs there directly apply.

\section{Hardware Experiments}
\label{sec:expts}

To demonstrate the effectiveness of our probabilistic readout error mitigation (\texttt{PROM}) strategy, we executed a series of experiments on $27$- and $127$-qubit IBM transmon-based superconducting quantum processors. These devices support mid-circuit measurements and feedforward, and have characteristic readout error rates of $0.5$--$3\%$ per measurement. As there is clear non-uniformity in readout error rates between qubits, we examine both general \texttt{PROM} and \texttt{PROM} assuming independent readout errors, but we omit further uniformity assumptions. We provide more details on the quantum hardware used in \cref{app-sec:methods/hardware}, with device specifications listed in \cref{app-tab:quantum-device-characteristics}. Our experiments span dynamic qubit reset, shallow-depth GHZ state preparation, and repeated quantum state teleportation (\cref{sec:expts/reset,sec:expts/ghz,sec:expts/teleport}), thus covering a variety of use-cases of practical interest.

Our experiments employ \texttt{PROM} for mid-circuit measurements and feedforward, in conjunction with standard REM for terminal measurements (see \cref{app-sec:methods/ro-terminal}). We clearly distinguish results with terminal REM only, and with both \texttt{PROM} and terminal REM. We implement dynamical decoupling on all circuits to suppress decoherence on idle qubits (see \cref{app-sec:methods/dynamical-decoupling}).

For comparison, we examine also a repetition-based REM strategy for feedforward. The strategy works by performing each mid-circuit measurement $r > 1$ times consecutively and taking a consensus outcome for feedforward, such that the readout error rate perceived by the feedforward is suppressed. This general approach is commonly invoked in quantum error correction and fault-tolerant quantum computation~\cite{nielsen2010quantum, gottesman2010introduction}. For our purposes, we consider two variants---\texttt{Rep-MAJ}, which takes the majority outcome for odd $r$, and \texttt{Rep-ALL}, which requires agreement between all measurement outcomes for feedforward to go through and discards the experiment shot otherwise (\ie~post-selection). 

Notably, \texttt{PROM} differs from these repetition-based methods in that \texttt{PROM} does not modify the quantum circuit to be executed\footnote{Other than $X$ twirling gates in BFA, which are of negligible quantum cost (in fidelity and gate time), and the classical bitmasks in feedforward.}. The repetition-based methods, on the other hand, increase the number of measurements and depth of the circuit. This difference has non-trivial implications in practice, as we shall observe.

To start, we demonstrate \texttt{PROM} in the context of dynamic qubit resets. A reset is a (non-unitary) operation that re-initializes a qubit to $\ket{0}$ regardless of the input state~\cite{divincenzo2000physical, rist2012initialization}; on digital quantum processors, it can be implemented as a mid-circuit measurement followed by an $X$ gate conditioned on a $1$ outcome. As they enable arbitrary re-use of qubits, resets are of pivotal utility across quantum algorithms, simulation, optimization and metrology domains~\cite{del2024robust, dibartolomeo2023efficient, hua2023exploiting}. Concretely, qubit resets have enabled long-time simulations of open quantum systems~\cite{rost2021demonstrating, han2021experimental}, execution of wide quantum approximate optimization algorithm (QAOA) circuits on present hardware~\cite{decross2023qubit}, and have been leveraged in phase estimation~\cite{corcoles2021exploiting} and spectroscopy techniques~\cite{yirka2021qubit}. More broadly, resets serve as an entropy sink in measurement-free quantum error correction protocols~\cite{heussen2024measurement, perlin2023fault}, and are generally useful in quantum algorithms that repeatedly access (and discard into) an ancillary reservoir~\cite{childs2012hamiltonian, berry2015simulating, gilyen2019quantum, dong2021success}.

\subsection{Mid-circuit dynamic qubit reset}
\label{sec:expts/reset}

\begin{figure}[!t]
    \centering
    \includegraphics[width=\linewidth]{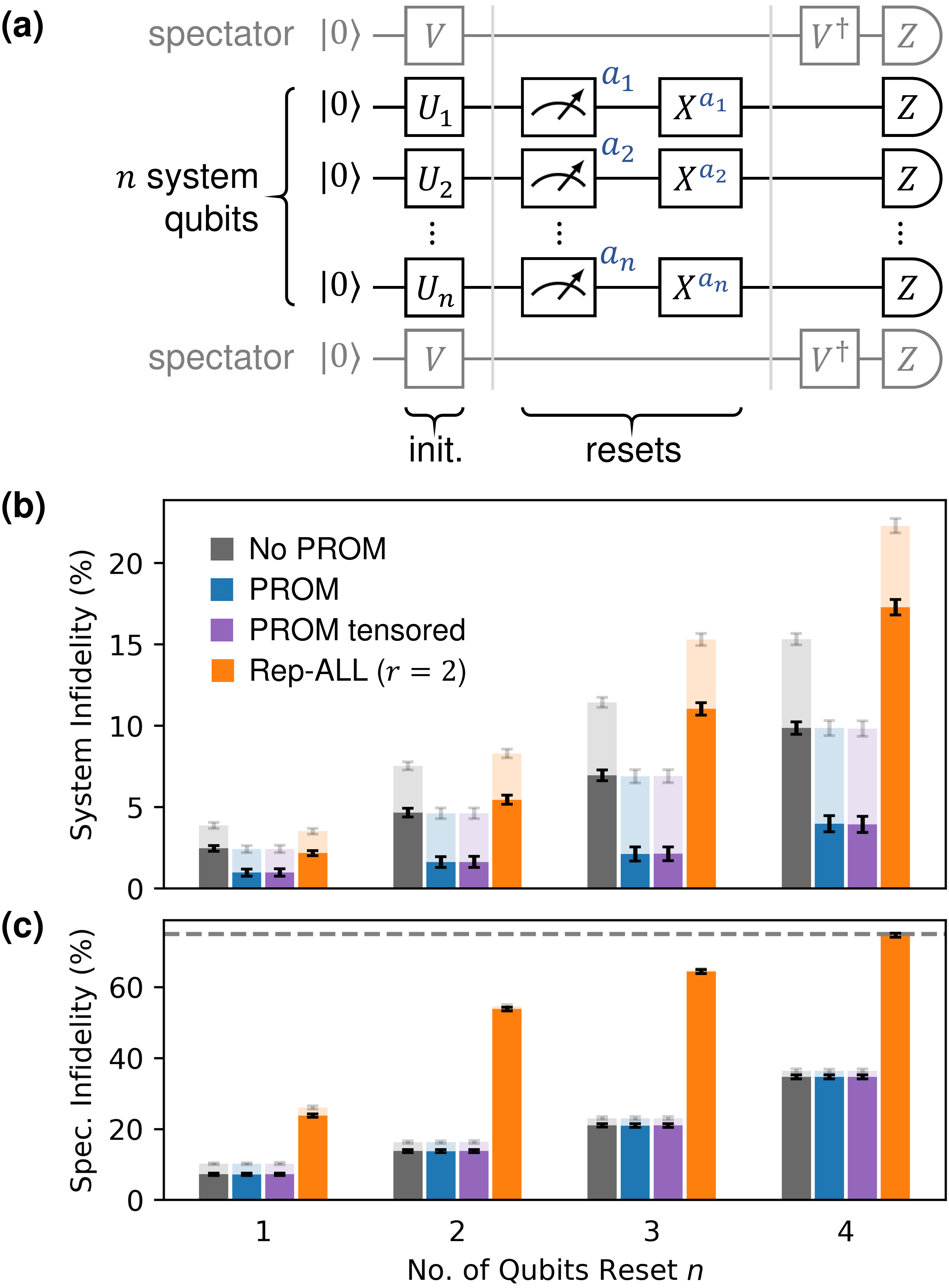}
    \phantomsubfloat{\label{fig:expts/reset/circuit-diagram}}
    \phantomsubfloat{\label{fig:expts/reset/fidelity-system}}
    \phantomsubfloat{\label{fig:expts/reset/fidelity-spectator}}
    \vspace{-2\baselineskip}
    \caption{\textbf{Dynamic qubit resets.} \textbf{(a)} Circuit structure to probe fidelity of qubit reset operations. A register of $n$ system qubits (black) undergo evolution by $U_1 \otimes \cdots \otimes U_n$ and are then reset into the $\smash{\ket{0}^{\otimes n}}$ state. A pair of adjacent spectator qubits (gray) sandwiching the system register witness the reset events. Dynamic resets are implemented as mid-circuit measurements followed by $X$ gates conditioned on outcomes (blue labels). \textbf{(b)} Infidelities of system qubits after dynamic resets against $\smash{\ket{0}^{\otimes n}}$, comparing \texttt{PROM}, \texttt{PROM} assuming tensored readout error channels, \texttt{Rep-ALL} at $r = 2$, and without readout error mitigation for mid-circuit measurements and feedforward. \textbf{(c)} Infidelities of spectator qubits against $\smash{\ket{0}^{\otimes 2}}$ after witnessing reset events, conditioned on the system qubits being reset correctly. Dashed line demarcates $75\%$ infidelity expected for a maximally mixed spectator state. In all plots, solid bars are with terminal readout error mitigation; translucent bars without. Experiments performed on superconducting quantum device \textit{ibmq\_kolkata}. Error bars are standard deviations across $1000$ trials per qubit chain, each with $10000$ shots; data averaged over ${\sim}8$ randomly chosen qubit chains.}
    \label{fig:expts/reset}
\end{figure}

The circuit structure we examine comprises $n$ system qubits that are prepared in $(U_1 \otimes U_2 \otimes \cdots \otimes U_n) \ket{0}^{\otimes n}$, dynamically reset, and then terminally measured to verify their post-reset state (see \cref{fig:expts/reset/circuit-diagram}). A pair of spectator qubits sandwiches the system register and are each prepared in $V \ket{0}$. After witnessing the resets on the system qubits, they are back-rotated by $V^\dag$ and measured. In an ideal noiseless setting, the terminal states of the system and spectator qubits should all be $\ket{0}$. To probe the fidelity of the reset operations, our observables $\vb{O}$ thus constitute $\smash{\ket{0}^{\otimes n} \bra{0}^{\otimes n}}$ on the system qubits and $\smash{\ket{0}^{\otimes 2} \bra{0}^{\otimes 2}}$ on the spectator qubits. In our experiments, we place all qubits on a contiguous chain on the quantum processor, and we take $U_1 = \cdots = U_n = V = H$. 

We present in \cref{fig:expts/reset/fidelity-system} system infidelities against the ideal $\ket{0}^{\otimes n}$ post-reset state, for $n \in \{1, 2, 3, 4\}$ qubits. We compare data without $\texttt{PROM}$, with general $\texttt{PROM}$ and $\texttt{PROM}$ under an assumption of tensored readout error channels, and with \texttt{Rep-ALL} at $r = 2$. The drastic effect of \texttt{PROM} is evident---our protocol reduces system infidelities to ${\sim} 30$--$40\%$ of their unmitigated values. In contrast to \texttt{PROM}, the repetition-based \texttt{Rep-ALL} scheme does not appreciably improve, and in fact worsens for $n > 1$, system fidelities across the board.

While \texttt{PROM} addresses readout errors affecting feedforward, on hardware a variety of other noise sources are present (\eg~gate errors and decoherence), which account for the observed residual infidelities. As discussed, in sensitive applications, \texttt{PROM} can be used in conjunction with other error mitigation techniques that complementarily address these other noise sources. In \cref{fig:expts/reset/fidelity-system} we have also reported data without terminal REM, drawn as translucent bars. The observed fidelity improvements provided by terminal REM alone (without \texttt{PROM}) and by the addition of \texttt{PROM} are comparable, indeed expected as there are equal numbers of mid-circuit and terminal measurements on the system qubits ($n$ each). 

An additional layer of contrast is unveiled when examining spectator qubit infidelities against their ideal $\smash{\ket{0}^{\otimes 2}}$ post-witness state, as presented in \cref{fig:expts/reset/fidelity-spectator}. Expectedly, \texttt{PROM} has negligible effect on spectator qubits, as the protocol does not modify the circuit structure. In comparison the repetition-based schemes introduce additional layers of mid-circuit measurements. On hardware, measurements are typically long operations---on the superconducting devices used here they are $2$--$8$ times the duration of a CX gate (see \cref{app-tab:quantum-device-characteristics})---thus waiting for mid-circuit measurements can incur considerable decoherence; moreover there may be measurement cross-talk, wherein the readout of a qubit disturbs the state of surrounding qubits. These effects manifest in the poor observed spectator fidelities of \texttt{Rep-ALL}. In particular, at $n = 4$ the spectator infidelity approaches that expected from a maximally mixed state, suggesting significant depolarization. This is despite the inclusion of dynamical decoupling in our experiments (see \cref{app-sec:methods/dynamical-decoupling}) that conventionally suppresses background decoherence.

Lastly we comment that, as observed in \cref{fig:expts/reset/fidelity-system,fig:expts/reset/fidelity-spectator}, \texttt{PROM} with and without independent readout error assumptions exhibit similar mitigation performance. The former generally produced system infidelities ${\sim}1$--$2\%$ worse relative to the latter. We checked that the total variation distance between characterized confusion vectors $\vb{q}$ and their tensored counterparts assuming independent readout errors are $\lesssim 1\%$, thus this similarity is consistent with the sensitivity properties of the protocol as stated in \cref{eq:theory-single/general/sensitivity-bounds-expval}. On hardware exhibiting larger error correlations, imposing independence assumptions will introduce a larger mitigation quality penalty than observed here.

We report additional data acquired on two other superconducting quantum devices in \cref{app-sec:data/reset/experiment}, which are qualitatively similar to those shown here in \cref{fig:expts/reset/fidelity-system,fig:expts/reset/fidelity-spectator}. There, we also show system and spectator infidelities for \texttt{Rep-ALL} and \texttt{Rep-MAJ} at $r = 3$---which, as could be expected from the $r = 2$ data, are much too large to be of practical relevance.

\subsection{Shallow-depth GHZ state preparation}
\label{sec:expts/ghz}

\begin{figure}[!t]
    \centering
    \includegraphics[width=\linewidth]{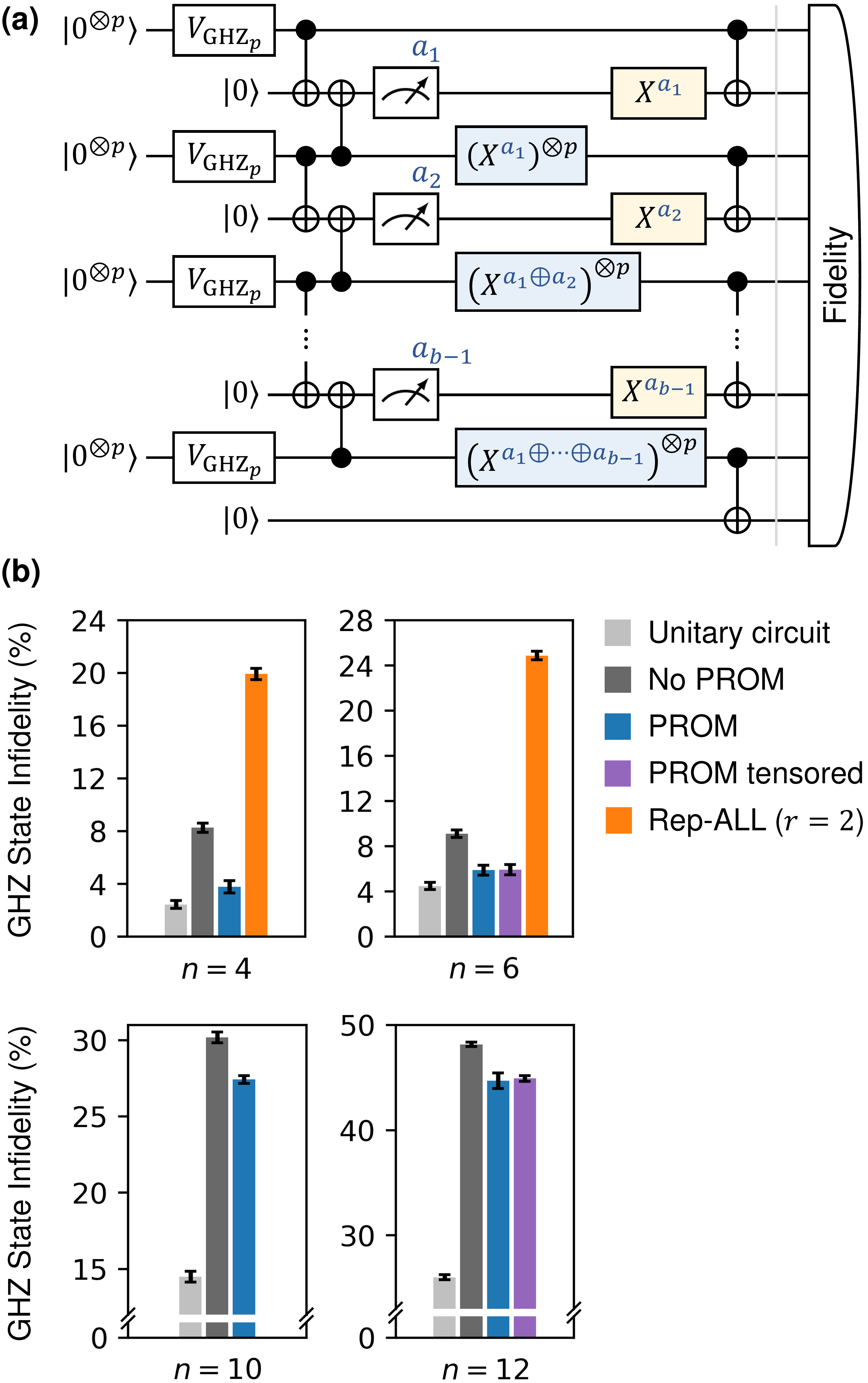}
    \phantomsubfloat{\label{fig:expts/ghz/circuit-diagram}}
    \phantomsubfloat{\label{fig:expts/ghz/fidelities-n-4-6}}
    \phantomsubfloat{\label{fig:expts/ghz/fidelities-n-10-12}}
    \vspace{-2\baselineskip}
    \caption{\textbf{GHZ state preparation using feedforward.} \textbf{(a)} Dynamic circuit to prepare an $n$-qubit GHZ state. The $n$ qubits are split into $b$ blocks each comprising $p$ qubits and an ancilla, the former prepared in a GHZ state through the standard unitary $\smash{V_{\textsc{ghz}_p}}$ comprising a fan-out of CX gates---see \cref{app-fig:methods/circuits/ghz-unitary}. Thereafter the small GHZ states are merged in constant depth through mid-circuit measurements and feedforward operations (shaded blue), and finally the ancillae are reset (shaded yellow) and absorbed into the state. \textbf{(b)} Infidelities of the prepared state against the ideal $n$-qubit GHZ state, comparing \texttt{PROM}, \texttt{PROM} assuming tensored readout channels, \texttt{Rep-ALL} at $r = 2$, and without readout error mitigation for mid-circuit measurements and feedforward. Also shown are infidelities on a unitary circuit ($\smash{V_{\textsc{ghz}_n}}$). Readout error mitigation for terminal measurements was applied for all experiments. The $n = 4, 6, 10, 12$ experiments employed $(b, p) = (2, 1)$, $(3, 1)$, $(2, 4)$, $(3, 3)$ circuit structures respectively, executed on the superconducting quantum device \textit{ibm\_osaka}. Error bars are standard deviations across $100$ trials each with $10000$ shots.}
    \label{fig:expts/ghz}
\end{figure}

Next we demonstrate \texttt{PROM} in the context of preparing GHZ states, which are entangled states of foundational importance across quantum information, metrology, and communications domains~\cite{greenberger1990bell, leibfried2004toward, liao2014dynamic, hahn2020anonymous}. On linear nearest-neighbor qubit connectivity, the unitary preparation of an $n$-qubit GHZ state ($n$ even) necessarily requires $n - 1$ CX gates with $n / 2$ depth~\cite{watts2019exponential}. The optimal unitary circuit comprises a qubit starting in the $\ket{+}$ state followed by a bidirectional fan-out of CX gates to grow the GHZ state (see \cref{app-fig:methods/circuits/ghz-unitary}).

With mid-circuit measurements and feedforward, however, a GHZ state can be prepared in constant depth independent of $n$~\cite{quek2024multivariate, baumer2023efficient, watts2019exponential}. Here we examine a circuit structure that is an interpolation~\cite{quek2024multivariate} between the unitary and feedforward circuit extremes (see \cref{fig:expts/ghz/circuit-diagram}), such that the number of CX gates and measurements in the circuit and circuit depth become adjustable. We consider splitting the $n$ qubits into $b$ blocks each containing $p$ system qubits and an ancillary qubit; smaller $p$-qubit GHZ states are unitarily prepared in each of the blocks, and the blocks are then merged in constant depth using mid-circuit measurements and feedforward. Conceptually, the ancillae perform parity checks on adjacent pairs of blocks, and detected domain walls in the quantum state are annihilated through feedforward (blue-shaded gates in \cref{fig:expts/ghz/circuit-diagram}). Lastly the ancillae are reset (yellow-shaded gates) and unitarily absorbed into the GHZ state. The constant-depth feedforward extreme discussed above corresponds to $(b, p) = (n / 2, 1)$. For $b \geq 2$, a $(b, p)$-circuit contains $b (p + 2) - 2$ CX gates, $b - 1$ mid-circuit measurements, and is of depth $\floor{p / 2} + 4$ including the layer of mid-circuit measurements on the ancillae.

First, we report results at $n = 4, 6$ in \cref{fig:expts/ghz/fidelities-n-4-6}, using $(b, p) = (2, 1), (3, 1)$ circuit structures respectively. We compare the observed infidelity of the prepared state against the ideal GHZ state without \texttt{PROM}, with general \texttt{PROM} and \texttt{PROM} assuming tensored readout error channels, and with \texttt{Rep-ALL} at $r = 2$. We include also for comparison the unitary GHZ state preparation circuit (drawn in \cref{app-fig:methods/circuits/ghz-unitary}) run on the same $n$-qubit chain. We used a stabilizer method to measure the fidelity of the prepared state against the GHZ state---see \cref{app-sec:methods/ghz-fidelity} for details. The application of \texttt{PROM} is evidently effective, reducing state infidelities to ${\sim}45$--$65\%$ of their unmitigated values. Residual infidelities are attributed to quantum errors and decoherence that are not targeted by \texttt{PROM}. Just as in the qubit reset experiments (\cref{sec:expts/reset}), the \texttt{Rep-ALL} mitigation scheme is not competitive.

The qualitative trend that unitary state preparation outperform unmitigated dynamic circuits is consistent with a prior study~\cite{baumer2023efficient}; as discussed there and in other dynamic circuit studies~\cite{baumer2024quantum}, this is generally attributable to decoherence, crosstalk, and imperfect quantum-nondemolition (QND) properties of measurements on hardware. In our experiments, with \texttt{PROM}, the infidelities on dynamic circuits become fairly competitive with but do not surpass unitary circuits.

We next examined larger $n = 10, 12$ settings in \cref{fig:expts/ghz/fidelities-n-10-12}, therein utilizing an interpolation between unitary and feedforward circuit extremes by choosing $(b, p) = (2, 4), (3, 3)$ circuit structures respectively. Here \texttt{PROM} provided reductions in state infidelities comparable to the $n = 4, 6$ experiments, but residual infidelities were expectedly higher. As it has become clear that the repetition-based REM schemes are uncompetitive, we have omitted them at $n = 10, 12$. On circuits with $b > 2$, in particular $n = 6, 12$, general \texttt{PROM} yielded relative state infidelities ${\sim}1$--$2\%$ lower than \texttt{PROM} assuming independent readout errors---similar to the qubit reset experiments in \cref{sec:expts/reset}, this small difference is consistent with our protocol sensitivity bounds as the characterized readout error channels were close to being independent. 

We report additional data on a second superconducting quantum device in \cref{app-sec:data/ghz/experiment} at $n = 4, 6$, which are qualitatively similar. There, we also report infidelities measured with \texttt{Rep-ALL} and \texttt{Rep-MAJ} at $r = 3$, which were entirely uncompetitive and omitted here.

\subsection{Staged quantum state teleportation}
\label{sec:expts/teleport}

\begin{figure}[!t]
    \centering
    \includegraphics[width=\linewidth]{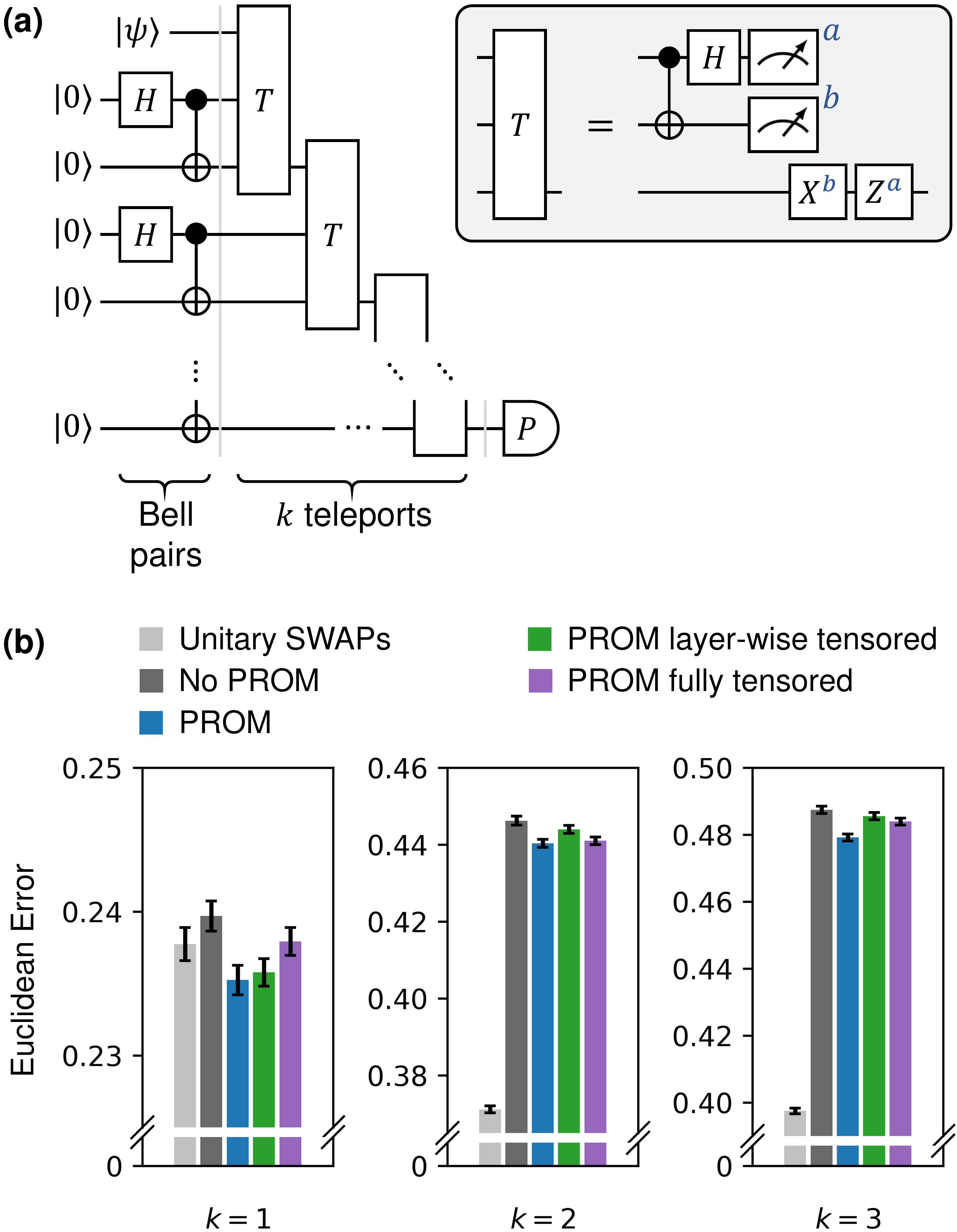}
    \phantomsubfloat{\label{fig:expts/teleport/circuit-diagram}}
    \phantomsubfloat{\label{fig:expts/teleport/expvals}}
    \vspace{-2\baselineskip}
    \caption{\textbf{Repeated quantum state teleportation.} \textbf{(a)} Circuit schematic. An input quantum state $\ket{\psi}$ is transported to a qubit a distance $2k$ away by $k$ teleportations, consuming $k$ shared Bell pairs. On the destination qubit, Pauli expectation values $\expval{P}$ are measured and the deviations $\delta P$ against ideal expectation values of the input state $\ket{\psi}$ assessed, for $P \in \{X, Y, Z\}$. \textbf{(b)} Euclidean errors $\smash{\Delta = (\delta X^2 + \delta Y^2 + \delta Z^2)^{1/2}}$ after $k \in \{1, 2, 3\}$ teleportations, comparing \texttt{PROM}, \texttt{PROM} assuming layer-wise and fully tensored readout channels, and without readout error mitigation for mid-circuit measurements and feedforward. Also shown are errors on a unitary circuit with SWAP gates that transport $\ket{\psi}$ across the same qubits. Readout error mitigation for terminal measurements was applied for all cases. Experiments were performed on the superconducting quantum device \textit{ibm\_osaka}. Error bars are standard deviations across $100$ trials each with $1000000$ shots.}
    \label{fig:expts/teleport}
\end{figure}

Lastly we investigate repeated quantum state teleportation (see \cref{fig:expts/teleport/circuit-diagram}). In standard state teleportation~\cite{bennett1993, bouwmeester1997experimental, ren2017ground}, a Bell pair is shared between a second and third qubit; the first qubit hosts the quantum state $\ket{\psi}$ to be transported. A Bell-basis measurement is performed on the first and second qubits and the outcomes feedforward into operations on the third qubit, which ends in state $\ket{\psi}$. As this procedure requires CX gates between adjacent pairs of qubits, the $3$ qubits must be connected to each other---thus $\ket{\psi}$ is transported a distance of $2$ qubits. Generically, repeating the teleportation $k$ times transports $\ket{\psi}$ a distance of $2k$ qubits (see \cref{fig:expts/teleport/circuit-diagram}). Scenarios that require the transport of a qubit state arise commonly in quantum computation and simulation, necessitated by device connectivity or classical control constraints~\cite{martiel2022architecture, wagner2023improving}. The straightforward alternative to teleportation is to connect starting and ending qubits with a path of SWAP gates, thus unitarily transporting the qubit state (see \cref{app-fig:methods/circuits/teleport-unitary}).

In our experiment, we prepare a generic single-qubit state $\smash{\ket{\psi} = e^{-i \phi_z Z} e^{-i \phi_x X} \ket{0}}$ with arbitrarily chosen $(\phi_x, \phi_z) = (\pi / 8, 3 \pi / 8)$, and teleport the state $k$ times, as drawn in \cref{fig:expts/teleport/circuit-diagram}. We terminally measure Pauli expectation values $\expval{X}, \expval{Y}, \expval{Z}$ on the ending qubit and assess their deviations $\delta X, \delta Y, \delta Z$ with respect to ideal values on the $\ket{\psi}$ starting state. To eliminate dependence on the measurement basis, we examine the Euclidean error $\smash{\Delta = (\delta X^2 + \delta Y^2 + \delta Z^2)^{1/2}}$. Thus $\Delta$ provides an averaged measure of the degradation of the state $\ket{\psi}$ as it undergoes transport. We employ the multi-layer formulation of \texttt{PROM}, with each teleportation stage placed in a layer (totalling $k$ layers). 

In \cref{fig:expts/teleport/expvals} we report results for $k \in \{1, 2, 3\}$ stages of teleportation, comparing the measured $\Delta$ without \texttt{PROM}, with general \texttt{PROM}, and with \texttt{PROM} assuming layer-wise and fully tensored readout error channels. We include also results on a unitary SWAP circuit transporting $\ket{\psi}$ across the same qubits. Across all $k$, \texttt{PROM} results in lower $\Delta$ than without mitigation. In particular at $k = 1$ this improvement is sufficient for teleportation to be comparable in error to unitary SWAPs, but at higher $k$ the unitary circuit is better. Similar to prior experiments (\cref{sec:expts/reset,sec:expts/ghz}), general \texttt{PROM} provides a relative advantage of ${\sim}1$--$2\%$ over \texttt{PROM} with independence assumptions, which is consistent with our sensitivity bounds.

We report additional data on a second quantum device in \cref{app-sec:data/teleport/experiment-data}, which are qualitatively similar. There we also show results with \texttt{Rep-ALL} at $r = 2$, which were uncompetitive and omitted here.

\section{Discussion \& Conclusion}
\label{sec:conclusion}

In this work we have described a general readout error mitigation (REM) method targeting mid-circuit measurements and feedforward in dynamic circuits, for which existing standard techniques for terminal measurements cannot address. In addition to laying the theoretical groundwork, we demonstrated the effectiveness of our method in a series of experiments on superconducting quantum processors. These experiments spanned multiple use cases of practical interest with several performance measures---in each case the observed error metric significantly decreased with our REM method.

To accommodate arbitrary correlations in readout errors, the application of our REM protocol requires resources scaling exponentially with the number of measurements. This is a consequence of universal lower bounds on the cost of quantum error mitigation~\cite{takagi2023universal, takagi2022fundamental, tsubouchi2023universal, quek2024exponentially}, applied here to readout errors instead of error channels acting directly on the quantum state. However, under an assumption that readout errors are independent, a reduction of classical resources required to construct the mitigated estimator to polynomial (in fact linear) scaling can be achieved. On quantum devices exhibiting small error correlations this cheaper scheme can suffice in providing high-quality mitigation. The sensitivity and overhead bounds we provide could aid in identifying the variant of the mitigation protocol to utilize in an experiment.

Lastly, we remark that quantum error mitigation can in general be recast into the broader frameworks of quantum channel and resource distillation~\cite{yuan2024virtual,takagi204virtual,regula2021fundamental,liu2020operational}. The present REM context for feedforward is no exception. The probabilistic sampling to correct the effects of broken measurement-feedforward correlations here, for example, is reminiscent of the classical sampling to restore coherence in distillation protocols. We point out related ideas in recent works which demonstrated noise protection on a memory qubit~\cite{hashim2024quasiprobabilistic} and considered building better positive operator-valued measures (POVMs) including accounting for quantum bit-flip errors during measurement~\cite{ivashkov2023highfidelity}. Our protocol primarily targets readout errors on the measurement outcomes; however, we could also treat coherent errors using a combination of these techniques and our method, or by applying probabilistic error cancellation~\cite{van2023probabilistic, gupta2023probabilistic}.

Our study paves the way for effectively utilizing dynamic circuit capabilities on both near-term and future hardware. This supports pivotal quantum resource advantages across a broad range of domains and is imperative for realizing useful quantum computation.

\section*{Acknowledgements}

The authors express gratitude to Lorcán O.~Conlon, Bujiao Wu, Jun Ye, Yunlong Xiao, Syed M.~Assad, Mile Gu, Tianqi Chen, and Ching Hua Lee for helpful discussions. The authors acknowledge the use of IBM Quantum services for this work. The views expressed are those of the authors, and do not reflect the official policy or position of IBM or the IBM Quantum team. Access to IBM quantum devices was enabled by the Quantum Engineering Programme (QEP) 2.0. This research is supported by the National Research Foundation, Singapore and the Agency for Science, Technology and Research (A*STAR) under its Quantum Engineering Programme (NRF2021-QEP2-02-P03), A*STAR C230917003, and A*STAR under the Central Research Fund (CRF) Award for Use-Inspired Basic Research (UIBR). J.T.~is supported by FQxI foundation under grant no.~FQXi-RFP-IPW-1903 (``Are quantum agents more energetically efficient at making predictions?'').

\clearpage
\pagebreak

\bibliography{references}

\begin{thebibliography}{124}%
\makeatletter
\providecommand \@ifxundefined [1]{%
 \@ifx{#1\undefined}
}%
\providecommand \@ifnum [1]{%
 \ifnum #1\expandafter \@firstoftwo
 \else \expandafter \@secondoftwo
 \fi
}%
\providecommand \@ifx [1]{%
 \ifx #1\expandafter \@firstoftwo
 \else \expandafter \@secondoftwo
 \fi
}%
\providecommand \natexlab [1]{#1}%
\providecommand \enquote  [1]{``#1''}%
\providecommand \bibnamefont  [1]{#1}%
\providecommand \bibfnamefont [1]{#1}%
\providecommand \citenamefont [1]{#1}%
\providecommand \href@noop [0]{\@secondoftwo}%
\providecommand \href [0]{\begingroup \@sanitize@url \@href}%
\providecommand \@href[1]{\@@startlink{#1}\@@href}%
\providecommand \@@href[1]{\endgroup#1\@@endlink}%
\providecommand \@sanitize@url [0]{\catcode `\\12\catcode `\$12\catcode `\&12\catcode `\#12\catcode `\^12\catcode `\_12\catcode `\%12\relax}%
\providecommand \@@startlink[1]{}%
\providecommand \@@endlink[0]{}%
\providecommand \url  [0]{\begingroup\@sanitize@url \@url }%
\providecommand \@url [1]{\endgroup\@href {#1}{\urlprefix }}%
\providecommand \urlprefix  [0]{URL }%
\providecommand \Eprint [0]{\href }%
\providecommand \doibase [0]{https://doi.org/}%
\providecommand \selectlanguage [0]{\@gobble}%
\providecommand \bibinfo  [0]{\@secondoftwo}%
\providecommand \bibfield  [0]{\@secondoftwo}%
\providecommand \translation [1]{[#1]}%
\providecommand \BibitemOpen [0]{}%
\providecommand \bibitemStop [0]{}%
\providecommand \bibitemNoStop [0]{.\EOS\space}%
\providecommand \EOS [0]{\spacefactor3000\relax}%
\providecommand \BibitemShut  [1]{\csname bibitem#1\endcsname}%
\let\auto@bib@innerbib\@empty
%</preamble>
\bibitem [{\citenamefont {DeCross}\ \emph {et~al.}(2023)\citenamefont {DeCross}, \citenamefont {Chertkov}, \citenamefont {Kohagen},\ and\ \citenamefont {Foss-Feig}}]{decross2023qubit}%
  \BibitemOpen
  \bibfield  {author} {\bibinfo {author} {\bibfnamefont {M.}~\bibnamefont {DeCross}}, \bibinfo {author} {\bibfnamefont {E.}~\bibnamefont {Chertkov}}, \bibinfo {author} {\bibfnamefont {M.}~\bibnamefont {Kohagen}},\ and\ \bibinfo {author} {\bibfnamefont {M.}~\bibnamefont {Foss-Feig}},\ }\bibfield  {title} {\bibinfo {title} {Qubit-reuse compilation with mid-circuit measurement and reset},\ }\href {https://doi.org/10.1103/PhysRevX.13.041057} {\bibfield  {journal} {\bibinfo  {journal} {Phys. Rev. X}\ }\textbf {\bibinfo {volume} {13}},\ \bibinfo {pages} {041057} (\bibinfo {year} {2023})}\BibitemShut {NoStop}%
\bibitem [{\citenamefont {Piveteau}\ and\ \citenamefont {Sutter}(2024)}]{piveteau2024circuit}%
  \BibitemOpen
  \bibfield  {author} {\bibinfo {author} {\bibfnamefont {C.}~\bibnamefont {Piveteau}}\ and\ \bibinfo {author} {\bibfnamefont {D.}~\bibnamefont {Sutter}},\ }\bibfield  {title} {\bibinfo {title} {Circuit knitting with classical communication},\ }\href {https://doi.org/10.1109/TIT.2023.3310797} {\bibfield  {journal} {\bibinfo  {journal} {IEEE Trans. Inf. Theory}\ }\textbf {\bibinfo {volume} {70}},\ \bibinfo {pages} {2734} (\bibinfo {year} {2024})}\BibitemShut {NoStop}%
\bibitem [{\citenamefont {Vazquez}\ \emph {et~al.}(2024)\citenamefont {Vazquez}, \citenamefont {Tornow}, \citenamefont {Riste}, \citenamefont {Woerner}, \citenamefont {Takita},\ and\ \citenamefont {Egger}}]{vazquez2024scaling}%
  \BibitemOpen
  \bibfield  {author} {\bibinfo {author} {\bibfnamefont {A.~C.}\ \bibnamefont {Vazquez}}, \bibinfo {author} {\bibfnamefont {C.}~\bibnamefont {Tornow}}, \bibinfo {author} {\bibfnamefont {D.}~\bibnamefont {Riste}}, \bibinfo {author} {\bibfnamefont {S.}~\bibnamefont {Woerner}}, \bibinfo {author} {\bibfnamefont {M.}~\bibnamefont {Takita}},\ and\ \bibinfo {author} {\bibfnamefont {D.~J.}\ \bibnamefont {Egger}},\ }\href {https://doi.org/10.48550/arXiv.2402.17833} {\bibinfo {title} {Scaling quantum computing with dynamic circuits}} (\bibinfo {year} {2024}),\ \Eprint {https://arxiv.org/abs/2402.17833} {arXiv:2402.17833 [quant-ph]} \BibitemShut {NoStop}%
\bibitem [{\citenamefont {Gottesman}(2010)}]{gottesman2010introduction}%
  \BibitemOpen
  \bibfield  {author} {\bibinfo {author} {\bibfnamefont {D.}~\bibnamefont {Gottesman}},\ }\bibfield  {title} {\bibinfo {title} {An introduction to quantum error correction and fault-tolerant quantum computation},\ }in\ \href@noop {} {\emph {\bibinfo {booktitle} {Quantum information science and its contributions to mathematics, Proceedings of Symposia in Applied Mathematics}}},\ Vol.~\bibinfo {volume} {68}\ (\bibinfo {year} {2010})\ pp.\ \bibinfo {pages} {13--58}\BibitemShut {NoStop}%
\bibitem [{\citenamefont {Acharya}\ \emph {et~al.}(2023)\citenamefont {Acharya}, \citenamefont {Aleiner}, \citenamefont {Allen}, \citenamefont {Andersen}, \citenamefont {Ansmann}, \citenamefont {Arute}, \citenamefont {Arya}, \citenamefont {Asfaw}, \citenamefont {Atalaya}, \citenamefont {Babbush} \emph {et~al.}}]{google2023suppressing}%
  \BibitemOpen
  \bibfield  {author} {\bibinfo {author} {\bibfnamefont {R.}~\bibnamefont {Acharya}}, \bibinfo {author} {\bibfnamefont {I.}~\bibnamefont {Aleiner}}, \bibinfo {author} {\bibfnamefont {R.}~\bibnamefont {Allen}}, \bibinfo {author} {\bibfnamefont {T.~I.}\ \bibnamefont {Andersen}}, \bibinfo {author} {\bibfnamefont {M.}~\bibnamefont {Ansmann}}, \bibinfo {author} {\bibfnamefont {F.}~\bibnamefont {Arute}}, \bibinfo {author} {\bibfnamefont {K.}~\bibnamefont {Arya}}, \bibinfo {author} {\bibfnamefont {A.}~\bibnamefont {Asfaw}}, \bibinfo {author} {\bibfnamefont {J.}~\bibnamefont {Atalaya}}, \bibinfo {author} {\bibfnamefont {R.}~\bibnamefont {Babbush}}, \emph {et~al.},\ }\bibfield  {title} {\bibinfo {title} {Suppressing quantum errors by scaling a surface code logical qubit},\ }\href {https://doi.org/10.1038/s41586-022-05434-1} {\bibfield  {journal} {\bibinfo  {journal} {Nature}\ }\textbf {\bibinfo {volume} {614}},\ \bibinfo {pages} {676} (\bibinfo {year} {2023})}\BibitemShut {NoStop}%
\bibitem [{\citenamefont {Sivak}\ \emph {et~al.}(2023)\citenamefont {Sivak}, \citenamefont {Eickbusch}, \citenamefont {Royer}, \citenamefont {Singh}, \citenamefont {Tsioutsios}, \citenamefont {Ganjam}, \citenamefont {Miano}, \citenamefont {Brock}, \citenamefont {Ding}, \citenamefont {Frunzio} \emph {et~al.}}]{sivak2023real}%
  \BibitemOpen
  \bibfield  {author} {\bibinfo {author} {\bibfnamefont {V.}~\bibnamefont {Sivak}}, \bibinfo {author} {\bibfnamefont {A.}~\bibnamefont {Eickbusch}}, \bibinfo {author} {\bibfnamefont {B.}~\bibnamefont {Royer}}, \bibinfo {author} {\bibfnamefont {S.}~\bibnamefont {Singh}}, \bibinfo {author} {\bibfnamefont {I.}~\bibnamefont {Tsioutsios}}, \bibinfo {author} {\bibfnamefont {S.}~\bibnamefont {Ganjam}}, \bibinfo {author} {\bibfnamefont {A.}~\bibnamefont {Miano}}, \bibinfo {author} {\bibfnamefont {B.}~\bibnamefont {Brock}}, \bibinfo {author} {\bibfnamefont {A.}~\bibnamefont {Ding}}, \bibinfo {author} {\bibfnamefont {L.}~\bibnamefont {Frunzio}}, \emph {et~al.},\ }\bibfield  {title} {\bibinfo {title} {Real-time quantum error correction beyond break-even},\ }\href {https://doi.org/10.1038/s41586-023-05782-6} {\bibfield  {journal} {\bibinfo  {journal} {Nature}\ }\textbf {\bibinfo {volume} {616}},\ \bibinfo {pages} {50} (\bibinfo {year} {2023})}\BibitemShut {NoStop}%
\bibitem [{\citenamefont {Sundaresan}\ \emph {et~al.}(2023)\citenamefont {Sundaresan}, \citenamefont {Yoder}, \citenamefont {Kim}, \citenamefont {Li}, \citenamefont {Chen}, \citenamefont {Harper}, \citenamefont {Thorbeck}, \citenamefont {Cross}, \citenamefont {C{\'o}rcoles},\ and\ \citenamefont {Takita}}]{sundaresan2023demonstrating}%
  \BibitemOpen
  \bibfield  {author} {\bibinfo {author} {\bibfnamefont {N.}~\bibnamefont {Sundaresan}}, \bibinfo {author} {\bibfnamefont {T.~J.}\ \bibnamefont {Yoder}}, \bibinfo {author} {\bibfnamefont {Y.}~\bibnamefont {Kim}}, \bibinfo {author} {\bibfnamefont {M.}~\bibnamefont {Li}}, \bibinfo {author} {\bibfnamefont {E.~H.}\ \bibnamefont {Chen}}, \bibinfo {author} {\bibfnamefont {G.}~\bibnamefont {Harper}}, \bibinfo {author} {\bibfnamefont {T.}~\bibnamefont {Thorbeck}}, \bibinfo {author} {\bibfnamefont {A.~W.}\ \bibnamefont {Cross}}, \bibinfo {author} {\bibfnamefont {A.~D.}\ \bibnamefont {C{\'o}rcoles}},\ and\ \bibinfo {author} {\bibfnamefont {M.}~\bibnamefont {Takita}},\ }\bibfield  {title} {\bibinfo {title} {Demonstrating multi-round subsystem quantum error correction using matching and maximum likelihood decoders},\ }\href {https://doi.org/10.1038/s41467-023-38247-5} {\bibfield  {journal} {\bibinfo  {journal} {Nat. Commun.}\ }\textbf {\bibinfo {volume} {14}},\ \bibinfo {pages} {2852} (\bibinfo {year}
  {2023})}\BibitemShut {NoStop}%
\bibitem [{\citenamefont {Knill}(2005)}]{knill2005quantum}%
  \BibitemOpen
  \bibfield  {author} {\bibinfo {author} {\bibfnamefont {E.}~\bibnamefont {Knill}},\ }\bibfield  {title} {\bibinfo {title} {Quantum computing with realistically noisy devices},\ }\href {https://doi.org/10.1038/nature03350} {\bibfield  {journal} {\bibinfo  {journal} {Nature}\ }\textbf {\bibinfo {volume} {434}},\ \bibinfo {pages} {39} (\bibinfo {year} {2005})}\BibitemShut {NoStop}%
\bibitem [{\citenamefont {Bartolucci}\ \emph {et~al.}(2023)\citenamefont {Bartolucci}, \citenamefont {Birchall}, \citenamefont {Bombin}, \citenamefont {Cable}, \citenamefont {Dawson}, \citenamefont {Gimeno-Segovia}, \citenamefont {Johnston}, \citenamefont {Kieling}, \citenamefont {Nickerson}, \citenamefont {Pant} \emph {et~al.}}]{bartolucci2023fusion}%
  \BibitemOpen
  \bibfield  {author} {\bibinfo {author} {\bibfnamefont {S.}~\bibnamefont {Bartolucci}}, \bibinfo {author} {\bibfnamefont {P.}~\bibnamefont {Birchall}}, \bibinfo {author} {\bibfnamefont {H.}~\bibnamefont {Bombin}}, \bibinfo {author} {\bibfnamefont {H.}~\bibnamefont {Cable}}, \bibinfo {author} {\bibfnamefont {C.}~\bibnamefont {Dawson}}, \bibinfo {author} {\bibfnamefont {M.}~\bibnamefont {Gimeno-Segovia}}, \bibinfo {author} {\bibfnamefont {E.}~\bibnamefont {Johnston}}, \bibinfo {author} {\bibfnamefont {K.}~\bibnamefont {Kieling}}, \bibinfo {author} {\bibfnamefont {N.}~\bibnamefont {Nickerson}}, \bibinfo {author} {\bibfnamefont {M.}~\bibnamefont {Pant}}, \emph {et~al.},\ }\bibfield  {title} {\bibinfo {title} {Fusion-based quantum computation},\ }\href {https://doi.org/10.1038/s41467-023-36493-1} {\bibfield  {journal} {\bibinfo  {journal} {Nat. Commun.}\ }\textbf {\bibinfo {volume} {14}},\ \bibinfo {pages} {912} (\bibinfo {year} {2023})}\BibitemShut {NoStop}%
\bibitem [{\citenamefont {Gupta}\ \emph {et~al.}(2024)\citenamefont {Gupta}, \citenamefont {Sundaresan}, \citenamefont {Alexander}, \citenamefont {Wood}, \citenamefont {Merkel}, \citenamefont {Healy}, \citenamefont {Hillenbrand}, \citenamefont {Jochym-O’Connor}, \citenamefont {Wootton}, \citenamefont {Yoder} \emph {et~al.}}]{gupta2024encoding}%
  \BibitemOpen
  \bibfield  {author} {\bibinfo {author} {\bibfnamefont {R.~S.}\ \bibnamefont {Gupta}}, \bibinfo {author} {\bibfnamefont {N.}~\bibnamefont {Sundaresan}}, \bibinfo {author} {\bibfnamefont {T.}~\bibnamefont {Alexander}}, \bibinfo {author} {\bibfnamefont {C.~J.}\ \bibnamefont {Wood}}, \bibinfo {author} {\bibfnamefont {S.~T.}\ \bibnamefont {Merkel}}, \bibinfo {author} {\bibfnamefont {M.~B.}\ \bibnamefont {Healy}}, \bibinfo {author} {\bibfnamefont {M.}~\bibnamefont {Hillenbrand}}, \bibinfo {author} {\bibfnamefont {T.}~\bibnamefont {Jochym-O’Connor}}, \bibinfo {author} {\bibfnamefont {J.~R.}\ \bibnamefont {Wootton}}, \bibinfo {author} {\bibfnamefont {T.~J.}\ \bibnamefont {Yoder}}, \emph {et~al.},\ }\bibfield  {title} {\bibinfo {title} {Encoding a magic state with beyond break-even fidelity},\ }\href {https://doi.org/10.1038/s41586-023-06846-3} {\bibfield  {journal} {\bibinfo  {journal} {Nature}\ }\textbf {\bibinfo {volume} {625}},\ \bibinfo {pages} {259} (\bibinfo {year} {2024})}\BibitemShut {NoStop}%
\bibitem [{\citenamefont {Pirandola}\ \emph {et~al.}(2015)\citenamefont {Pirandola}, \citenamefont {Eisert}, \citenamefont {Weedbrook}, \citenamefont {Furusawa},\ and\ \citenamefont {Braunstein}}]{pirandola2015advances}%
  \BibitemOpen
  \bibfield  {author} {\bibinfo {author} {\bibfnamefont {S.}~\bibnamefont {Pirandola}}, \bibinfo {author} {\bibfnamefont {J.}~\bibnamefont {Eisert}}, \bibinfo {author} {\bibfnamefont {C.}~\bibnamefont {Weedbrook}}, \bibinfo {author} {\bibfnamefont {A.}~\bibnamefont {Furusawa}},\ and\ \bibinfo {author} {\bibfnamefont {S.~L.}\ \bibnamefont {Braunstein}},\ }\bibfield  {title} {\bibinfo {title} {Advances in quantum teleportation},\ }\href {https://doi.org/10.1038/nphoton.2015.154} {\bibfield  {journal} {\bibinfo  {journal} {Nat. Photonics}\ }\textbf {\bibinfo {volume} {9}},\ \bibinfo {pages} {641} (\bibinfo {year} {2015})}\BibitemShut {NoStop}%
\bibitem [{\citenamefont {Wan}\ \emph {et~al.}(2019)\citenamefont {Wan}, \citenamefont {Kienzler}, \citenamefont {Erickson}, \citenamefont {Mayer}, \citenamefont {Tan}, \citenamefont {Wu}, \citenamefont {Vasconcelos}, \citenamefont {Glancy}, \citenamefont {Knill}, \citenamefont {Wineland} \emph {et~al.}}]{wan2019quantum}%
  \BibitemOpen
  \bibfield  {author} {\bibinfo {author} {\bibfnamefont {Y.}~\bibnamefont {Wan}}, \bibinfo {author} {\bibfnamefont {D.}~\bibnamefont {Kienzler}}, \bibinfo {author} {\bibfnamefont {S.~D.}\ \bibnamefont {Erickson}}, \bibinfo {author} {\bibfnamefont {K.~H.}\ \bibnamefont {Mayer}}, \bibinfo {author} {\bibfnamefont {T.~R.}\ \bibnamefont {Tan}}, \bibinfo {author} {\bibfnamefont {J.~J.}\ \bibnamefont {Wu}}, \bibinfo {author} {\bibfnamefont {H.~M.}\ \bibnamefont {Vasconcelos}}, \bibinfo {author} {\bibfnamefont {S.}~\bibnamefont {Glancy}}, \bibinfo {author} {\bibfnamefont {E.}~\bibnamefont {Knill}}, \bibinfo {author} {\bibfnamefont {D.~J.}\ \bibnamefont {Wineland}}, \emph {et~al.},\ }\bibfield  {title} {\bibinfo {title} {Quantum gate teleportation between separated qubits in a trapped-ion processor},\ }\href {https://doi.org/10.1126/science.aaw9415} {\bibfield  {journal} {\bibinfo  {journal} {Science}\ }\textbf {\bibinfo {volume} {364}},\ \bibinfo {pages} {875} (\bibinfo {year} {2019})}\BibitemShut {NoStop}%
\bibitem [{\citenamefont {Hu}\ \emph {et~al.}(2023)\citenamefont {Hu}, \citenamefont {Guo}, \citenamefont {Liu}, \citenamefont {Li},\ and\ \citenamefont {Guo}}]{hu2023progress}%
  \BibitemOpen
  \bibfield  {author} {\bibinfo {author} {\bibfnamefont {X.-M.}\ \bibnamefont {Hu}}, \bibinfo {author} {\bibfnamefont {Y.}~\bibnamefont {Guo}}, \bibinfo {author} {\bibfnamefont {B.-H.}\ \bibnamefont {Liu}}, \bibinfo {author} {\bibfnamefont {C.-F.}\ \bibnamefont {Li}},\ and\ \bibinfo {author} {\bibfnamefont {G.-C.}\ \bibnamefont {Guo}},\ }\bibfield  {title} {\bibinfo {title} {Progress in quantum teleportation},\ }\href {https://doi.org/10.1038/s42254-023-00588-x} {\bibfield  {journal} {\bibinfo  {journal} {Nat. Rev. Phys.}\ }\textbf {\bibinfo {volume} {5}},\ \bibinfo {pages} {339} (\bibinfo {year} {2023})}\BibitemShut {NoStop}%
\bibitem [{\citenamefont {Rost}\ \emph {et~al.}(2021)\citenamefont {Rost}, \citenamefont {Re}, \citenamefont {Earnest}, \citenamefont {Kemper}, \citenamefont {Jones},\ and\ \citenamefont {Freericks}}]{rost2021demonstrating}%
  \BibitemOpen
  \bibfield  {author} {\bibinfo {author} {\bibfnamefont {B.}~\bibnamefont {Rost}}, \bibinfo {author} {\bibfnamefont {L.~D.}\ \bibnamefont {Re}}, \bibinfo {author} {\bibfnamefont {N.}~\bibnamefont {Earnest}}, \bibinfo {author} {\bibfnamefont {A.~F.}\ \bibnamefont {Kemper}}, \bibinfo {author} {\bibfnamefont {B.}~\bibnamefont {Jones}},\ and\ \bibinfo {author} {\bibfnamefont {J.~K.}\ \bibnamefont {Freericks}},\ }\href {https://doi.org/10.48550/arXiv.2108.01183} {\bibinfo {title} {Demonstrating robust simulation of driven-dissipative problems on near-term quantum computers}} (\bibinfo {year} {2021}),\ \Eprint {https://arxiv.org/abs/2108.01183} {arXiv:2108.01183 [quant-ph]} \BibitemShut {NoStop}%
\bibitem [{\citenamefont {Han}\ \emph {et~al.}(2021)\citenamefont {Han}, \citenamefont {Cai}, \citenamefont {Hu}, \citenamefont {Mu}, \citenamefont {Ma}, \citenamefont {Xu}, \citenamefont {Wang}, \citenamefont {Wang}, \citenamefont {Song}, \citenamefont {Zou},\ and\ \citenamefont {Sun}}]{han2021experimental}%
  \BibitemOpen
  \bibfield  {author} {\bibinfo {author} {\bibfnamefont {J.}~\bibnamefont {Han}}, \bibinfo {author} {\bibfnamefont {W.}~\bibnamefont {Cai}}, \bibinfo {author} {\bibfnamefont {L.}~\bibnamefont {Hu}}, \bibinfo {author} {\bibfnamefont {X.}~\bibnamefont {Mu}}, \bibinfo {author} {\bibfnamefont {Y.}~\bibnamefont {Ma}}, \bibinfo {author} {\bibfnamefont {Y.}~\bibnamefont {Xu}}, \bibinfo {author} {\bibfnamefont {W.}~\bibnamefont {Wang}}, \bibinfo {author} {\bibfnamefont {H.}~\bibnamefont {Wang}}, \bibinfo {author} {\bibfnamefont {Y.~P.}\ \bibnamefont {Song}}, \bibinfo {author} {\bibfnamefont {C.-L.}\ \bibnamefont {Zou}},\ and\ \bibinfo {author} {\bibfnamefont {L.}~\bibnamefont {Sun}},\ }\bibfield  {title} {\bibinfo {title} {Experimental simulation of open quantum system dynamics via {T}rotterization},\ }\href {https://doi.org/10.1103/PhysRevLett.127.020504} {\bibfield  {journal} {\bibinfo  {journal} {Phys. Rev. Lett.}\ }\textbf {\bibinfo {volume} {127}},\ \bibinfo {pages} {020504} (\bibinfo {year} {2021})}\BibitemShut
  {NoStop}%
\bibitem [{\citenamefont {Del~Re}\ \emph {et~al.}(2024)\citenamefont {Del~Re}, \citenamefont {Rost}, \citenamefont {Foss-Feig}, \citenamefont {Kemper},\ and\ \citenamefont {Freericks}}]{del2024robust}%
  \BibitemOpen
  \bibfield  {author} {\bibinfo {author} {\bibfnamefont {L.}~\bibnamefont {Del~Re}}, \bibinfo {author} {\bibfnamefont {B.}~\bibnamefont {Rost}}, \bibinfo {author} {\bibfnamefont {M.}~\bibnamefont {Foss-Feig}}, \bibinfo {author} {\bibfnamefont {A.~F.}\ \bibnamefont {Kemper}},\ and\ \bibinfo {author} {\bibfnamefont {J.~K.}\ \bibnamefont {Freericks}},\ }\bibfield  {title} {\bibinfo {title} {Robust measurements of $n$-point correlation functions of driven-dissipative quantum systems on a digital quantum computer},\ }\href {https://doi.org/10.1103/PhysRevLett.132.100601} {\bibfield  {journal} {\bibinfo  {journal} {Phys. Rev. Lett.}\ }\textbf {\bibinfo {volume} {132}},\ \bibinfo {pages} {100601} (\bibinfo {year} {2024})}\BibitemShut {NoStop}%
\bibitem [{\citenamefont {Bäumer}\ \emph {et~al.}(2023)\citenamefont {Bäumer}, \citenamefont {Tripathi}, \citenamefont {Wang}, \citenamefont {Rall}, \citenamefont {Chen}, \citenamefont {Majumder}, \citenamefont {Seif},\ and\ \citenamefont {Minev}}]{baumer2023efficient}%
  \BibitemOpen
  \bibfield  {author} {\bibinfo {author} {\bibfnamefont {E.}~\bibnamefont {Bäumer}}, \bibinfo {author} {\bibfnamefont {V.}~\bibnamefont {Tripathi}}, \bibinfo {author} {\bibfnamefont {D.~S.}\ \bibnamefont {Wang}}, \bibinfo {author} {\bibfnamefont {P.}~\bibnamefont {Rall}}, \bibinfo {author} {\bibfnamefont {E.~H.}\ \bibnamefont {Chen}}, \bibinfo {author} {\bibfnamefont {S.}~\bibnamefont {Majumder}}, \bibinfo {author} {\bibfnamefont {A.}~\bibnamefont {Seif}},\ and\ \bibinfo {author} {\bibfnamefont {Z.~K.}\ \bibnamefont {Minev}},\ }\href {https://doi.org/10.48550/arXiv.2308.13065} {\bibinfo {title} {Efficient long-range entanglement using dynamic circuits}} (\bibinfo {year} {2023}),\ \Eprint {https://arxiv.org/abs/2308.13065} {arXiv:2308.13065 [quant-ph]} \BibitemShut {NoStop}%
\bibitem [{\citenamefont {Zhu}\ \emph {et~al.}(2023)\citenamefont {Zhu}, \citenamefont {Tantivasadakarn}, \citenamefont {Vishwanath}, \citenamefont {Trebst},\ and\ \citenamefont {Verresen}}]{zhu2023nishimori}%
  \BibitemOpen
  \bibfield  {author} {\bibinfo {author} {\bibfnamefont {G.-Y.}\ \bibnamefont {Zhu}}, \bibinfo {author} {\bibfnamefont {N.}~\bibnamefont {Tantivasadakarn}}, \bibinfo {author} {\bibfnamefont {A.}~\bibnamefont {Vishwanath}}, \bibinfo {author} {\bibfnamefont {S.}~\bibnamefont {Trebst}},\ and\ \bibinfo {author} {\bibfnamefont {R.}~\bibnamefont {Verresen}},\ }\bibfield  {title} {\bibinfo {title} {{N}ishimori's cat: Stable long-range entanglement from finite-depth unitaries and weak measurements},\ }\href {https://doi.org/10.1103/PhysRevLett.131.200201} {\bibfield  {journal} {\bibinfo  {journal} {Phys. Rev. Lett.}\ }\textbf {\bibinfo {volume} {131}},\ \bibinfo {pages} {200201} (\bibinfo {year} {2023})}\BibitemShut {NoStop}%
\bibitem [{\citenamefont {Tantivasadakarn}\ \emph {et~al.}(2023)\citenamefont {Tantivasadakarn}, \citenamefont {Verresen},\ and\ \citenamefont {Vishwanath}}]{tantivasadakarn2023shortest}%
  \BibitemOpen
  \bibfield  {author} {\bibinfo {author} {\bibfnamefont {N.}~\bibnamefont {Tantivasadakarn}}, \bibinfo {author} {\bibfnamefont {R.}~\bibnamefont {Verresen}},\ and\ \bibinfo {author} {\bibfnamefont {A.}~\bibnamefont {Vishwanath}},\ }\bibfield  {title} {\bibinfo {title} {Shortest route to non-{A}belian topological order on a quantum processor},\ }\href {https://doi.org/10.1103/PhysRevLett.131.060405} {\bibfield  {journal} {\bibinfo  {journal} {Phys. Rev. Lett.}\ }\textbf {\bibinfo {volume} {131}},\ \bibinfo {pages} {060405} (\bibinfo {year} {2023})}\BibitemShut {NoStop}%
\bibitem [{\citenamefont {Foss-Feig}\ \emph {et~al.}(2023)\citenamefont {Foss-Feig}, \citenamefont {Tikku}, \citenamefont {Lu}, \citenamefont {Mayer}, \citenamefont {Iqbal}, \citenamefont {Gatterman}, \citenamefont {Gerber}, \citenamefont {Gilmore}, \citenamefont {Gresh}, \citenamefont {Hankin}, \citenamefont {Hewitt}, \citenamefont {Horst}, \citenamefont {Matheny}, \citenamefont {Mengle}, \citenamefont {Neyenhuis}, \citenamefont {Dreyer}, \citenamefont {Hayes}, \citenamefont {Hsieh},\ and\ \citenamefont {Kim}}]{fossfeig2023experimental}%
  \BibitemOpen
  \bibfield  {author} {\bibinfo {author} {\bibfnamefont {M.}~\bibnamefont {Foss-Feig}}, \bibinfo {author} {\bibfnamefont {A.}~\bibnamefont {Tikku}}, \bibinfo {author} {\bibfnamefont {T.-C.}\ \bibnamefont {Lu}}, \bibinfo {author} {\bibfnamefont {K.}~\bibnamefont {Mayer}}, \bibinfo {author} {\bibfnamefont {M.}~\bibnamefont {Iqbal}}, \bibinfo {author} {\bibfnamefont {T.~M.}\ \bibnamefont {Gatterman}}, \bibinfo {author} {\bibfnamefont {J.~A.}\ \bibnamefont {Gerber}}, \bibinfo {author} {\bibfnamefont {K.}~\bibnamefont {Gilmore}}, \bibinfo {author} {\bibfnamefont {D.}~\bibnamefont {Gresh}}, \bibinfo {author} {\bibfnamefont {A.}~\bibnamefont {Hankin}}, \bibinfo {author} {\bibfnamefont {N.}~\bibnamefont {Hewitt}}, \bibinfo {author} {\bibfnamefont {C.~V.}\ \bibnamefont {Horst}}, \bibinfo {author} {\bibfnamefont {M.}~\bibnamefont {Matheny}}, \bibinfo {author} {\bibfnamefont {T.}~\bibnamefont {Mengle}}, \bibinfo {author} {\bibfnamefont {B.}~\bibnamefont {Neyenhuis}}, \bibinfo {author} {\bibfnamefont {H.}~\bibnamefont
  {Dreyer}}, \bibinfo {author} {\bibfnamefont {D.}~\bibnamefont {Hayes}}, \bibinfo {author} {\bibfnamefont {T.~H.}\ \bibnamefont {Hsieh}},\ and\ \bibinfo {author} {\bibfnamefont {I.~H.}\ \bibnamefont {Kim}},\ }\href {https://doi.org/10.48550/arXiv.2302.03029} {\bibinfo {title} {Experimental demonstration of the advantage of adaptive quantum circuits}} (\bibinfo {year} {2023}),\ \Eprint {https://arxiv.org/abs/2302.03029} {arXiv:2302.03029 [quant-ph]} \BibitemShut {NoStop}%
\bibitem [{\citenamefont {Smith}\ \emph {et~al.}(2023)\citenamefont {Smith}, \citenamefont {Crane}, \citenamefont {Wiebe},\ and\ \citenamefont {Girvin}}]{smith2023deterministic}%
  \BibitemOpen
  \bibfield  {author} {\bibinfo {author} {\bibfnamefont {K.~C.}\ \bibnamefont {Smith}}, \bibinfo {author} {\bibfnamefont {E.}~\bibnamefont {Crane}}, \bibinfo {author} {\bibfnamefont {N.}~\bibnamefont {Wiebe}},\ and\ \bibinfo {author} {\bibfnamefont {S.}~\bibnamefont {Girvin}},\ }\bibfield  {title} {\bibinfo {title} {Deterministic constant-depth preparation of the {AKLT} state on a quantum processor using fusion measurements},\ }\href {https://doi.org/10.1103/PRXQuantum.4.020315} {\bibfield  {journal} {\bibinfo  {journal} {PRX Quantum}\ }\textbf {\bibinfo {volume} {4}},\ \bibinfo {pages} {020315} (\bibinfo {year} {2023})}\BibitemShut {NoStop}%
\bibitem [{\citenamefont {Iqbal}\ \emph {et~al.}(2024)\citenamefont {Iqbal}, \citenamefont {Tantivasadakarn}, \citenamefont {Verresen}, \citenamefont {Campbell}, \citenamefont {Dreiling}, \citenamefont {Figgatt}, \citenamefont {Gaebler}, \citenamefont {Johansen}, \citenamefont {Mills}, \citenamefont {Moses} \emph {et~al.}}]{iqbal2024non}%
  \BibitemOpen
  \bibfield  {author} {\bibinfo {author} {\bibfnamefont {M.}~\bibnamefont {Iqbal}}, \bibinfo {author} {\bibfnamefont {N.}~\bibnamefont {Tantivasadakarn}}, \bibinfo {author} {\bibfnamefont {R.}~\bibnamefont {Verresen}}, \bibinfo {author} {\bibfnamefont {S.~L.}\ \bibnamefont {Campbell}}, \bibinfo {author} {\bibfnamefont {J.~M.}\ \bibnamefont {Dreiling}}, \bibinfo {author} {\bibfnamefont {C.}~\bibnamefont {Figgatt}}, \bibinfo {author} {\bibfnamefont {J.~P.}\ \bibnamefont {Gaebler}}, \bibinfo {author} {\bibfnamefont {J.}~\bibnamefont {Johansen}}, \bibinfo {author} {\bibfnamefont {M.}~\bibnamefont {Mills}}, \bibinfo {author} {\bibfnamefont {S.~A.}\ \bibnamefont {Moses}}, \emph {et~al.},\ }\bibfield  {title} {\bibinfo {title} {Non-{A}belian topological order and anyons on a trapped-ion processor},\ }\href {https://doi.org/10.1038/s41586-023-06934-4} {\bibfield  {journal} {\bibinfo  {journal} {Nature}\ }\textbf {\bibinfo {volume} {626}},\ \bibinfo {pages} {505} (\bibinfo {year} {2024})}\BibitemShut {NoStop}%
\bibitem [{\citenamefont {Cai}\ \emph {et~al.}(2023)\citenamefont {Cai}, \citenamefont {Babbush}, \citenamefont {Benjamin}, \citenamefont {Endo}, \citenamefont {Huggins}, \citenamefont {Li}, \citenamefont {McClean},\ and\ \citenamefont {O'Brien}}]{cai2023quantum}%
  \BibitemOpen
  \bibfield  {author} {\bibinfo {author} {\bibfnamefont {Z.}~\bibnamefont {Cai}}, \bibinfo {author} {\bibfnamefont {R.}~\bibnamefont {Babbush}}, \bibinfo {author} {\bibfnamefont {S.~C.}\ \bibnamefont {Benjamin}}, \bibinfo {author} {\bibfnamefont {S.}~\bibnamefont {Endo}}, \bibinfo {author} {\bibfnamefont {W.~J.}\ \bibnamefont {Huggins}}, \bibinfo {author} {\bibfnamefont {Y.}~\bibnamefont {Li}}, \bibinfo {author} {\bibfnamefont {J.~R.}\ \bibnamefont {McClean}},\ and\ \bibinfo {author} {\bibfnamefont {T.~E.}\ \bibnamefont {O'Brien}},\ }\bibfield  {title} {\bibinfo {title} {Quantum error mitigation},\ }\href {https://doi.org/10.1103/RevModPhys.95.045005} {\bibfield  {journal} {\bibinfo  {journal} {Rev. Mod. Phys.}\ }\textbf {\bibinfo {volume} {95}},\ \bibinfo {pages} {045005} (\bibinfo {year} {2023})}\BibitemShut {NoStop}%
\bibitem [{\citenamefont {Lin}\ \emph {et~al.}(2021)\citenamefont {Lin}, \citenamefont {Wallman}, \citenamefont {Hincks},\ and\ \citenamefont {Laflamme}}]{lin2021independent}%
  \BibitemOpen
  \bibfield  {author} {\bibinfo {author} {\bibfnamefont {J.}~\bibnamefont {Lin}}, \bibinfo {author} {\bibfnamefont {J.~J.}\ \bibnamefont {Wallman}}, \bibinfo {author} {\bibfnamefont {I.}~\bibnamefont {Hincks}},\ and\ \bibinfo {author} {\bibfnamefont {R.}~\bibnamefont {Laflamme}},\ }\bibfield  {title} {\bibinfo {title} {Independent state and measurement characterization for quantum computers},\ }\href {https://doi.org/10.1103/PhysRevResearch.3.033285} {\bibfield  {journal} {\bibinfo  {journal} {Phys. Rev. Res.}\ }\textbf {\bibinfo {volume} {3}},\ \bibinfo {pages} {033285} (\bibinfo {year} {2021})}\BibitemShut {NoStop}%
\bibitem [{\citenamefont {Kandala}\ \emph {et~al.}(2017)\citenamefont {Kandala}, \citenamefont {Mezzacapo}, \citenamefont {Temme}, \citenamefont {Takita}, \citenamefont {Brink}, \citenamefont {Chow},\ and\ \citenamefont {Gambetta}}]{kandala2017hardware}%
  \BibitemOpen
  \bibfield  {author} {\bibinfo {author} {\bibfnamefont {A.}~\bibnamefont {Kandala}}, \bibinfo {author} {\bibfnamefont {A.}~\bibnamefont {Mezzacapo}}, \bibinfo {author} {\bibfnamefont {K.}~\bibnamefont {Temme}}, \bibinfo {author} {\bibfnamefont {M.}~\bibnamefont {Takita}}, \bibinfo {author} {\bibfnamefont {M.}~\bibnamefont {Brink}}, \bibinfo {author} {\bibfnamefont {J.~M.}\ \bibnamefont {Chow}},\ and\ \bibinfo {author} {\bibfnamefont {J.~M.}\ \bibnamefont {Gambetta}},\ }\bibfield  {title} {\bibinfo {title} {Hardware-efficient variational quantum eigensolver for small molecules and quantum magnets},\ }\href {https://doi.org/10.1038/nature23879} {\bibfield  {journal} {\bibinfo  {journal} {Nature}\ }\textbf {\bibinfo {volume} {549}},\ \bibinfo {pages} {242} (\bibinfo {year} {2017})}\BibitemShut {NoStop}%
\bibitem [{\citenamefont {Kandala}\ \emph {et~al.}(2019)\citenamefont {Kandala}, \citenamefont {Temme}, \citenamefont {C{\'o}rcoles}, \citenamefont {Mezzacapo}, \citenamefont {Chow},\ and\ \citenamefont {Gambetta}}]{kandala2019error}%
  \BibitemOpen
  \bibfield  {author} {\bibinfo {author} {\bibfnamefont {A.}~\bibnamefont {Kandala}}, \bibinfo {author} {\bibfnamefont {K.}~\bibnamefont {Temme}}, \bibinfo {author} {\bibfnamefont {A.~D.}\ \bibnamefont {C{\'o}rcoles}}, \bibinfo {author} {\bibfnamefont {A.}~\bibnamefont {Mezzacapo}}, \bibinfo {author} {\bibfnamefont {J.~M.}\ \bibnamefont {Chow}},\ and\ \bibinfo {author} {\bibfnamefont {J.~M.}\ \bibnamefont {Gambetta}},\ }\bibfield  {title} {\bibinfo {title} {Error mitigation extends the computational reach of a noisy quantum processor},\ }\href {https://doi.org/10.1038/s41586-019-1040-7} {\bibfield  {journal} {\bibinfo  {journal} {Nature}\ }\textbf {\bibinfo {volume} {567}},\ \bibinfo {pages} {491} (\bibinfo {year} {2019})}\BibitemShut {NoStop}%
\bibitem [{\citenamefont {Jurcevic}\ \emph {et~al.}(2021)\citenamefont {Jurcevic}, \citenamefont {Javadi-Abhari}, \citenamefont {Bishop}, \citenamefont {Lauer}, \citenamefont {Bogorin}, \citenamefont {Brink}, \citenamefont {Capelluto}, \citenamefont {G{\"u}nl{\"u}k}, \citenamefont {Itoko}, \citenamefont {Kanazawa} \emph {et~al.}}]{jurcevic2021demonstration}%
  \BibitemOpen
  \bibfield  {author} {\bibinfo {author} {\bibfnamefont {P.}~\bibnamefont {Jurcevic}}, \bibinfo {author} {\bibfnamefont {A.}~\bibnamefont {Javadi-Abhari}}, \bibinfo {author} {\bibfnamefont {L.~S.}\ \bibnamefont {Bishop}}, \bibinfo {author} {\bibfnamefont {I.}~\bibnamefont {Lauer}}, \bibinfo {author} {\bibfnamefont {D.~F.}\ \bibnamefont {Bogorin}}, \bibinfo {author} {\bibfnamefont {M.}~\bibnamefont {Brink}}, \bibinfo {author} {\bibfnamefont {L.}~\bibnamefont {Capelluto}}, \bibinfo {author} {\bibfnamefont {O.}~\bibnamefont {G{\"u}nl{\"u}k}}, \bibinfo {author} {\bibfnamefont {T.}~\bibnamefont {Itoko}}, \bibinfo {author} {\bibfnamefont {N.}~\bibnamefont {Kanazawa}}, \emph {et~al.},\ }\bibfield  {title} {\bibinfo {title} {Demonstration of quantum volume 64 on a superconducting quantum computing system},\ }\href {https://doi.org/10.1088/2058-9565/abe519} {\bibfield  {journal} {\bibinfo  {journal} {Quantum Sci. Technol.}\ }\textbf {\bibinfo {volume} {6}},\ \bibinfo {pages} {025020} (\bibinfo {year}
  {2021})}\BibitemShut {NoStop}%
\bibitem [{\citenamefont {Van Den~Berg}\ \emph {et~al.}(2023)\citenamefont {Van Den~Berg}, \citenamefont {Minev}, \citenamefont {Kandala},\ and\ \citenamefont {Temme}}]{van2023probabilistic}%
  \BibitemOpen
  \bibfield  {author} {\bibinfo {author} {\bibfnamefont {E.}~\bibnamefont {Van Den~Berg}}, \bibinfo {author} {\bibfnamefont {Z.~K.}\ \bibnamefont {Minev}}, \bibinfo {author} {\bibfnamefont {A.}~\bibnamefont {Kandala}},\ and\ \bibinfo {author} {\bibfnamefont {K.}~\bibnamefont {Temme}},\ }\bibfield  {title} {\bibinfo {title} {Probabilistic error cancellation with sparse {P}auli--{L}indblad models on noisy quantum processors},\ }\href {https://doi.org/10.1038/s41567-023-02042-2} {\bibfield  {journal} {\bibinfo  {journal} {Nat. Phys.}\ }\textbf {\bibinfo {volume} {19}},\ \bibinfo {pages} {1116} (\bibinfo {year} {2023})}\BibitemShut {NoStop}%
\bibitem [{\citenamefont {Gupta}\ \emph {et~al.}(2023)\citenamefont {Gupta}, \citenamefont {van~den Berg}, \citenamefont {Takita}, \citenamefont {Riste}, \citenamefont {Temme},\ and\ \citenamefont {Kandala}}]{gupta2023probabilistic}%
  \BibitemOpen
  \bibfield  {author} {\bibinfo {author} {\bibfnamefont {R.~S.}\ \bibnamefont {Gupta}}, \bibinfo {author} {\bibfnamefont {E.}~\bibnamefont {van~den Berg}}, \bibinfo {author} {\bibfnamefont {M.}~\bibnamefont {Takita}}, \bibinfo {author} {\bibfnamefont {D.}~\bibnamefont {Riste}}, \bibinfo {author} {\bibfnamefont {K.}~\bibnamefont {Temme}},\ and\ \bibinfo {author} {\bibfnamefont {A.}~\bibnamefont {Kandala}},\ }\href {https://doi.org/10.48550/arXiv.2310.07825} {\bibinfo {title} {Probabilistic error cancellation for dynamic quantum circuits}} (\bibinfo {year} {2023}),\ \Eprint {https://arxiv.org/abs/2310.07825} {arXiv:2310.07825 [quant-ph]} \BibitemShut {NoStop}%
\bibitem [{\citenamefont {Li}\ and\ \citenamefont {Benjamin}(2017)}]{li2017efficient}%
  \BibitemOpen
  \bibfield  {author} {\bibinfo {author} {\bibfnamefont {Y.}~\bibnamefont {Li}}\ and\ \bibinfo {author} {\bibfnamefont {S.~C.}\ \bibnamefont {Benjamin}},\ }\bibfield  {title} {\bibinfo {title} {Efficient variational quantum simulator incorporating active error minimization},\ }\href {https://doi.org/10.1103/PhysRevX.7.021050} {\bibfield  {journal} {\bibinfo  {journal} {Phys. Rev. X}\ }\textbf {\bibinfo {volume} {7}},\ \bibinfo {pages} {021050} (\bibinfo {year} {2017})}\BibitemShut {NoStop}%
\bibitem [{\citenamefont {Temme}\ \emph {et~al.}(2017)\citenamefont {Temme}, \citenamefont {Bravyi},\ and\ \citenamefont {Gambetta}}]{temme2017error}%
  \BibitemOpen
  \bibfield  {author} {\bibinfo {author} {\bibfnamefont {K.}~\bibnamefont {Temme}}, \bibinfo {author} {\bibfnamefont {S.}~\bibnamefont {Bravyi}},\ and\ \bibinfo {author} {\bibfnamefont {J.~M.}\ \bibnamefont {Gambetta}},\ }\bibfield  {title} {\bibinfo {title} {Error mitigation for short-depth quantum circuits},\ }\href {https://doi.org/10.1103/PhysRevLett.119.180509} {\bibfield  {journal} {\bibinfo  {journal} {Phys. Rev. Lett.}\ }\textbf {\bibinfo {volume} {119}},\ \bibinfo {pages} {180509} (\bibinfo {year} {2017})}\BibitemShut {NoStop}%
\bibitem [{\citenamefont {Giurgica-Tiron}\ \emph {et~al.}(2020)\citenamefont {Giurgica-Tiron}, \citenamefont {Hindy}, \citenamefont {LaRose}, \citenamefont {Mari},\ and\ \citenamefont {Zeng}}]{giurgica2020digital}%
  \BibitemOpen
  \bibfield  {author} {\bibinfo {author} {\bibfnamefont {T.}~\bibnamefont {Giurgica-Tiron}}, \bibinfo {author} {\bibfnamefont {Y.}~\bibnamefont {Hindy}}, \bibinfo {author} {\bibfnamefont {R.}~\bibnamefont {LaRose}}, \bibinfo {author} {\bibfnamefont {A.}~\bibnamefont {Mari}},\ and\ \bibinfo {author} {\bibfnamefont {W.~J.}\ \bibnamefont {Zeng}},\ }\bibfield  {title} {\bibinfo {title} {Digital zero noise extrapolation for quantum error mitigation},\ }in\ \href {https://doi.org/10.1109/QCE49297.2020.00045} {\emph {\bibinfo {booktitle} {2020 IEEE International Conference on Quantum Computing and Engineering (QCE)}}}\ (\bibinfo {organization} {IEEE},\ \bibinfo {year} {2020})\ pp.\ \bibinfo {pages} {306--316}\BibitemShut {NoStop}%
\bibitem [{\citenamefont {Czarnik}\ \emph {et~al.}(2021)\citenamefont {Czarnik}, \citenamefont {Arrasmith}, \citenamefont {Coles},\ and\ \citenamefont {Cincio}}]{czarnik2021error}%
  \BibitemOpen
  \bibfield  {author} {\bibinfo {author} {\bibfnamefont {P.}~\bibnamefont {Czarnik}}, \bibinfo {author} {\bibfnamefont {A.}~\bibnamefont {Arrasmith}}, \bibinfo {author} {\bibfnamefont {P.~J.}\ \bibnamefont {Coles}},\ and\ \bibinfo {author} {\bibfnamefont {L.}~\bibnamefont {Cincio}},\ }\bibfield  {title} {\bibinfo {title} {Error mitigation with {C}lifford quantum-circuit data},\ }\href {https://doi.org/10.22331/q-2021-11-26-592} {\bibfield  {journal} {\bibinfo  {journal} {Quantum}\ }\textbf {\bibinfo {volume} {5}},\ \bibinfo {pages} {592} (\bibinfo {year} {2021})}\BibitemShut {NoStop}%
\bibitem [{\citenamefont {Liao}\ \emph {et~al.}(2023)\citenamefont {Liao}, \citenamefont {Wang}, \citenamefont {Sitdikov}, \citenamefont {Salcedo}, \citenamefont {Seif},\ and\ \citenamefont {Minev}}]{liao2023machine}%
  \BibitemOpen
  \bibfield  {author} {\bibinfo {author} {\bibfnamefont {H.}~\bibnamefont {Liao}}, \bibinfo {author} {\bibfnamefont {D.~S.}\ \bibnamefont {Wang}}, \bibinfo {author} {\bibfnamefont {I.}~\bibnamefont {Sitdikov}}, \bibinfo {author} {\bibfnamefont {C.}~\bibnamefont {Salcedo}}, \bibinfo {author} {\bibfnamefont {A.}~\bibnamefont {Seif}},\ and\ \bibinfo {author} {\bibfnamefont {Z.~K.}\ \bibnamefont {Minev}},\ }\href {https://doi.org/10.48550/arXiv.2309.17368} {\bibinfo {title} {Machine learning for practical quantum error mitigation}} (\bibinfo {year} {2023}),\ \Eprint {https://arxiv.org/abs/2309.17368} {arXiv:2309.17368 [quant-ph]} \BibitemShut {NoStop}%
\bibitem [{\citenamefont {Wallman}\ and\ \citenamefont {Emerson}(2016)}]{wallman2016noise}%
  \BibitemOpen
  \bibfield  {author} {\bibinfo {author} {\bibfnamefont {J.~J.}\ \bibnamefont {Wallman}}\ and\ \bibinfo {author} {\bibfnamefont {J.}~\bibnamefont {Emerson}},\ }\bibfield  {title} {\bibinfo {title} {Noise tailoring for scalable quantum computation via randomized compiling},\ }\href {https://doi.org/10.1103/PhysRevA.94.052325} {\bibfield  {journal} {\bibinfo  {journal} {Phys. Rev. A}\ }\textbf {\bibinfo {volume} {94}},\ \bibinfo {pages} {052325} (\bibinfo {year} {2016})}\BibitemShut {NoStop}%
\bibitem [{\citenamefont {Hashim}\ \emph {et~al.}(2021)\citenamefont {Hashim}, \citenamefont {Naik}, \citenamefont {Morvan}, \citenamefont {Ville}, \citenamefont {Mitchell}, \citenamefont {Kreikebaum}, \citenamefont {Davis}, \citenamefont {Smith}, \citenamefont {Iancu}, \citenamefont {O'Brien}, \citenamefont {Hincks}, \citenamefont {Wallman}, \citenamefont {Emerson},\ and\ \citenamefont {Siddiqi}}]{hashim2021randomized}%
  \BibitemOpen
  \bibfield  {author} {\bibinfo {author} {\bibfnamefont {A.}~\bibnamefont {Hashim}}, \bibinfo {author} {\bibfnamefont {R.~K.}\ \bibnamefont {Naik}}, \bibinfo {author} {\bibfnamefont {A.}~\bibnamefont {Morvan}}, \bibinfo {author} {\bibfnamefont {J.-L.}\ \bibnamefont {Ville}}, \bibinfo {author} {\bibfnamefont {B.}~\bibnamefont {Mitchell}}, \bibinfo {author} {\bibfnamefont {J.~M.}\ \bibnamefont {Kreikebaum}}, \bibinfo {author} {\bibfnamefont {M.}~\bibnamefont {Davis}}, \bibinfo {author} {\bibfnamefont {E.}~\bibnamefont {Smith}}, \bibinfo {author} {\bibfnamefont {C.}~\bibnamefont {Iancu}}, \bibinfo {author} {\bibfnamefont {K.~P.}\ \bibnamefont {O'Brien}}, \bibinfo {author} {\bibfnamefont {I.}~\bibnamefont {Hincks}}, \bibinfo {author} {\bibfnamefont {J.~J.}\ \bibnamefont {Wallman}}, \bibinfo {author} {\bibfnamefont {J.}~\bibnamefont {Emerson}},\ and\ \bibinfo {author} {\bibfnamefont {I.}~\bibnamefont {Siddiqi}},\ }\bibfield  {title} {\bibinfo {title} {Randomized compiling for scalable quantum computing on a noisy
  superconducting quantum processor},\ }\href {https://doi.org/10.1103/PhysRevX.11.041039} {\bibfield  {journal} {\bibinfo  {journal} {Phys. Rev. X}\ }\textbf {\bibinfo {volume} {11}},\ \bibinfo {pages} {041039} (\bibinfo {year} {2021})}\BibitemShut {NoStop}%
\bibitem [{\citenamefont {Sun}\ \emph {et~al.}(2024)\citenamefont {Sun}, \citenamefont {Marinelli}, \citenamefont {Koh}, \citenamefont {Kim}, \citenamefont {Nguyen}, \citenamefont {Chen}, \citenamefont {Kreikebaum}, \citenamefont {Santiago}, \citenamefont {Siddiqi},\ and\ \citenamefont {Minnich}}]{sun2024quantum}%
  \BibitemOpen
  \bibfield  {author} {\bibinfo {author} {\bibfnamefont {S.-N.}\ \bibnamefont {Sun}}, \bibinfo {author} {\bibfnamefont {B.}~\bibnamefont {Marinelli}}, \bibinfo {author} {\bibfnamefont {J.~M.}\ \bibnamefont {Koh}}, \bibinfo {author} {\bibfnamefont {Y.}~\bibnamefont {Kim}}, \bibinfo {author} {\bibfnamefont {L.~B.}\ \bibnamefont {Nguyen}}, \bibinfo {author} {\bibfnamefont {L.}~\bibnamefont {Chen}}, \bibinfo {author} {\bibfnamefont {J.~M.}\ \bibnamefont {Kreikebaum}}, \bibinfo {author} {\bibfnamefont {D.~I.}\ \bibnamefont {Santiago}}, \bibinfo {author} {\bibfnamefont {I.}~\bibnamefont {Siddiqi}},\ and\ \bibinfo {author} {\bibfnamefont {A.~J.}\ \bibnamefont {Minnich}},\ }\bibfield  {title} {\bibinfo {title} {Quantum computation of frequency-domain molecular response properties using a three-qubit itoffoli gate},\ }\href {https://doi.org/10.1038/s41534-024-00850-9} {\bibfield  {journal} {\bibinfo  {journal} {npj Quantum Inf.}\ }\textbf {\bibinfo {volume} {10}},\ \bibinfo {pages} {55} (\bibinfo {year}
  {2024})}\BibitemShut {NoStop}%
\bibitem [{\citenamefont {Smith}\ \emph {et~al.}(2021)\citenamefont {Smith}, \citenamefont {Khosla}, \citenamefont {Self},\ and\ \citenamefont {Kim}}]{smith2021qubit}%
  \BibitemOpen
  \bibfield  {author} {\bibinfo {author} {\bibfnamefont {A.~W.}\ \bibnamefont {Smith}}, \bibinfo {author} {\bibfnamefont {K.~E.}\ \bibnamefont {Khosla}}, \bibinfo {author} {\bibfnamefont {C.~N.}\ \bibnamefont {Self}},\ and\ \bibinfo {author} {\bibfnamefont {M.}~\bibnamefont {Kim}},\ }\bibfield  {title} {\bibinfo {title} {Qubit readout error mitigation with bit-flip averaging},\ }\href {https://doi.org/10.1126/sciadv.abi8009} {\bibfield  {journal} {\bibinfo  {journal} {Sci. Adv.}\ }\textbf {\bibinfo {volume} {7}},\ \bibinfo {pages} {eabi8009} (\bibinfo {year} {2021})}\BibitemShut {NoStop}%
\bibitem [{\citenamefont {Hicks}\ \emph {et~al.}(2021)\citenamefont {Hicks}, \citenamefont {Bauer},\ and\ \citenamefont {Nachman}}]{hicks2021readout}%
  \BibitemOpen
  \bibfield  {author} {\bibinfo {author} {\bibfnamefont {R.}~\bibnamefont {Hicks}}, \bibinfo {author} {\bibfnamefont {C.~W.}\ \bibnamefont {Bauer}},\ and\ \bibinfo {author} {\bibfnamefont {B.}~\bibnamefont {Nachman}},\ }\bibfield  {title} {\bibinfo {title} {Readout rebalancing for near-term quantum computers},\ }\href {https://doi.org/10.1103/PhysRevA.103.022407} {\bibfield  {journal} {\bibinfo  {journal} {Phys. Rev. A}\ }\textbf {\bibinfo {volume} {103}},\ \bibinfo {pages} {022407} (\bibinfo {year} {2021})}\BibitemShut {NoStop}%
\bibitem [{\citenamefont {Chow}\ \emph {et~al.}(2010)\citenamefont {Chow}, \citenamefont {DiCarlo}, \citenamefont {Gambetta}, \citenamefont {Nunnenkamp}, \citenamefont {Bishop}, \citenamefont {Frunzio}, \citenamefont {Devoret}, \citenamefont {Girvin},\ and\ \citenamefont {Schoelkopf}}]{chow2010detecting}%
  \BibitemOpen
  \bibfield  {author} {\bibinfo {author} {\bibfnamefont {J.~M.}\ \bibnamefont {Chow}}, \bibinfo {author} {\bibfnamefont {L.}~\bibnamefont {DiCarlo}}, \bibinfo {author} {\bibfnamefont {J.~M.}\ \bibnamefont {Gambetta}}, \bibinfo {author} {\bibfnamefont {A.}~\bibnamefont {Nunnenkamp}}, \bibinfo {author} {\bibfnamefont {L.~S.}\ \bibnamefont {Bishop}}, \bibinfo {author} {\bibfnamefont {L.}~\bibnamefont {Frunzio}}, \bibinfo {author} {\bibfnamefont {M.~H.}\ \bibnamefont {Devoret}}, \bibinfo {author} {\bibfnamefont {S.~M.}\ \bibnamefont {Girvin}},\ and\ \bibinfo {author} {\bibfnamefont {R.~J.}\ \bibnamefont {Schoelkopf}},\ }\bibfield  {title} {\bibinfo {title} {Detecting highly entangled states with a joint qubit readout},\ }\href {https://doi.org/10.1103/PhysRevA.81.062325} {\bibfield  {journal} {\bibinfo  {journal} {Phys. Rev. A}\ }\textbf {\bibinfo {volume} {81}},\ \bibinfo {pages} {062325} (\bibinfo {year} {2010})}\BibitemShut {NoStop}%
\bibitem [{\citenamefont {Bravyi}\ \emph {et~al.}(2021)\citenamefont {Bravyi}, \citenamefont {Sheldon}, \citenamefont {Kandala}, \citenamefont {Mckay},\ and\ \citenamefont {Gambetta}}]{bravyi2021mitigating}%
  \BibitemOpen
  \bibfield  {author} {\bibinfo {author} {\bibfnamefont {S.}~\bibnamefont {Bravyi}}, \bibinfo {author} {\bibfnamefont {S.}~\bibnamefont {Sheldon}}, \bibinfo {author} {\bibfnamefont {A.}~\bibnamefont {Kandala}}, \bibinfo {author} {\bibfnamefont {D.~C.}\ \bibnamefont {Mckay}},\ and\ \bibinfo {author} {\bibfnamefont {J.~M.}\ \bibnamefont {Gambetta}},\ }\bibfield  {title} {\bibinfo {title} {Mitigating measurement errors in multiqubit experiments},\ }\href {https://doi.org/10.1103/PhysRevA.103.042605} {\bibfield  {journal} {\bibinfo  {journal} {Phys. Rev. A}\ }\textbf {\bibinfo {volume} {103}},\ \bibinfo {pages} {042605} (\bibinfo {year} {2021})}\BibitemShut {NoStop}%
\bibitem [{\citenamefont {Barron}\ and\ \citenamefont {Wood}(2020)}]{barron2020measurement}%
  \BibitemOpen
  \bibfield  {author} {\bibinfo {author} {\bibfnamefont {G.~S.}\ \bibnamefont {Barron}}\ and\ \bibinfo {author} {\bibfnamefont {C.~J.}\ \bibnamefont {Wood}},\ }\href@noop {} {\bibinfo {title} {Measurement error mitigation for variational quantum algorithms}} (\bibinfo {year} {2020}),\ \Eprint {https://arxiv.org/abs/2010.08520} {arXiv:2010.08520 [quant-ph]} \BibitemShut {NoStop}%
\bibitem [{\citenamefont {Nation}\ \emph {et~al.}(2021)\citenamefont {Nation}, \citenamefont {Kang}, \citenamefont {Sundaresan},\ and\ \citenamefont {Gambetta}}]{nation2021scalable}%
  \BibitemOpen
  \bibfield  {author} {\bibinfo {author} {\bibfnamefont {P.~D.}\ \bibnamefont {Nation}}, \bibinfo {author} {\bibfnamefont {H.}~\bibnamefont {Kang}}, \bibinfo {author} {\bibfnamefont {N.}~\bibnamefont {Sundaresan}},\ and\ \bibinfo {author} {\bibfnamefont {J.~M.}\ \bibnamefont {Gambetta}},\ }\bibfield  {title} {\bibinfo {title} {Scalable mitigation of measurement errors on quantum computers},\ }\href {https://doi.org/10.1103/PRXQuantum.2.040326} {\bibfield  {journal} {\bibinfo  {journal} {PRX Quantum}\ }\textbf {\bibinfo {volume} {2}},\ \bibinfo {pages} {040326} (\bibinfo {year} {2021})}\BibitemShut {NoStop}%
\bibitem [{\citenamefont {Ahmed}\ and\ \citenamefont {Rao}(1975)}]{ahmed1975walsh}%
  \BibitemOpen
  \bibfield  {author} {\bibinfo {author} {\bibfnamefont {N.}~\bibnamefont {Ahmed}}\ and\ \bibinfo {author} {\bibfnamefont {K.~R.}\ \bibnamefont {Rao}},\ }\bibinfo {title} {Walsh-{H}adamard transform},\ in\ \href {https://doi.org/10.1007/978-3-642-45450-9_6} {\emph {\bibinfo {booktitle} {Orthogonal Transforms for Digital Signal Processing}}}\ (\bibinfo  {publisher} {Springer Berlin Heidelberg},\ \bibinfo {address} {Berlin, Heidelberg},\ \bibinfo {year} {1975})\ pp.\ \bibinfo {pages} {99--152}\BibitemShut {NoStop}%
\bibitem [{\citenamefont {Beer}(1981)}]{beer1981walsh}%
  \BibitemOpen
  \bibfield  {author} {\bibinfo {author} {\bibfnamefont {T.}~\bibnamefont {Beer}},\ }\bibfield  {title} {\bibinfo {title} {Walsh transforms},\ }\href {https://doi.org/10.1119/1.12714} {\bibfield  {journal} {\bibinfo  {journal} {Am. J. Phys.}\ }\textbf {\bibinfo {volume} {49}},\ \bibinfo {pages} {466} (\bibinfo {year} {1981})}\BibitemShut {NoStop}%
\bibitem [{\citenamefont {Marsaglia}\ \emph {et~al.}(2004)\citenamefont {Marsaglia}, \citenamefont {Tsang},\ and\ \citenamefont {Wang}}]{marsaglia2004fast}%
  \BibitemOpen
  \bibfield  {author} {\bibinfo {author} {\bibfnamefont {G.}~\bibnamefont {Marsaglia}}, \bibinfo {author} {\bibfnamefont {W.~W.}\ \bibnamefont {Tsang}},\ and\ \bibinfo {author} {\bibfnamefont {J.}~\bibnamefont {Wang}},\ }\bibfield  {title} {\bibinfo {title} {Fast generation of discrete random variables},\ }\href {https://doi.org/10.18637/jss.v011.i03} {\bibfield  {journal} {\bibinfo  {journal} {J. Stat. Softw.}\ }\textbf {\bibinfo {volume} {11}},\ \bibinfo {pages} {1} (\bibinfo {year} {2004})}\BibitemShut {NoStop}%
\bibitem [{\citenamefont {Vose}(1991)}]{vose1991linear}%
  \BibitemOpen
  \bibfield  {author} {\bibinfo {author} {\bibfnamefont {M.~D.}\ \bibnamefont {Vose}},\ }\bibfield  {title} {\bibinfo {title} {A linear algorithm for generating random numbers with a given distribution},\ }\href {https://doi.org/10.1109/32.92917} {\bibfield  {journal} {\bibinfo  {journal} {IEEE Trans. Softw. Eng.}\ }\textbf {\bibinfo {volume} {17}},\ \bibinfo {pages} {972} (\bibinfo {year} {1991})}\BibitemShut {NoStop}%
\bibitem [{\citenamefont {Walker}(1977)}]{walker1977efficient}%
  \BibitemOpen
  \bibfield  {author} {\bibinfo {author} {\bibfnamefont {A.~J.}\ \bibnamefont {Walker}},\ }\bibfield  {title} {\bibinfo {title} {An efficient method for generating discrete random variables with general distributions},\ }\href {https://doi.org/10.1145/355744.355749} {\bibfield  {journal} {\bibinfo  {journal} {ACM Trans. Math. Softw.}\ }\textbf {\bibinfo {volume} {3}},\ \bibinfo {pages} {253} (\bibinfo {year} {1977})}\BibitemShut {NoStop}%
\bibitem [{\citenamefont {Suter}\ and\ \citenamefont {\'Alvarez}(2016)}]{suter2016colloquium}%
  \BibitemOpen
  \bibfield  {author} {\bibinfo {author} {\bibfnamefont {D.}~\bibnamefont {Suter}}\ and\ \bibinfo {author} {\bibfnamefont {G.~A.}\ \bibnamefont {\'Alvarez}},\ }\bibfield  {title} {\bibinfo {title} {Colloquium: Protecting quantum information against environmental noise},\ }\href {https://doi.org/10.1103/RevModPhys.88.041001} {\bibfield  {journal} {\bibinfo  {journal} {Rev. Mod. Phys.}\ }\textbf {\bibinfo {volume} {88}},\ \bibinfo {pages} {041001} (\bibinfo {year} {2016})}\BibitemShut {NoStop}%
\bibitem [{\citenamefont {Clerk}\ \emph {et~al.}(2010)\citenamefont {Clerk}, \citenamefont {Devoret}, \citenamefont {Girvin}, \citenamefont {Marquardt},\ and\ \citenamefont {Schoelkopf}}]{clerk2010intro}%
  \BibitemOpen
  \bibfield  {author} {\bibinfo {author} {\bibfnamefont {A.~A.}\ \bibnamefont {Clerk}}, \bibinfo {author} {\bibfnamefont {M.~H.}\ \bibnamefont {Devoret}}, \bibinfo {author} {\bibfnamefont {S.~M.}\ \bibnamefont {Girvin}}, \bibinfo {author} {\bibfnamefont {F.}~\bibnamefont {Marquardt}},\ and\ \bibinfo {author} {\bibfnamefont {R.~J.}\ \bibnamefont {Schoelkopf}},\ }\bibfield  {title} {\bibinfo {title} {Introduction to quantum noise, measurement, and amplification},\ }\href {https://doi.org/10.1103/RevModPhys.82.1155} {\bibfield  {journal} {\bibinfo  {journal} {Rev. Mod. Phys.}\ }\textbf {\bibinfo {volume} {82}},\ \bibinfo {pages} {1155} (\bibinfo {year} {2010})}\BibitemShut {NoStop}%
\bibitem [{\citenamefont {Nielsen}\ and\ \citenamefont {Chuang}(2010)}]{nielsen2010quantum}%
  \BibitemOpen
  \bibfield  {author} {\bibinfo {author} {\bibfnamefont {M.~A.}\ \bibnamefont {Nielsen}}\ and\ \bibinfo {author} {\bibfnamefont {I.~L.}\ \bibnamefont {Chuang}},\ }\href {https://doi.org/10.1017/CBO9780511976667} {\emph {\bibinfo {title} {Quantum computation and quantum information}}}\ (\bibinfo  {publisher} {Cambridge university press},\ \bibinfo {year} {2010})\BibitemShut {NoStop}%
\bibitem [{\citenamefont {DiVincenzo}(2000)}]{divincenzo2000physical}%
  \BibitemOpen
  \bibfield  {author} {\bibinfo {author} {\bibfnamefont {D.~P.}\ \bibnamefont {DiVincenzo}},\ }\bibfield  {title} {\bibinfo {title} {The physical implementation of quantum computation},\ }\href {https://doi.org/10.1002/1521-3978(200009)48:9/11<771::AID-PROP771>3.0.CO;2-E} {\bibfield  {journal} {\bibinfo  {journal} {Fortschr. Phys.}\ }\textbf {\bibinfo {volume} {48}},\ \bibinfo {pages} {771} (\bibinfo {year} {2000})}\BibitemShut {NoStop}%
\bibitem [{\citenamefont {Rist\`e}\ \emph {et~al.}(2012)\citenamefont {Rist\`e}, \citenamefont {van Leeuwen}, \citenamefont {Ku}, \citenamefont {Lehnert},\ and\ \citenamefont {DiCarlo}}]{rist2012initialization}%
  \BibitemOpen
  \bibfield  {author} {\bibinfo {author} {\bibfnamefont {D.}~\bibnamefont {Rist\`e}}, \bibinfo {author} {\bibfnamefont {J.~G.}\ \bibnamefont {van Leeuwen}}, \bibinfo {author} {\bibfnamefont {H.-S.}\ \bibnamefont {Ku}}, \bibinfo {author} {\bibfnamefont {K.~W.}\ \bibnamefont {Lehnert}},\ and\ \bibinfo {author} {\bibfnamefont {L.}~\bibnamefont {DiCarlo}},\ }\bibfield  {title} {\bibinfo {title} {Initialization by measurement of a superconducting quantum bit circuit},\ }\href {https://doi.org/10.1103/PhysRevLett.109.050507} {\bibfield  {journal} {\bibinfo  {journal} {Phys. Rev. Lett.}\ }\textbf {\bibinfo {volume} {109}},\ \bibinfo {pages} {050507} (\bibinfo {year} {2012})}\BibitemShut {NoStop}%
\bibitem [{\citenamefont {Bartolomeo}\ \emph {et~al.}(2023)\citenamefont {Bartolomeo}, \citenamefont {Vischi}, \citenamefont {Feri}, \citenamefont {Bassi},\ and\ \citenamefont {Donadi}}]{dibartolomeo2023efficient}%
  \BibitemOpen
  \bibfield  {author} {\bibinfo {author} {\bibfnamefont {G.~D.}\ \bibnamefont {Bartolomeo}}, \bibinfo {author} {\bibfnamefont {M.}~\bibnamefont {Vischi}}, \bibinfo {author} {\bibfnamefont {T.}~\bibnamefont {Feri}}, \bibinfo {author} {\bibfnamefont {A.}~\bibnamefont {Bassi}},\ and\ \bibinfo {author} {\bibfnamefont {S.}~\bibnamefont {Donadi}},\ }\href {https://doi.org/10.48550/arXiv.2311.10009} {\bibinfo {title} {Efficient quantum algorithm to simulate open systems through the quantum noise formalism}} (\bibinfo {year} {2023}),\ \Eprint {https://arxiv.org/abs/2311.10009} {arXiv:2311.10009 [quant-ph]} \BibitemShut {NoStop}%
\bibitem [{\citenamefont {Hua}\ \emph {et~al.}(2023)\citenamefont {Hua}, \citenamefont {Jin}, \citenamefont {Chen}, \citenamefont {Vittal}, \citenamefont {Krsulich}, \citenamefont {Bishop}, \citenamefont {Lapeyre}, \citenamefont {Javadi-Abhari},\ and\ \citenamefont {Zhang}}]{hua2023exploiting}%
  \BibitemOpen
  \bibfield  {author} {\bibinfo {author} {\bibfnamefont {F.}~\bibnamefont {Hua}}, \bibinfo {author} {\bibfnamefont {Y.}~\bibnamefont {Jin}}, \bibinfo {author} {\bibfnamefont {Y.}~\bibnamefont {Chen}}, \bibinfo {author} {\bibfnamefont {S.}~\bibnamefont {Vittal}}, \bibinfo {author} {\bibfnamefont {K.}~\bibnamefont {Krsulich}}, \bibinfo {author} {\bibfnamefont {L.~S.}\ \bibnamefont {Bishop}}, \bibinfo {author} {\bibfnamefont {J.}~\bibnamefont {Lapeyre}}, \bibinfo {author} {\bibfnamefont {A.}~\bibnamefont {Javadi-Abhari}},\ and\ \bibinfo {author} {\bibfnamefont {E.~Z.}\ \bibnamefont {Zhang}},\ }\href {https://doi.org/10.48550/arXiv.2211.01925} {\bibinfo {title} {Exploiting qubit reuse through mid-circuit measurement and reset}} (\bibinfo {year} {2023}),\ \Eprint {https://arxiv.org/abs/2211.01925} {arXiv:2211.01925 [quant-ph]} \BibitemShut {NoStop}%
\bibitem [{\citenamefont {C\'orcoles}\ \emph {et~al.}(2021)\citenamefont {C\'orcoles}, \citenamefont {Takita}, \citenamefont {Inoue}, \citenamefont {Lekuch}, \citenamefont {Minev}, \citenamefont {Chow},\ and\ \citenamefont {Gambetta}}]{corcoles2021exploiting}%
  \BibitemOpen
  \bibfield  {author} {\bibinfo {author} {\bibfnamefont {A.~D.}\ \bibnamefont {C\'orcoles}}, \bibinfo {author} {\bibfnamefont {M.}~\bibnamefont {Takita}}, \bibinfo {author} {\bibfnamefont {K.}~\bibnamefont {Inoue}}, \bibinfo {author} {\bibfnamefont {S.}~\bibnamefont {Lekuch}}, \bibinfo {author} {\bibfnamefont {Z.~K.}\ \bibnamefont {Minev}}, \bibinfo {author} {\bibfnamefont {J.~M.}\ \bibnamefont {Chow}},\ and\ \bibinfo {author} {\bibfnamefont {J.~M.}\ \bibnamefont {Gambetta}},\ }\bibfield  {title} {\bibinfo {title} {Exploiting dynamic quantum circuits in a quantum algorithm with superconducting qubits},\ }\href {https://doi.org/10.1103/PhysRevLett.127.100501} {\bibfield  {journal} {\bibinfo  {journal} {Phys. Rev. Lett.}\ }\textbf {\bibinfo {volume} {127}},\ \bibinfo {pages} {100501} (\bibinfo {year} {2021})}\BibitemShut {NoStop}%
\bibitem [{\citenamefont {Yirka}\ and\ \citenamefont {Suba{\c{s}}{\i}}(2021)}]{yirka2021qubit}%
  \BibitemOpen
  \bibfield  {author} {\bibinfo {author} {\bibfnamefont {J.}~\bibnamefont {Yirka}}\ and\ \bibinfo {author} {\bibfnamefont {Y.}~\bibnamefont {Suba{\c{s}}{\i}}},\ }\bibfield  {title} {\bibinfo {title} {Qubit-efficient entanglement spectroscopy using qubit resets},\ }\href {https://doi.org/10.22331/q-2021-09-02-535} {\bibfield  {journal} {\bibinfo  {journal} {Quantum}\ }\textbf {\bibinfo {volume} {5}},\ \bibinfo {pages} {535} (\bibinfo {year} {2021})}\BibitemShut {NoStop}%
\bibitem [{\citenamefont {Heu\ss{}en}\ \emph {et~al.}(2024)\citenamefont {Heu\ss{}en}, \citenamefont {Locher},\ and\ \citenamefont {M\"uller}}]{heussen2024measurement}%
  \BibitemOpen
  \bibfield  {author} {\bibinfo {author} {\bibfnamefont {S.}~\bibnamefont {Heu\ss{}en}}, \bibinfo {author} {\bibfnamefont {D.~F.}\ \bibnamefont {Locher}},\ and\ \bibinfo {author} {\bibfnamefont {M.}~\bibnamefont {M\"uller}},\ }\bibfield  {title} {\bibinfo {title} {Measurement-free fault-tolerant quantum error correction in near-term devices},\ }\href {https://doi.org/10.1103/PRXQuantum.5.010333} {\bibfield  {journal} {\bibinfo  {journal} {PRX Quantum}\ }\textbf {\bibinfo {volume} {5}},\ \bibinfo {pages} {010333} (\bibinfo {year} {2024})}\BibitemShut {NoStop}%
\bibitem [{\citenamefont {Perlin}\ \emph {et~al.}(2023)\citenamefont {Perlin}, \citenamefont {Premakumar}, \citenamefont {Wang}, \citenamefont {Saffman},\ and\ \citenamefont {Joynt}}]{perlin2023fault}%
  \BibitemOpen
  \bibfield  {author} {\bibinfo {author} {\bibfnamefont {M.~A.}\ \bibnamefont {Perlin}}, \bibinfo {author} {\bibfnamefont {V.~N.}\ \bibnamefont {Premakumar}}, \bibinfo {author} {\bibfnamefont {J.}~\bibnamefont {Wang}}, \bibinfo {author} {\bibfnamefont {M.}~\bibnamefont {Saffman}},\ and\ \bibinfo {author} {\bibfnamefont {R.}~\bibnamefont {Joynt}},\ }\bibfield  {title} {\bibinfo {title} {Fault-tolerant measurement-free quantum error correction with multiqubit gates},\ }\href {https://doi.org/10.1103/PhysRevA.108.062426} {\bibfield  {journal} {\bibinfo  {journal} {Phys. Rev. A}\ }\textbf {\bibinfo {volume} {108}},\ \bibinfo {pages} {062426} (\bibinfo {year} {2023})}\BibitemShut {NoStop}%
\bibitem [{\citenamefont {Childs}\ and\ \citenamefont {Wiebe}(2012)}]{childs2012hamiltonian}%
  \BibitemOpen
  \bibfield  {author} {\bibinfo {author} {\bibfnamefont {A.~M.}\ \bibnamefont {Childs}}\ and\ \bibinfo {author} {\bibfnamefont {N.}~\bibnamefont {Wiebe}},\ }\bibfield  {title} {\bibinfo {title} {Hamiltonian simulation using linear combinations of unitary operations},\ }\href {https://doi.org/10.26421/QIC12.11-12-1} {\bibfield  {journal} {\bibinfo  {journal} {Quantum Inf. Comput.}\ }\textbf {\bibinfo {volume} {12}},\ \bibinfo {pages} {0901} (\bibinfo {year} {2012})}\BibitemShut {NoStop}%
\bibitem [{\citenamefont {Berry}\ \emph {et~al.}(2015)\citenamefont {Berry}, \citenamefont {Childs}, \citenamefont {Cleve}, \citenamefont {Kothari},\ and\ \citenamefont {Somma}}]{berry2015simulating}%
  \BibitemOpen
  \bibfield  {author} {\bibinfo {author} {\bibfnamefont {D.~W.}\ \bibnamefont {Berry}}, \bibinfo {author} {\bibfnamefont {A.~M.}\ \bibnamefont {Childs}}, \bibinfo {author} {\bibfnamefont {R.}~\bibnamefont {Cleve}}, \bibinfo {author} {\bibfnamefont {R.}~\bibnamefont {Kothari}},\ and\ \bibinfo {author} {\bibfnamefont {R.~D.}\ \bibnamefont {Somma}},\ }\bibfield  {title} {\bibinfo {title} {Simulating {H}amiltonian dynamics with a truncated {T}aylor series},\ }\href {https://doi.org/10.1103/PhysRevLett.114.090502} {\bibfield  {journal} {\bibinfo  {journal} {Phys. Rev. Lett.}\ }\textbf {\bibinfo {volume} {114}},\ \bibinfo {pages} {090502} (\bibinfo {year} {2015})}\BibitemShut {NoStop}%
\bibitem [{\citenamefont {Gily{\'e}n}\ \emph {et~al.}(2019)\citenamefont {Gily{\'e}n}, \citenamefont {Su}, \citenamefont {Low},\ and\ \citenamefont {Wiebe}}]{gilyen2019quantum}%
  \BibitemOpen
  \bibfield  {author} {\bibinfo {author} {\bibfnamefont {A.}~\bibnamefont {Gily{\'e}n}}, \bibinfo {author} {\bibfnamefont {Y.}~\bibnamefont {Su}}, \bibinfo {author} {\bibfnamefont {G.~H.}\ \bibnamefont {Low}},\ and\ \bibinfo {author} {\bibfnamefont {N.}~\bibnamefont {Wiebe}},\ }\bibfield  {title} {\bibinfo {title} {Quantum singular value transformation and beyond: exponential improvements for quantum matrix arithmetics},\ }in\ \href {https://doi.org/10.1145/3313276.3316366} {\emph {\bibinfo {booktitle} {Proceedings of the 51st Annual ACM SIGACT Symposium on Theory of Computing}}}\ (\bibinfo {year} {2019})\ pp.\ \bibinfo {pages} {193--204}\BibitemShut {NoStop}%
\bibitem [{\citenamefont {Dong}\ \emph {et~al.}(2021)\citenamefont {Dong}, \citenamefont {Quintino}, \citenamefont {Soeda},\ and\ \citenamefont {Murao}}]{dong2021success}%
  \BibitemOpen
  \bibfield  {author} {\bibinfo {author} {\bibfnamefont {Q.}~\bibnamefont {Dong}}, \bibinfo {author} {\bibfnamefont {M.~T.}\ \bibnamefont {Quintino}}, \bibinfo {author} {\bibfnamefont {A.}~\bibnamefont {Soeda}},\ and\ \bibinfo {author} {\bibfnamefont {M.}~\bibnamefont {Murao}},\ }\bibfield  {title} {\bibinfo {title} {Success-or-draw: A strategy allowing repeat-until-success in quantum computation},\ }\href {https://doi.org/10.1103/PhysRevLett.126.150504} {\bibfield  {journal} {\bibinfo  {journal} {Phys. Rev. Lett.}\ }\textbf {\bibinfo {volume} {126}},\ \bibinfo {pages} {150504} (\bibinfo {year} {2021})}\BibitemShut {NoStop}%
\bibitem [{\citenamefont {Greenberger}\ \emph {et~al.}(1990)\citenamefont {Greenberger}, \citenamefont {Horne}, \citenamefont {Shimony},\ and\ \citenamefont {Zeilinger}}]{greenberger1990bell}%
  \BibitemOpen
  \bibfield  {author} {\bibinfo {author} {\bibfnamefont {D.~M.}\ \bibnamefont {Greenberger}}, \bibinfo {author} {\bibfnamefont {M.~A.}\ \bibnamefont {Horne}}, \bibinfo {author} {\bibfnamefont {A.}~\bibnamefont {Shimony}},\ and\ \bibinfo {author} {\bibfnamefont {A.}~\bibnamefont {Zeilinger}},\ }\bibfield  {title} {\bibinfo {title} {{B}ell’s theorem without inequalities},\ }\href {https://doi.org/10.1119/1.16243} {\bibfield  {journal} {\bibinfo  {journal} {Am. J. Phys.}\ }\textbf {\bibinfo {volume} {58}},\ \bibinfo {pages} {1131} (\bibinfo {year} {1990})}\BibitemShut {NoStop}%
\bibitem [{\citenamefont {Leibfried}\ \emph {et~al.}(2004)\citenamefont {Leibfried}, \citenamefont {Barrett}, \citenamefont {Schaetz}, \citenamefont {Britton}, \citenamefont {Chiaverini}, \citenamefont {Itano}, \citenamefont {Jost}, \citenamefont {Langer},\ and\ \citenamefont {Wineland}}]{leibfried2004toward}%
  \BibitemOpen
  \bibfield  {author} {\bibinfo {author} {\bibfnamefont {D.}~\bibnamefont {Leibfried}}, \bibinfo {author} {\bibfnamefont {M.~D.}\ \bibnamefont {Barrett}}, \bibinfo {author} {\bibfnamefont {T.}~\bibnamefont {Schaetz}}, \bibinfo {author} {\bibfnamefont {J.}~\bibnamefont {Britton}}, \bibinfo {author} {\bibfnamefont {J.}~\bibnamefont {Chiaverini}}, \bibinfo {author} {\bibfnamefont {W.~M.}\ \bibnamefont {Itano}}, \bibinfo {author} {\bibfnamefont {J.~D.}\ \bibnamefont {Jost}}, \bibinfo {author} {\bibfnamefont {C.}~\bibnamefont {Langer}},\ and\ \bibinfo {author} {\bibfnamefont {D.~J.}\ \bibnamefont {Wineland}},\ }\bibfield  {title} {\bibinfo {title} {Toward {H}eisenberg-limited spectroscopy with multiparticle entangled states},\ }\href {https://doi.org/10.1126/science.1097576} {\bibfield  {journal} {\bibinfo  {journal} {Science}\ }\textbf {\bibinfo {volume} {304}},\ \bibinfo {pages} {1476} (\bibinfo {year} {2004})}\BibitemShut {NoStop}%
\bibitem [{\citenamefont {Liao}\ \emph {et~al.}(2014)\citenamefont {Liao}, \citenamefont {Yang},\ and\ \citenamefont {Hwang}}]{liao2014dynamic}%
  \BibitemOpen
  \bibfield  {author} {\bibinfo {author} {\bibfnamefont {C.-H.}\ \bibnamefont {Liao}}, \bibinfo {author} {\bibfnamefont {C.-W.}\ \bibnamefont {Yang}},\ and\ \bibinfo {author} {\bibfnamefont {T.}~\bibnamefont {Hwang}},\ }\bibfield  {title} {\bibinfo {title} {Dynamic quantum secret sharing protocol based on {GHZ} state},\ }\href {https://doi.org/10.1007/s11128-014-0779-x} {\bibfield  {journal} {\bibinfo  {journal} {Quantum Inf. Process.}\ }\textbf {\bibinfo {volume} {13}},\ \bibinfo {pages} {1907} (\bibinfo {year} {2014})}\BibitemShut {NoStop}%
\bibitem [{\citenamefont {Hahn}\ \emph {et~al.}(2020)\citenamefont {Hahn}, \citenamefont {de~Jong},\ and\ \citenamefont {Pappa}}]{hahn2020anonymous}%
  \BibitemOpen
  \bibfield  {author} {\bibinfo {author} {\bibfnamefont {F.}~\bibnamefont {Hahn}}, \bibinfo {author} {\bibfnamefont {J.}~\bibnamefont {de~Jong}},\ and\ \bibinfo {author} {\bibfnamefont {A.}~\bibnamefont {Pappa}},\ }\bibfield  {title} {\bibinfo {title} {Anonymous quantum conference key agreement},\ }\href {https://doi.org/10.1103/PRXQuantum.1.020325} {\bibfield  {journal} {\bibinfo  {journal} {PRX Quantum}\ }\textbf {\bibinfo {volume} {1}},\ \bibinfo {pages} {020325} (\bibinfo {year} {2020})}\BibitemShut {NoStop}%
\bibitem [{\citenamefont {Watts}\ \emph {et~al.}(2019)\citenamefont {Watts}, \citenamefont {Kothari}, \citenamefont {Schaeffer},\ and\ \citenamefont {Tal}}]{watts2019exponential}%
  \BibitemOpen
  \bibfield  {author} {\bibinfo {author} {\bibfnamefont {A.~B.}\ \bibnamefont {Watts}}, \bibinfo {author} {\bibfnamefont {R.}~\bibnamefont {Kothari}}, \bibinfo {author} {\bibfnamefont {L.}~\bibnamefont {Schaeffer}},\ and\ \bibinfo {author} {\bibfnamefont {A.}~\bibnamefont {Tal}},\ }\bibfield  {title} {\bibinfo {title} {Exponential separation between shallow quantum circuits and unbounded fan-in shallow classical circuits},\ }in\ \href {https://doi.org/10.1145/3313276.3316404} {\emph {\bibinfo {booktitle} {Proceedings of the 51st Annual ACM SIGACT Symposium on Theory of Computing}}}\ (\bibinfo {year} {2019})\ pp.\ \bibinfo {pages} {515--526}\BibitemShut {NoStop}%
\bibitem [{\citenamefont {Quek}\ \emph {et~al.}(2024{\natexlab{a}})\citenamefont {Quek}, \citenamefont {Kaur},\ and\ \citenamefont {Wilde}}]{quek2024multivariate}%
  \BibitemOpen
  \bibfield  {author} {\bibinfo {author} {\bibfnamefont {Y.}~\bibnamefont {Quek}}, \bibinfo {author} {\bibfnamefont {E.}~\bibnamefont {Kaur}},\ and\ \bibinfo {author} {\bibfnamefont {M.~M.}\ \bibnamefont {Wilde}},\ }\bibfield  {title} {\bibinfo {title} {Multivariate trace estimation in constant quantum depth},\ }\href {https://doi.org/10.22331/q-2024-01-10-1220} {\bibfield  {journal} {\bibinfo  {journal} {Quantum}\ }\textbf {\bibinfo {volume} {8}},\ \bibinfo {pages} {1220} (\bibinfo {year} {2024}{\natexlab{a}})}\BibitemShut {NoStop}%
\bibitem [{\citenamefont {Bäumer}\ \emph {et~al.}(2024)\citenamefont {Bäumer}, \citenamefont {Tripathi}, \citenamefont {Seif}, \citenamefont {Lidar},\ and\ \citenamefont {Wang}}]{baumer2024quantum}%
  \BibitemOpen
  \bibfield  {author} {\bibinfo {author} {\bibfnamefont {E.}~\bibnamefont {Bäumer}}, \bibinfo {author} {\bibfnamefont {V.}~\bibnamefont {Tripathi}}, \bibinfo {author} {\bibfnamefont {A.}~\bibnamefont {Seif}}, \bibinfo {author} {\bibfnamefont {D.}~\bibnamefont {Lidar}},\ and\ \bibinfo {author} {\bibfnamefont {D.~S.}\ \bibnamefont {Wang}},\ }\href {https://doi.org/10.48550/arXiv.2403.09514} {\bibinfo {title} {Quantum {F}ourier transform using dynamic circuits}} (\bibinfo {year} {2024}),\ \Eprint {https://arxiv.org/abs/2403.09514} {arXiv:2403.09514 [quant-ph]} \BibitemShut {NoStop}%
\bibitem [{\citenamefont {Bennett}\ \emph {et~al.}(1993)\citenamefont {Bennett}, \citenamefont {Brassard}, \citenamefont {Cr\'epeau}, \citenamefont {Jozsa}, \citenamefont {Peres},\ and\ \citenamefont {Wootters}}]{bennett1993}%
  \BibitemOpen
  \bibfield  {author} {\bibinfo {author} {\bibfnamefont {C.~H.}\ \bibnamefont {Bennett}}, \bibinfo {author} {\bibfnamefont {G.}~\bibnamefont {Brassard}}, \bibinfo {author} {\bibfnamefont {C.}~\bibnamefont {Cr\'epeau}}, \bibinfo {author} {\bibfnamefont {R.}~\bibnamefont {Jozsa}}, \bibinfo {author} {\bibfnamefont {A.}~\bibnamefont {Peres}},\ and\ \bibinfo {author} {\bibfnamefont {W.~K.}\ \bibnamefont {Wootters}},\ }\bibfield  {title} {\bibinfo {title} {Teleporting an unknown quantum state via dual classical and einstein-podolsky-rosen channels},\ }\href {https://doi.org/10.1103/PhysRevLett.70.1895} {\bibfield  {journal} {\bibinfo  {journal} {Phys. Rev. Lett.}\ }\textbf {\bibinfo {volume} {70}},\ \bibinfo {pages} {1895} (\bibinfo {year} {1993})}\BibitemShut {NoStop}%
\bibitem [{\citenamefont {Bouwmeester}\ \emph {et~al.}(1997)\citenamefont {Bouwmeester}, \citenamefont {Pan}, \citenamefont {Mattle}, \citenamefont {Eibl}, \citenamefont {Weinfurter},\ and\ \citenamefont {Zeilinger}}]{bouwmeester1997experimental}%
  \BibitemOpen
  \bibfield  {author} {\bibinfo {author} {\bibfnamefont {D.}~\bibnamefont {Bouwmeester}}, \bibinfo {author} {\bibfnamefont {J.-W.}\ \bibnamefont {Pan}}, \bibinfo {author} {\bibfnamefont {K.}~\bibnamefont {Mattle}}, \bibinfo {author} {\bibfnamefont {M.}~\bibnamefont {Eibl}}, \bibinfo {author} {\bibfnamefont {H.}~\bibnamefont {Weinfurter}},\ and\ \bibinfo {author} {\bibfnamefont {A.}~\bibnamefont {Zeilinger}},\ }\bibfield  {title} {\bibinfo {title} {Experimental quantum teleportation},\ }\href {https://doi.org/10.1038/37539} {\bibfield  {journal} {\bibinfo  {journal} {Nature}\ }\textbf {\bibinfo {volume} {390}},\ \bibinfo {pages} {575} (\bibinfo {year} {1997})}\BibitemShut {NoStop}%
\bibitem [{\citenamefont {Ren}\ \emph {et~al.}(2017)\citenamefont {Ren}, \citenamefont {Xu}, \citenamefont {Yong}, \citenamefont {Zhang}, \citenamefont {Liao}, \citenamefont {Yin}, \citenamefont {Liu}, \citenamefont {Cai}, \citenamefont {Yang}, \citenamefont {Li} \emph {et~al.}}]{ren2017ground}%
  \BibitemOpen
  \bibfield  {author} {\bibinfo {author} {\bibfnamefont {J.-G.}\ \bibnamefont {Ren}}, \bibinfo {author} {\bibfnamefont {P.}~\bibnamefont {Xu}}, \bibinfo {author} {\bibfnamefont {H.-L.}\ \bibnamefont {Yong}}, \bibinfo {author} {\bibfnamefont {L.}~\bibnamefont {Zhang}}, \bibinfo {author} {\bibfnamefont {S.-K.}\ \bibnamefont {Liao}}, \bibinfo {author} {\bibfnamefont {J.}~\bibnamefont {Yin}}, \bibinfo {author} {\bibfnamefont {W.-Y.}\ \bibnamefont {Liu}}, \bibinfo {author} {\bibfnamefont {W.-Q.}\ \bibnamefont {Cai}}, \bibinfo {author} {\bibfnamefont {M.}~\bibnamefont {Yang}}, \bibinfo {author} {\bibfnamefont {L.}~\bibnamefont {Li}}, \emph {et~al.},\ }\bibfield  {title} {\bibinfo {title} {Ground-to-satellite quantum teleportation},\ }\href {https://doi.org/10.1038/nature23675} {\bibfield  {journal} {\bibinfo  {journal} {Nature}\ }\textbf {\bibinfo {volume} {549}},\ \bibinfo {pages} {70} (\bibinfo {year} {2017})}\BibitemShut {NoStop}%
\bibitem [{\citenamefont {Martiel}\ and\ \citenamefont {de~Brugi{\`e}re}(2022)}]{martiel2022architecture}%
  \BibitemOpen
  \bibfield  {author} {\bibinfo {author} {\bibfnamefont {S.}~\bibnamefont {Martiel}}\ and\ \bibinfo {author} {\bibfnamefont {T.~G.}\ \bibnamefont {de~Brugi{\`e}re}},\ }\bibfield  {title} {\bibinfo {title} {Architecture aware compilation of quantum circuits via lazy synthesis},\ }\href {https://doi.org/10.22331/q-2022-06-07-729} {\bibfield  {journal} {\bibinfo  {journal} {Quantum}\ }\textbf {\bibinfo {volume} {6}},\ \bibinfo {pages} {729} (\bibinfo {year} {2022})}\BibitemShut {NoStop}%
\bibitem [{\citenamefont {Wagner}\ \emph {et~al.}(2023)\citenamefont {Wagner}, \citenamefont {B{\"a}rmann}, \citenamefont {Liers},\ and\ \citenamefont {Weissenb{\"a}ck}}]{wagner2023improving}%
  \BibitemOpen
  \bibfield  {author} {\bibinfo {author} {\bibfnamefont {F.}~\bibnamefont {Wagner}}, \bibinfo {author} {\bibfnamefont {A.}~\bibnamefont {B{\"a}rmann}}, \bibinfo {author} {\bibfnamefont {F.}~\bibnamefont {Liers}},\ and\ \bibinfo {author} {\bibfnamefont {M.}~\bibnamefont {Weissenb{\"a}ck}},\ }\bibfield  {title} {\bibinfo {title} {Improving quantum computation by optimized qubit routing},\ }\href {https://doi.org/10.1007/s10957-023-02229-w} {\bibfield  {journal} {\bibinfo  {journal} {J. Optim. Theory Appl.}\ }\textbf {\bibinfo {volume} {197}},\ \bibinfo {pages} {1161} (\bibinfo {year} {2023})}\BibitemShut {NoStop}%
\bibitem [{\citenamefont {Takagi}\ \emph {et~al.}(2023)\citenamefont {Takagi}, \citenamefont {Tajima},\ and\ \citenamefont {Gu}}]{takagi2023universal}%
  \BibitemOpen
  \bibfield  {author} {\bibinfo {author} {\bibfnamefont {R.}~\bibnamefont {Takagi}}, \bibinfo {author} {\bibfnamefont {H.}~\bibnamefont {Tajima}},\ and\ \bibinfo {author} {\bibfnamefont {M.}~\bibnamefont {Gu}},\ }\bibfield  {title} {\bibinfo {title} {Universal sampling lower bounds for quantum error mitigation},\ }\href {https://doi.org/10.1103/PhysRevLett.131.210602} {\bibfield  {journal} {\bibinfo  {journal} {Phys. Rev. Lett.}\ }\textbf {\bibinfo {volume} {131}},\ \bibinfo {pages} {210602} (\bibinfo {year} {2023})}\BibitemShut {NoStop}%
\bibitem [{\citenamefont {Takagi}\ \emph {et~al.}(2022)\citenamefont {Takagi}, \citenamefont {Endo}, \citenamefont {Minagawa},\ and\ \citenamefont {Gu}}]{takagi2022fundamental}%
  \BibitemOpen
  \bibfield  {author} {\bibinfo {author} {\bibfnamefont {R.}~\bibnamefont {Takagi}}, \bibinfo {author} {\bibfnamefont {S.}~\bibnamefont {Endo}}, \bibinfo {author} {\bibfnamefont {S.}~\bibnamefont {Minagawa}},\ and\ \bibinfo {author} {\bibfnamefont {M.}~\bibnamefont {Gu}},\ }\bibfield  {title} {\bibinfo {title} {Fundamental limits of quantum error mitigation},\ }\href {https://doi.org/10.1038/s41534-022-00618-z} {\bibfield  {journal} {\bibinfo  {journal} {npj Quantum Inf.}\ }\textbf {\bibinfo {volume} {8}},\ \bibinfo {pages} {114} (\bibinfo {year} {2022})}\BibitemShut {NoStop}%
\bibitem [{\citenamefont {Tsubouchi}\ \emph {et~al.}(2023)\citenamefont {Tsubouchi}, \citenamefont {Sagawa},\ and\ \citenamefont {Yoshioka}}]{tsubouchi2023universal}%
  \BibitemOpen
  \bibfield  {author} {\bibinfo {author} {\bibfnamefont {K.}~\bibnamefont {Tsubouchi}}, \bibinfo {author} {\bibfnamefont {T.}~\bibnamefont {Sagawa}},\ and\ \bibinfo {author} {\bibfnamefont {N.}~\bibnamefont {Yoshioka}},\ }\bibfield  {title} {\bibinfo {title} {Universal cost bound of quantum error mitigation based on quantum estimation theory},\ }\href {https://doi.org/10.1103/PhysRevLett.131.210601} {\bibfield  {journal} {\bibinfo  {journal} {Phys. Rev. Lett.}\ }\textbf {\bibinfo {volume} {131}},\ \bibinfo {pages} {210601} (\bibinfo {year} {2023})}\BibitemShut {NoStop}%
\bibitem [{\citenamefont {Quek}\ \emph {et~al.}(2024{\natexlab{b}})\citenamefont {Quek}, \citenamefont {França}, \citenamefont {Khatri}, \citenamefont {Meyer},\ and\ \citenamefont {Eisert}}]{quek2024exponentially}%
  \BibitemOpen
  \bibfield  {author} {\bibinfo {author} {\bibfnamefont {Y.}~\bibnamefont {Quek}}, \bibinfo {author} {\bibfnamefont {D.~S.}\ \bibnamefont {França}}, \bibinfo {author} {\bibfnamefont {S.}~\bibnamefont {Khatri}}, \bibinfo {author} {\bibfnamefont {J.~J.}\ \bibnamefont {Meyer}},\ and\ \bibinfo {author} {\bibfnamefont {J.}~\bibnamefont {Eisert}},\ }\href {https://doi.org/10.48550/arXiv.2210.11505} {\bibinfo {title} {Exponentially tighter bounds on limitations of quantum error mitigation}} (\bibinfo {year} {2024}{\natexlab{b}}),\ \Eprint {https://arxiv.org/abs/2210.11505} {arXiv:2210.11505 [quant-ph]} \BibitemShut {NoStop}%
\bibitem [{\citenamefont {Yuan}\ \emph {et~al.}(2024)\citenamefont {Yuan}, \citenamefont {Regula}, \citenamefont {Takagi},\ and\ \citenamefont {Gu}}]{yuan2024virtual}%
  \BibitemOpen
  \bibfield  {author} {\bibinfo {author} {\bibfnamefont {X.}~\bibnamefont {Yuan}}, \bibinfo {author} {\bibfnamefont {B.}~\bibnamefont {Regula}}, \bibinfo {author} {\bibfnamefont {R.}~\bibnamefont {Takagi}},\ and\ \bibinfo {author} {\bibfnamefont {M.}~\bibnamefont {Gu}},\ }\bibfield  {title} {\bibinfo {title} {Virtual quantum resource distillation},\ }\href {https://doi.org/10.1103/PhysRevLett.132.050203} {\bibfield  {journal} {\bibinfo  {journal} {Phys. Rev. Lett.}\ }\textbf {\bibinfo {volume} {132}},\ \bibinfo {pages} {050203} (\bibinfo {year} {2024})}\BibitemShut {NoStop}%
\bibitem [{\citenamefont {Takagi}\ \emph {et~al.}(2024)\citenamefont {Takagi}, \citenamefont {Yuan}, \citenamefont {Regula},\ and\ \citenamefont {Gu}}]{takagi204virtual}%
  \BibitemOpen
  \bibfield  {author} {\bibinfo {author} {\bibfnamefont {R.}~\bibnamefont {Takagi}}, \bibinfo {author} {\bibfnamefont {X.}~\bibnamefont {Yuan}}, \bibinfo {author} {\bibfnamefont {B.}~\bibnamefont {Regula}},\ and\ \bibinfo {author} {\bibfnamefont {M.}~\bibnamefont {Gu}},\ }\bibfield  {title} {\bibinfo {title} {Virtual quantum resource distillation: General framework and applications},\ }\href {https://doi.org/10.1103/PhysRevA.109.022403} {\bibfield  {journal} {\bibinfo  {journal} {Phys. Rev. A}\ }\textbf {\bibinfo {volume} {109}},\ \bibinfo {pages} {022403} (\bibinfo {year} {2024})}\BibitemShut {NoStop}%
\bibitem [{\citenamefont {Regula}\ and\ \citenamefont {Takagi}(2021)}]{regula2021fundamental}%
  \BibitemOpen
  \bibfield  {author} {\bibinfo {author} {\bibfnamefont {B.}~\bibnamefont {Regula}}\ and\ \bibinfo {author} {\bibfnamefont {R.}~\bibnamefont {Takagi}},\ }\bibfield  {title} {\bibinfo {title} {Fundamental limitations on distillation of quantum channel resources},\ }\href {https://doi.org/10.1038/s41467-021-24699-0} {\bibfield  {journal} {\bibinfo  {journal} {Nat. Commun.}\ }\textbf {\bibinfo {volume} {12}},\ \bibinfo {pages} {4411} (\bibinfo {year} {2021})}\BibitemShut {NoStop}%
\bibitem [{\citenamefont {Liu}\ and\ \citenamefont {Yuan}(2020)}]{liu2020operational}%
  \BibitemOpen
  \bibfield  {author} {\bibinfo {author} {\bibfnamefont {Y.}~\bibnamefont {Liu}}\ and\ \bibinfo {author} {\bibfnamefont {X.}~\bibnamefont {Yuan}},\ }\bibfield  {title} {\bibinfo {title} {Operational resource theory of quantum channels},\ }\href {https://doi.org/10.1103/PhysRevResearch.2.012035} {\bibfield  {journal} {\bibinfo  {journal} {Phys. Rev. Res.}\ }\textbf {\bibinfo {volume} {2}},\ \bibinfo {pages} {012035} (\bibinfo {year} {2020})}\BibitemShut {NoStop}%
\bibitem [{\citenamefont {Hashim}\ \emph {et~al.}(2024)\citenamefont {Hashim}, \citenamefont {Carignan-Dugas}, \citenamefont {Chen}, \citenamefont {Juenger}, \citenamefont {Fruitwala}, \citenamefont {Xu}, \citenamefont {Huang}, \citenamefont {Wallman},\ and\ \citenamefont {Siddiqi}}]{hashim2024quasiprobabilistic}%
  \BibitemOpen
  \bibfield  {author} {\bibinfo {author} {\bibfnamefont {A.}~\bibnamefont {Hashim}}, \bibinfo {author} {\bibfnamefont {A.}~\bibnamefont {Carignan-Dugas}}, \bibinfo {author} {\bibfnamefont {L.}~\bibnamefont {Chen}}, \bibinfo {author} {\bibfnamefont {C.}~\bibnamefont {Juenger}}, \bibinfo {author} {\bibfnamefont {N.}~\bibnamefont {Fruitwala}}, \bibinfo {author} {\bibfnamefont {Y.}~\bibnamefont {Xu}}, \bibinfo {author} {\bibfnamefont {G.}~\bibnamefont {Huang}}, \bibinfo {author} {\bibfnamefont {J.~J.}\ \bibnamefont {Wallman}},\ and\ \bibinfo {author} {\bibfnamefont {I.}~\bibnamefont {Siddiqi}},\ }\href {https://doi.org/10.48550/arXiv.2312.14139} {\bibinfo {title} {Quasi-probabilistic readout correction of mid-circuit measurements for adaptive feedback via measurement randomized compiling}} (\bibinfo {year} {2024}),\ \Eprint {https://arxiv.org/abs/2312.14139} {arXiv:2312.14139 [quant-ph]} \BibitemShut {NoStop}%
\bibitem [{\citenamefont {Ivashkov}\ \emph {et~al.}(2023)\citenamefont {Ivashkov}, \citenamefont {Uchehara}, \citenamefont {Jiang}, \citenamefont {Wang},\ and\ \citenamefont {Seif}}]{ivashkov2023highfidelity}%
  \BibitemOpen
  \bibfield  {author} {\bibinfo {author} {\bibfnamefont {P.}~\bibnamefont {Ivashkov}}, \bibinfo {author} {\bibfnamefont {G.}~\bibnamefont {Uchehara}}, \bibinfo {author} {\bibfnamefont {L.}~\bibnamefont {Jiang}}, \bibinfo {author} {\bibfnamefont {D.~S.}\ \bibnamefont {Wang}},\ and\ \bibinfo {author} {\bibfnamefont {A.}~\bibnamefont {Seif}},\ }\href {https://doi.org/10.48550/arXiv.2312.14087} {\bibinfo {title} {High-fidelity, multi-qubit generalized measurements with dynamic circuits}} (\bibinfo {year} {2023}),\ \Eprint {https://arxiv.org/abs/2312.14087} {arXiv:2312.14087 [quant-ph]} \BibitemShut {NoStop}%
\bibitem [{\citenamefont {Fino}\ and\ \citenamefont {Algazi}(1976)}]{fino1976unified}%
  \BibitemOpen
  \bibfield  {author} {\bibinfo {author} {\bibnamefont {Fino}}\ and\ \bibinfo {author} {\bibnamefont {Algazi}},\ }\bibfield  {title} {\bibinfo {title} {Unified matrix treatment of the fast walsh-hadamard transform},\ }\href {https://doi.org/10.1109/TC.1976.1674569} {\bibfield  {journal} {\bibinfo  {journal} {IEEE Trans. Comput.}\ }\textbf {\bibinfo {volume} {100}},\ \bibinfo {pages} {1142} (\bibinfo {year} {1976})}\BibitemShut {NoStop}%
\bibitem [{\citenamefont {Walker}(1974)}]{walker1974new}%
  \BibitemOpen
  \bibfield  {author} {\bibinfo {author} {\bibfnamefont {A.~J.}\ \bibnamefont {Walker}},\ }\bibfield  {title} {\bibinfo {title} {New fast method for generating discrete random numbers with arbitrary frequency distributions},\ }\href {https://doi.org/10.1049/el:19740097} {\bibfield  {journal} {\bibinfo  {journal} {Electron. Lett.}\ }\textbf {\bibinfo {volume} {8}},\ \bibinfo {pages} {127} (\bibinfo {year} {1974})}\BibitemShut {NoStop}%
\bibitem [{\citenamefont {Yuan}\ \emph {et~al.}(2021)\citenamefont {Yuan}, \citenamefont {Liu}, \citenamefont {Zhao}, \citenamefont {Regula}, \citenamefont {Thompson},\ and\ \citenamefont {Gu}}]{yuan2021universal}%
  \BibitemOpen
  \bibfield  {author} {\bibinfo {author} {\bibfnamefont {X.}~\bibnamefont {Yuan}}, \bibinfo {author} {\bibfnamefont {Y.}~\bibnamefont {Liu}}, \bibinfo {author} {\bibfnamefont {Q.}~\bibnamefont {Zhao}}, \bibinfo {author} {\bibfnamefont {B.}~\bibnamefont {Regula}}, \bibinfo {author} {\bibfnamefont {J.}~\bibnamefont {Thompson}},\ and\ \bibinfo {author} {\bibfnamefont {M.}~\bibnamefont {Gu}},\ }\bibfield  {title} {\bibinfo {title} {Universal and operational benchmarking of quantum memories},\ }\href {https://doi.org/10.1038/s41534-021-00444-9} {\bibfield  {journal} {\bibinfo  {journal} {npj Quantum Inf.}\ }\textbf {\bibinfo {volume} {7}},\ \bibinfo {pages} {108} (\bibinfo {year} {2021})}\BibitemShut {NoStop}%
\bibitem [{\citenamefont {Deif}(1986)}]{deif1986sensitivity}%
  \BibitemOpen
  \bibfield  {author} {\bibinfo {author} {\bibfnamefont {A.}~\bibnamefont {Deif}},\ }\bibinfo {title} {Perturbation of linear equations},\ in\ \href {https://doi.org/10.1007/978-3-642-82739-6_1} {\emph {\bibinfo {booktitle} {Sensitivity Analysis in Linear Systems}}}\ (\bibinfo  {publisher} {Springer Berlin Heidelberg},\ \bibinfo {address} {Berlin, Heidelberg},\ \bibinfo {year} {1986})\ pp.\ \bibinfo {pages} {1--43}\BibitemShut {NoStop}%
\bibitem [{\citenamefont {Rohn}(1989)}]{rohn1989new}%
  \BibitemOpen
  \bibfield  {author} {\bibinfo {author} {\bibfnamefont {J.}~\bibnamefont {Rohn}},\ }\bibfield  {title} {\bibinfo {title} {New condition numbers for matrices and linear systems},\ }\href {https://doi.org/10.1007/BF02238741} {\bibfield  {journal} {\bibinfo  {journal} {Computing}\ }\textbf {\bibinfo {volume} {41}},\ \bibinfo {pages} {167} (\bibinfo {year} {1989})}\BibitemShut {NoStop}%
\bibitem [{\citenamefont {McKay}\ \emph {et~al.}(2017)\citenamefont {McKay}, \citenamefont {Wood}, \citenamefont {Sheldon}, \citenamefont {Chow},\ and\ \citenamefont {Gambetta}}]{mckay2017efficient}%
  \BibitemOpen
  \bibfield  {author} {\bibinfo {author} {\bibfnamefont {D.~C.}\ \bibnamefont {McKay}}, \bibinfo {author} {\bibfnamefont {C.~J.}\ \bibnamefont {Wood}}, \bibinfo {author} {\bibfnamefont {S.}~\bibnamefont {Sheldon}}, \bibinfo {author} {\bibfnamefont {J.~M.}\ \bibnamefont {Chow}},\ and\ \bibinfo {author} {\bibfnamefont {J.~M.}\ \bibnamefont {Gambetta}},\ }\bibfield  {title} {\bibinfo {title} {Efficient {$Z$} gates for quantum computing},\ }\href {https://doi.org/10.1103/PhysRevA.96.022330} {\bibfield  {journal} {\bibinfo  {journal} {Phys. Rev. A}\ }\textbf {\bibinfo {volume} {96}},\ \bibinfo {pages} {022330} (\bibinfo {year} {2017})}\BibitemShut {NoStop}%
\bibitem [{\citenamefont {Sundaresan}\ \emph {et~al.}(2020)\citenamefont {Sundaresan}, \citenamefont {Lauer}, \citenamefont {Pritchett}, \citenamefont {Magesan}, \citenamefont {Jurcevic},\ and\ \citenamefont {Gambetta}}]{sundaresan2020reducing}%
  \BibitemOpen
  \bibfield  {author} {\bibinfo {author} {\bibfnamefont {N.}~\bibnamefont {Sundaresan}}, \bibinfo {author} {\bibfnamefont {I.}~\bibnamefont {Lauer}}, \bibinfo {author} {\bibfnamefont {E.}~\bibnamefont {Pritchett}}, \bibinfo {author} {\bibfnamefont {E.}~\bibnamefont {Magesan}}, \bibinfo {author} {\bibfnamefont {P.}~\bibnamefont {Jurcevic}},\ and\ \bibinfo {author} {\bibfnamefont {J.~M.}\ \bibnamefont {Gambetta}},\ }\bibfield  {title} {\bibinfo {title} {Reducing unitary and spectator errors in cross resonance with optimized rotary echoes},\ }\href {https://doi.org/10.1103/PRXQuantum.1.020318} {\bibfield  {journal} {\bibinfo  {journal} {PRX Quantum}\ }\textbf {\bibinfo {volume} {1}},\ \bibinfo {pages} {020318} (\bibinfo {year} {2020})}\BibitemShut {NoStop}%
\bibitem [{\citenamefont {Smolin}\ \emph {et~al.}(2012)\citenamefont {Smolin}, \citenamefont {Gambetta},\ and\ \citenamefont {Smith}}]{smolin2012efficient}%
  \BibitemOpen
  \bibfield  {author} {\bibinfo {author} {\bibfnamefont {J.~A.}\ \bibnamefont {Smolin}}, \bibinfo {author} {\bibfnamefont {J.~M.}\ \bibnamefont {Gambetta}},\ and\ \bibinfo {author} {\bibfnamefont {G.}~\bibnamefont {Smith}},\ }\bibfield  {title} {\bibinfo {title} {Efficient method for computing the maximum-likelihood quantum state from measurements with additive gaussian noise},\ }\href {https://doi.org/10.1103/PhysRevLett.108.070502} {\bibfield  {journal} {\bibinfo  {journal} {Phys. Rev. Lett.}\ }\textbf {\bibinfo {volume} {108}},\ \bibinfo {pages} {070502} (\bibinfo {year} {2012})}\BibitemShut {NoStop}%
\bibitem [{\citenamefont {Geller}(2020)}]{geller2020rigorous}%
  \BibitemOpen
  \bibfield  {author} {\bibinfo {author} {\bibfnamefont {M.~R.}\ \bibnamefont {Geller}},\ }\bibfield  {title} {\bibinfo {title} {Rigorous measurement error correction},\ }\href {https://doi.org/10.1088/2058-9565/ab9591} {\bibfield  {journal} {\bibinfo  {journal} {Quantum Sci. Technol.}\ }\textbf {\bibinfo {volume} {5}},\ \bibinfo {pages} {03LT01} (\bibinfo {year} {2020})}\BibitemShut {NoStop}%
\bibitem [{\citenamefont {Maciejewski}\ \emph {et~al.}(2020)\citenamefont {Maciejewski}, \citenamefont {Zimbor{\'a}s},\ and\ \citenamefont {Oszmaniec}}]{maciejewski2020mitigation}%
  \BibitemOpen
  \bibfield  {author} {\bibinfo {author} {\bibfnamefont {F.~B.}\ \bibnamefont {Maciejewski}}, \bibinfo {author} {\bibfnamefont {Z.}~\bibnamefont {Zimbor{\'a}s}},\ and\ \bibinfo {author} {\bibfnamefont {M.}~\bibnamefont {Oszmaniec}},\ }\bibfield  {title} {\bibinfo {title} {Mitigation of readout noise in near-term quantum devices by classical post-processing based on detector tomography},\ }\href {https://doi.org/10.22331/q-2020-04-24-257} {\bibfield  {journal} {\bibinfo  {journal} {Quantum}\ }\textbf {\bibinfo {volume} {4}},\ \bibinfo {pages} {257} (\bibinfo {year} {2020})}\BibitemShut {NoStop}%
\bibitem [{\citenamefont {Nachman}\ \emph {et~al.}(2020)\citenamefont {Nachman}, \citenamefont {Urbanek}, \citenamefont {de~Jong},\ and\ \citenamefont {Bauer}}]{nachman2020unfolding}%
  \BibitemOpen
  \bibfield  {author} {\bibinfo {author} {\bibfnamefont {B.}~\bibnamefont {Nachman}}, \bibinfo {author} {\bibfnamefont {M.}~\bibnamefont {Urbanek}}, \bibinfo {author} {\bibfnamefont {W.~A.}\ \bibnamefont {de~Jong}},\ and\ \bibinfo {author} {\bibfnamefont {C.~W.}\ \bibnamefont {Bauer}},\ }\bibfield  {title} {\bibinfo {title} {Unfolding quantum computer readout noise},\ }\href {https://doi.org/10.1038/s41534-020-00309-7} {\bibfield  {journal} {\bibinfo  {journal} {npj Quantum Inf.}\ }\textbf {\bibinfo {volume} {6}},\ \bibinfo {pages} {84} (\bibinfo {year} {2020})}\BibitemShut {NoStop}%
\bibitem [{\citenamefont {Koh}\ \emph {et~al.}(2022{\natexlab{a}})\citenamefont {Koh}, \citenamefont {Tai},\ and\ \citenamefont {Lee}}]{koh2022simulation}%
  \BibitemOpen
  \bibfield  {author} {\bibinfo {author} {\bibfnamefont {J.~M.}\ \bibnamefont {Koh}}, \bibinfo {author} {\bibfnamefont {T.}~\bibnamefont {Tai}},\ and\ \bibinfo {author} {\bibfnamefont {C.~H.}\ \bibnamefont {Lee}},\ }\bibfield  {title} {\bibinfo {title} {Simulation of interaction-induced chiral topological dynamics on a digital quantum computer},\ }\href {https://doi.org/10.1103/PhysRevLett.129.140502} {\bibfield  {journal} {\bibinfo  {journal} {Phys. Rev. Lett.}\ }\textbf {\bibinfo {volume} {129}},\ \bibinfo {pages} {140502} (\bibinfo {year} {2022}{\natexlab{a}})}\BibitemShut {NoStop}%
\bibitem [{\citenamefont {Koh}\ \emph {et~al.}(2022{\natexlab{b}})\citenamefont {Koh}, \citenamefont {Tai}, \citenamefont {Phee}, \citenamefont {Ng},\ and\ \citenamefont {Lee}}]{koh2022stabilizing}%
  \BibitemOpen
  \bibfield  {author} {\bibinfo {author} {\bibfnamefont {J.~M.}\ \bibnamefont {Koh}}, \bibinfo {author} {\bibfnamefont {T.}~\bibnamefont {Tai}}, \bibinfo {author} {\bibfnamefont {Y.~H.}\ \bibnamefont {Phee}}, \bibinfo {author} {\bibfnamefont {W.~E.}\ \bibnamefont {Ng}},\ and\ \bibinfo {author} {\bibfnamefont {C.~H.}\ \bibnamefont {Lee}},\ }\bibfield  {title} {\bibinfo {title} {Stabilizing multiple topological fermions on a quantum computer},\ }\href {https://doi.org/10.1038/s41534-022-00527-1} {\bibfield  {journal} {\bibinfo  {journal} {npj Quantum Inf.}\ }\textbf {\bibinfo {volume} {8}},\ \bibinfo {pages} {16} (\bibinfo {year} {2022}{\natexlab{b}})}\BibitemShut {NoStop}%
\bibitem [{\citenamefont {Koh}\ \emph {et~al.}(2023{\natexlab{a}})\citenamefont {Koh}, \citenamefont {Sun}, \citenamefont {Motta},\ and\ \citenamefont {Minnich}}]{koh2023measurement}%
  \BibitemOpen
  \bibfield  {author} {\bibinfo {author} {\bibfnamefont {J.~M.}\ \bibnamefont {Koh}}, \bibinfo {author} {\bibfnamefont {S.-N.}\ \bibnamefont {Sun}}, \bibinfo {author} {\bibfnamefont {M.}~\bibnamefont {Motta}},\ and\ \bibinfo {author} {\bibfnamefont {A.~J.}\ \bibnamefont {Minnich}},\ }\bibfield  {title} {\bibinfo {title} {Measurement-induced entanglement phase transition on a superconducting quantum processor with mid-circuit readout},\ }\href {https://doi.org/10.1038/s41567-023-02076-6} {\bibfield  {journal} {\bibinfo  {journal} {Nat. Phys.}\ }\textbf {\bibinfo {volume} {19}},\ \bibinfo {pages} {1314} (\bibinfo {year} {2023}{\natexlab{a}})}\BibitemShut {NoStop}%
\bibitem [{\citenamefont {Koh}\ \emph {et~al.}(2023{\natexlab{b}})\citenamefont {Koh}, \citenamefont {Tai},\ and\ \citenamefont {Lee}}]{koh2023observation}%
  \BibitemOpen
  \bibfield  {author} {\bibinfo {author} {\bibfnamefont {J.~M.}\ \bibnamefont {Koh}}, \bibinfo {author} {\bibfnamefont {T.}~\bibnamefont {Tai}},\ and\ \bibinfo {author} {\bibfnamefont {C.~H.}\ \bibnamefont {Lee}},\ }\href {https://doi.org/10.48550/arXiv.2303.02179} {\bibinfo {title} {Observation of higher-order topological states on a quantum computer}} (\bibinfo {year} {2023}{\natexlab{b}}),\ \Eprint {https://arxiv.org/abs/2303.02179} {arXiv:2303.02179 [cond-mat.str-el]} \BibitemShut {NoStop}%
\bibitem [{\citenamefont {Viola}\ \emph {et~al.}(1999)\citenamefont {Viola}, \citenamefont {Knill},\ and\ \citenamefont {Lloyd}}]{viola1999dynamical}%
  \BibitemOpen
  \bibfield  {author} {\bibinfo {author} {\bibfnamefont {L.}~\bibnamefont {Viola}}, \bibinfo {author} {\bibfnamefont {E.}~\bibnamefont {Knill}},\ and\ \bibinfo {author} {\bibfnamefont {S.}~\bibnamefont {Lloyd}},\ }\bibfield  {title} {\bibinfo {title} {Dynamical decoupling of open quantum systems},\ }\href {https://doi.org/10.1103/PhysRevLett.82.2417} {\bibfield  {journal} {\bibinfo  {journal} {Phys. Rev. Lett.}\ }\textbf {\bibinfo {volume} {82}},\ \bibinfo {pages} {2417} (\bibinfo {year} {1999})}\BibitemShut {NoStop}%
\bibitem [{\citenamefont {Kim}\ \emph {et~al.}(2023{\natexlab{a}})\citenamefont {Kim}, \citenamefont {Wood}, \citenamefont {Yoder}, \citenamefont {Merkel}, \citenamefont {Gambetta}, \citenamefont {Temme},\ and\ \citenamefont {Kandala}}]{kim2023scalable}%
  \BibitemOpen
  \bibfield  {author} {\bibinfo {author} {\bibfnamefont {Y.}~\bibnamefont {Kim}}, \bibinfo {author} {\bibfnamefont {C.~J.}\ \bibnamefont {Wood}}, \bibinfo {author} {\bibfnamefont {T.~J.}\ \bibnamefont {Yoder}}, \bibinfo {author} {\bibfnamefont {S.~T.}\ \bibnamefont {Merkel}}, \bibinfo {author} {\bibfnamefont {J.~M.}\ \bibnamefont {Gambetta}}, \bibinfo {author} {\bibfnamefont {K.}~\bibnamefont {Temme}},\ and\ \bibinfo {author} {\bibfnamefont {A.}~\bibnamefont {Kandala}},\ }\bibfield  {title} {\bibinfo {title} {Scalable error mitigation for noisy quantum circuits produces competitive expectation values},\ }\href {https://doi.org/10.1038/s41567-022-01914-3} {\bibfield  {journal} {\bibinfo  {journal} {Nat. Phys.}\ }\textbf {\bibinfo {volume} {19}},\ \bibinfo {pages} {752} (\bibinfo {year} {2023}{\natexlab{a}})}\BibitemShut {NoStop}%
\bibitem [{\citenamefont {Pokharel}\ \emph {et~al.}(2018)\citenamefont {Pokharel}, \citenamefont {Anand}, \citenamefont {Fortman},\ and\ \citenamefont {Lidar}}]{pokharel2018demonstration}%
  \BibitemOpen
  \bibfield  {author} {\bibinfo {author} {\bibfnamefont {B.}~\bibnamefont {Pokharel}}, \bibinfo {author} {\bibfnamefont {N.}~\bibnamefont {Anand}}, \bibinfo {author} {\bibfnamefont {B.}~\bibnamefont {Fortman}},\ and\ \bibinfo {author} {\bibfnamefont {D.~A.}\ \bibnamefont {Lidar}},\ }\bibfield  {title} {\bibinfo {title} {Demonstration of fidelity improvement using dynamical decoupling with superconducting qubits},\ }\href {https://doi.org/10.1103/PhysRevLett.121.220502} {\bibfield  {journal} {\bibinfo  {journal} {Phys. Rev. Lett.}\ }\textbf {\bibinfo {volume} {121}},\ \bibinfo {pages} {220502} (\bibinfo {year} {2018})}\BibitemShut {NoStop}%
\bibitem [{\citenamefont {Khodjasteh}\ and\ \citenamefont {Lidar}(2005)}]{khodjasteh2005fault}%
  \BibitemOpen
  \bibfield  {author} {\bibinfo {author} {\bibfnamefont {K.}~\bibnamefont {Khodjasteh}}\ and\ \bibinfo {author} {\bibfnamefont {D.~A.}\ \bibnamefont {Lidar}},\ }\bibfield  {title} {\bibinfo {title} {Fault-tolerant quantum dynamical decoupling},\ }\href {https://doi.org/10.1103/PhysRevLett.95.180501} {\bibfield  {journal} {\bibinfo  {journal} {Phys. Rev. Lett.}\ }\textbf {\bibinfo {volume} {95}},\ \bibinfo {pages} {180501} (\bibinfo {year} {2005})}\BibitemShut {NoStop}%
\bibitem [{\citenamefont {Uhrig}(2007)}]{uhrig2007keeping}%
  \BibitemOpen
  \bibfield  {author} {\bibinfo {author} {\bibfnamefont {G.~S.}\ \bibnamefont {Uhrig}},\ }\bibfield  {title} {\bibinfo {title} {Keeping a quantum bit alive by optimized $\ensuremath{\pi}$-pulse sequences},\ }\href {https://doi.org/10.1103/PhysRevLett.98.100504} {\bibfield  {journal} {\bibinfo  {journal} {Phys. Rev. Lett.}\ }\textbf {\bibinfo {volume} {98}},\ \bibinfo {pages} {100504} (\bibinfo {year} {2007})}\BibitemShut {NoStop}%
\bibitem [{\citenamefont {Ezzell}\ \emph {et~al.}(2023)\citenamefont {Ezzell}, \citenamefont {Pokharel}, \citenamefont {Tewala}, \citenamefont {Quiroz},\ and\ \citenamefont {Lidar}}]{ezzell2023dynamical}%
  \BibitemOpen
  \bibfield  {author} {\bibinfo {author} {\bibfnamefont {N.}~\bibnamefont {Ezzell}}, \bibinfo {author} {\bibfnamefont {B.}~\bibnamefont {Pokharel}}, \bibinfo {author} {\bibfnamefont {L.}~\bibnamefont {Tewala}}, \bibinfo {author} {\bibfnamefont {G.}~\bibnamefont {Quiroz}},\ and\ \bibinfo {author} {\bibfnamefont {D.~A.}\ \bibnamefont {Lidar}},\ }\bibfield  {title} {\bibinfo {title} {Dynamical decoupling for superconducting qubits: A performance survey},\ }\href {https://doi.org/10.1103/PhysRevApplied.20.064027} {\bibfield  {journal} {\bibinfo  {journal} {Phys. Rev. Appl.}\ }\textbf {\bibinfo {volume} {20}},\ \bibinfo {pages} {064027} (\bibinfo {year} {2023})}\BibitemShut {NoStop}%
\bibitem [{\citenamefont {Pokharel}\ and\ \citenamefont {Lidar}(2024)}]{pokharel2024better}%
  \BibitemOpen
  \bibfield  {author} {\bibinfo {author} {\bibfnamefont {B.}~\bibnamefont {Pokharel}}\ and\ \bibinfo {author} {\bibfnamefont {D.~A.}\ \bibnamefont {Lidar}},\ }\bibfield  {title} {\bibinfo {title} {Better-than-classical grover search via quantum error detection and suppression},\ }\href {https://doi.org/10.1038/s41534-023-00794-6} {\bibfield  {journal} {\bibinfo  {journal} {npj Quantum Inf.}\ }\textbf {\bibinfo {volume} {10}},\ \bibinfo {pages} {23} (\bibinfo {year} {2024})}\BibitemShut {NoStop}%
\bibitem [{\citenamefont {Pokharel}\ and\ \citenamefont {Lidar}(2023)}]{pokharel2023demonstration}%
  \BibitemOpen
  \bibfield  {author} {\bibinfo {author} {\bibfnamefont {B.}~\bibnamefont {Pokharel}}\ and\ \bibinfo {author} {\bibfnamefont {D.~A.}\ \bibnamefont {Lidar}},\ }\bibfield  {title} {\bibinfo {title} {Demonstration of algorithmic quantum speedup},\ }\href {https://doi.org/10.1103/PhysRevLett.130.210602} {\bibfield  {journal} {\bibinfo  {journal} {Phys. Rev. Lett.}\ }\textbf {\bibinfo {volume} {130}},\ \bibinfo {pages} {210602} (\bibinfo {year} {2023})}\BibitemShut {NoStop}%
\bibitem [{\citenamefont {Kim}\ \emph {et~al.}(2023{\natexlab{b}})\citenamefont {Kim}, \citenamefont {Eddins}, \citenamefont {Anand}, \citenamefont {Wei}, \citenamefont {Van Den~Berg}, \citenamefont {Rosenblatt}, \citenamefont {Nayfeh}, \citenamefont {Wu}, \citenamefont {Zaletel}, \citenamefont {Temme} \emph {et~al.}}]{kim2023evidence}%
  \BibitemOpen
  \bibfield  {author} {\bibinfo {author} {\bibfnamefont {Y.}~\bibnamefont {Kim}}, \bibinfo {author} {\bibfnamefont {A.}~\bibnamefont {Eddins}}, \bibinfo {author} {\bibfnamefont {S.}~\bibnamefont {Anand}}, \bibinfo {author} {\bibfnamefont {K.~X.}\ \bibnamefont {Wei}}, \bibinfo {author} {\bibfnamefont {E.}~\bibnamefont {Van Den~Berg}}, \bibinfo {author} {\bibfnamefont {S.}~\bibnamefont {Rosenblatt}}, \bibinfo {author} {\bibfnamefont {H.}~\bibnamefont {Nayfeh}}, \bibinfo {author} {\bibfnamefont {Y.}~\bibnamefont {Wu}}, \bibinfo {author} {\bibfnamefont {M.}~\bibnamefont {Zaletel}}, \bibinfo {author} {\bibfnamefont {K.}~\bibnamefont {Temme}}, \emph {et~al.},\ }\bibfield  {title} {\bibinfo {title} {Evidence for the utility of quantum computing before fault tolerance},\ }\href {https://doi.org/10.1038/s41586-023-06096-3} {\bibfield  {journal} {\bibinfo  {journal} {Nature}\ }\textbf {\bibinfo {volume} {618}},\ \bibinfo {pages} {500} (\bibinfo {year} {2023}{\natexlab{b}})}\BibitemShut {NoStop}%
\bibitem [{\citenamefont {Tong}\ \emph {et~al.}(2024)\citenamefont {Tong}, \citenamefont {Zhang},\ and\ \citenamefont {Pokharel}}]{tong2024empirical}%
  \BibitemOpen
  \bibfield  {author} {\bibinfo {author} {\bibfnamefont {C.}~\bibnamefont {Tong}}, \bibinfo {author} {\bibfnamefont {H.}~\bibnamefont {Zhang}},\ and\ \bibinfo {author} {\bibfnamefont {B.}~\bibnamefont {Pokharel}},\ }\href {https://doi.org/10.48550/arXiv.2403.02294} {\bibinfo {title} {Empirical learning of dynamical decoupling on quantum processors}} (\bibinfo {year} {2024}),\ \Eprint {https://arxiv.org/abs/2403.02294} {arXiv:2403.02294 [quant-ph]} \BibitemShut {NoStop}%
\bibitem [{\citenamefont {Buhrman}\ \emph {et~al.}(2001)\citenamefont {Buhrman}, \citenamefont {Cleve}, \citenamefont {Watrous},\ and\ \citenamefont {de~Wolf}}]{buhrman2001quantum}%
  \BibitemOpen
  \bibfield  {author} {\bibinfo {author} {\bibfnamefont {H.}~\bibnamefont {Buhrman}}, \bibinfo {author} {\bibfnamefont {R.}~\bibnamefont {Cleve}}, \bibinfo {author} {\bibfnamefont {J.}~\bibnamefont {Watrous}},\ and\ \bibinfo {author} {\bibfnamefont {R.}~\bibnamefont {de~Wolf}},\ }\bibfield  {title} {\bibinfo {title} {Quantum fingerprinting},\ }\href {https://doi.org/10.1103/PhysRevLett.87.167902} {\bibfield  {journal} {\bibinfo  {journal} {Phys. Rev. Lett.}\ }\textbf {\bibinfo {volume} {87}},\ \bibinfo {pages} {167902} (\bibinfo {year} {2001})}\BibitemShut {NoStop}%
\bibitem [{\citenamefont {Barenco}\ \emph {et~al.}(1997)\citenamefont {Barenco}, \citenamefont {Berthiaume}, \citenamefont {Deutsch}, \citenamefont {Ekert}, \citenamefont {Jozsa},\ and\ \citenamefont {Macchiavello}}]{barenco1997stabilization}%
  \BibitemOpen
  \bibfield  {author} {\bibinfo {author} {\bibfnamefont {A.}~\bibnamefont {Barenco}}, \bibinfo {author} {\bibfnamefont {A.}~\bibnamefont {Berthiaume}}, \bibinfo {author} {\bibfnamefont {D.}~\bibnamefont {Deutsch}}, \bibinfo {author} {\bibfnamefont {A.}~\bibnamefont {Ekert}}, \bibinfo {author} {\bibfnamefont {R.}~\bibnamefont {Jozsa}},\ and\ \bibinfo {author} {\bibfnamefont {C.}~\bibnamefont {Macchiavello}},\ }\bibfield  {title} {\bibinfo {title} {Stabilization of quantum computations by symmetrization},\ }\href {https://doi.org/10.1137/S0097539796302452} {\bibfield  {journal} {\bibinfo  {journal} {SIAM J. Comput.}\ }\textbf {\bibinfo {volume} {26}},\ \bibinfo {pages} {1541} (\bibinfo {year} {1997})}\BibitemShut {NoStop}%
\bibitem [{\citenamefont {Flammia}\ and\ \citenamefont {Liu}(2011)}]{flammia2011direct}%
  \BibitemOpen
  \bibfield  {author} {\bibinfo {author} {\bibfnamefont {S.~T.}\ \bibnamefont {Flammia}}\ and\ \bibinfo {author} {\bibfnamefont {Y.-K.}\ \bibnamefont {Liu}},\ }\bibfield  {title} {\bibinfo {title} {Direct fidelity estimation from few {P}auli measurements},\ }\href {https://doi.org/10.1103/PhysRevLett.106.230501} {\bibfield  {journal} {\bibinfo  {journal} {Phys. Rev. Lett.}\ }\textbf {\bibinfo {volume} {106}},\ \bibinfo {pages} {230501} (\bibinfo {year} {2011})}\BibitemShut {NoStop}%
\bibitem [{\citenamefont {da~Silva}\ \emph {et~al.}(2011)\citenamefont {da~Silva}, \citenamefont {Landon-Cardinal},\ and\ \citenamefont {Poulin}}]{silva2011practical}%
  \BibitemOpen
  \bibfield  {author} {\bibinfo {author} {\bibfnamefont {M.~P.}\ \bibnamefont {da~Silva}}, \bibinfo {author} {\bibfnamefont {O.}~\bibnamefont {Landon-Cardinal}},\ and\ \bibinfo {author} {\bibfnamefont {D.}~\bibnamefont {Poulin}},\ }\bibfield  {title} {\bibinfo {title} {Practical characterization of quantum devices without tomography},\ }\href {https://doi.org/10.1103/PhysRevLett.107.210404} {\bibfield  {journal} {\bibinfo  {journal} {Phys. Rev. Lett.}\ }\textbf {\bibinfo {volume} {107}},\ \bibinfo {pages} {210404} (\bibinfo {year} {2011})}\BibitemShut {NoStop}%
\bibitem [{\citenamefont {Cao}\ \emph {et~al.}(2023)\citenamefont {Cao}, \citenamefont {Wu}, \citenamefont {Chen}, \citenamefont {Gong}, \citenamefont {Wu}, \citenamefont {Ye}, \citenamefont {Zha}, \citenamefont {Qian}, \citenamefont {Ying}, \citenamefont {Guo} \emph {et~al.}}]{cao2023generation}%
  \BibitemOpen
  \bibfield  {author} {\bibinfo {author} {\bibfnamefont {S.}~\bibnamefont {Cao}}, \bibinfo {author} {\bibfnamefont {B.}~\bibnamefont {Wu}}, \bibinfo {author} {\bibfnamefont {F.}~\bibnamefont {Chen}}, \bibinfo {author} {\bibfnamefont {M.}~\bibnamefont {Gong}}, \bibinfo {author} {\bibfnamefont {Y.}~\bibnamefont {Wu}}, \bibinfo {author} {\bibfnamefont {Y.}~\bibnamefont {Ye}}, \bibinfo {author} {\bibfnamefont {C.}~\bibnamefont {Zha}}, \bibinfo {author} {\bibfnamefont {H.}~\bibnamefont {Qian}}, \bibinfo {author} {\bibfnamefont {C.}~\bibnamefont {Ying}}, \bibinfo {author} {\bibfnamefont {S.}~\bibnamefont {Guo}}, \emph {et~al.},\ }\bibfield  {title} {\bibinfo {title} {Generation of genuine entanglement up to 51 superconducting qubits},\ }\href {https://doi.org/10.1038/s41586-023-06195-1} {\bibfield  {journal} {\bibinfo  {journal} {Nature}\ }\textbf {\bibinfo {volume} {619}},\ \bibinfo {pages} {738} (\bibinfo {year} {2023})}\BibitemShut {NoStop}%
\bibitem [{\citenamefont {Gokhale}\ \emph {et~al.}(2019)\citenamefont {Gokhale}, \citenamefont {Angiuli}, \citenamefont {Ding}, \citenamefont {Gui}, \citenamefont {Tomesh}, \citenamefont {Suchara}, \citenamefont {Martonosi},\ and\ \citenamefont {Chong}}]{gokhale2019minimizing}%
  \BibitemOpen
  \bibfield  {author} {\bibinfo {author} {\bibfnamefont {P.}~\bibnamefont {Gokhale}}, \bibinfo {author} {\bibfnamefont {O.}~\bibnamefont {Angiuli}}, \bibinfo {author} {\bibfnamefont {Y.}~\bibnamefont {Ding}}, \bibinfo {author} {\bibfnamefont {K.}~\bibnamefont {Gui}}, \bibinfo {author} {\bibfnamefont {T.}~\bibnamefont {Tomesh}}, \bibinfo {author} {\bibfnamefont {M.}~\bibnamefont {Suchara}}, \bibinfo {author} {\bibfnamefont {M.}~\bibnamefont {Martonosi}},\ and\ \bibinfo {author} {\bibfnamefont {F.~T.}\ \bibnamefont {Chong}},\ }\href {https://doi.org/10.48550/arXiv.1907.13623} {\bibinfo {title} {Minimizing state preparations in variational quantum eigensolver by partitioning into commuting families}} (\bibinfo {year} {2019}),\ \Eprint {https://arxiv.org/abs/1907.13623} {arXiv:1907.13623 [quant-ph]} \BibitemShut {NoStop}%
\bibitem [{\citenamefont {Sackett}\ \emph {et~al.}(2000)\citenamefont {Sackett}, \citenamefont {Kielpinski}, \citenamefont {King}, \citenamefont {Langer}, \citenamefont {Meyer}, \citenamefont {Myatt}, \citenamefont {Rowe}, \citenamefont {Turchette}, \citenamefont {Itano}, \citenamefont {Wineland} \emph {et~al.}}]{sackett2000experimental}%
  \BibitemOpen
  \bibfield  {author} {\bibinfo {author} {\bibfnamefont {C.~A.}\ \bibnamefont {Sackett}}, \bibinfo {author} {\bibfnamefont {D.}~\bibnamefont {Kielpinski}}, \bibinfo {author} {\bibfnamefont {B.~E.}\ \bibnamefont {King}}, \bibinfo {author} {\bibfnamefont {C.}~\bibnamefont {Langer}}, \bibinfo {author} {\bibfnamefont {V.}~\bibnamefont {Meyer}}, \bibinfo {author} {\bibfnamefont {C.~J.}\ \bibnamefont {Myatt}}, \bibinfo {author} {\bibfnamefont {M.}~\bibnamefont {Rowe}}, \bibinfo {author} {\bibfnamefont {Q.}~\bibnamefont {Turchette}}, \bibinfo {author} {\bibfnamefont {W.~M.}\ \bibnamefont {Itano}}, \bibinfo {author} {\bibfnamefont {D.~J.}\ \bibnamefont {Wineland}}, \emph {et~al.},\ }\bibfield  {title} {\bibinfo {title} {Experimental entanglement of four particles},\ }\href {https://doi.org/10.1038/35005011} {\bibfield  {journal} {\bibinfo  {journal} {Nature}\ }\textbf {\bibinfo {volume} {404}},\ \bibinfo {pages} {256} (\bibinfo {year} {2000})}\BibitemShut {NoStop}%
\bibitem [{\citenamefont {Leibfried}\ \emph {et~al.}(2005)\citenamefont {Leibfried}, \citenamefont {Knill}, \citenamefont {Seidelin}, \citenamefont {Britton}, \citenamefont {Blakestad}, \citenamefont {Chiaverini}, \citenamefont {Hume}, \citenamefont {Itano}, \citenamefont {Jost}, \citenamefont {Langer} \emph {et~al.}}]{leibfried2005creation}%
  \BibitemOpen
  \bibfield  {author} {\bibinfo {author} {\bibfnamefont {D.}~\bibnamefont {Leibfried}}, \bibinfo {author} {\bibfnamefont {E.}~\bibnamefont {Knill}}, \bibinfo {author} {\bibfnamefont {S.}~\bibnamefont {Seidelin}}, \bibinfo {author} {\bibfnamefont {J.}~\bibnamefont {Britton}}, \bibinfo {author} {\bibfnamefont {R.~B.}\ \bibnamefont {Blakestad}}, \bibinfo {author} {\bibfnamefont {J.}~\bibnamefont {Chiaverini}}, \bibinfo {author} {\bibfnamefont {D.~B.}\ \bibnamefont {Hume}}, \bibinfo {author} {\bibfnamefont {W.~M.}\ \bibnamefont {Itano}}, \bibinfo {author} {\bibfnamefont {J.~D.}\ \bibnamefont {Jost}}, \bibinfo {author} {\bibfnamefont {C.}~\bibnamefont {Langer}}, \emph {et~al.},\ }\bibfield  {title} {\bibinfo {title} {Creation of a six-atom `{S}chr{\"o}dinger cat' state},\ }\href {https://doi.org/10.1038/nature04251} {\bibfield  {journal} {\bibinfo  {journal} {Nature}\ }\textbf {\bibinfo {volume} {438}},\ \bibinfo {pages} {639} (\bibinfo {year} {2005})}\BibitemShut {NoStop}%
\bibitem [{\citenamefont {Monz}\ \emph {et~al.}(2011)\citenamefont {Monz}, \citenamefont {Schindler}, \citenamefont {Barreiro}, \citenamefont {Chwalla}, \citenamefont {Nigg}, \citenamefont {Coish}, \citenamefont {Harlander}, \citenamefont {H\"ansel}, \citenamefont {Hennrich},\ and\ \citenamefont {Blatt}}]{monz2011fourteen}%
  \BibitemOpen
  \bibfield  {author} {\bibinfo {author} {\bibfnamefont {T.}~\bibnamefont {Monz}}, \bibinfo {author} {\bibfnamefont {P.}~\bibnamefont {Schindler}}, \bibinfo {author} {\bibfnamefont {J.~T.}\ \bibnamefont {Barreiro}}, \bibinfo {author} {\bibfnamefont {M.}~\bibnamefont {Chwalla}}, \bibinfo {author} {\bibfnamefont {D.}~\bibnamefont {Nigg}}, \bibinfo {author} {\bibfnamefont {W.~A.}\ \bibnamefont {Coish}}, \bibinfo {author} {\bibfnamefont {M.}~\bibnamefont {Harlander}}, \bibinfo {author} {\bibfnamefont {W.}~\bibnamefont {H\"ansel}}, \bibinfo {author} {\bibfnamefont {M.}~\bibnamefont {Hennrich}},\ and\ \bibinfo {author} {\bibfnamefont {R.}~\bibnamefont {Blatt}},\ }\bibfield  {title} {\bibinfo {title} {14-qubit entanglement: Creation and coherence},\ }\href {https://doi.org/10.1103/PhysRevLett.106.130506} {\bibfield  {journal} {\bibinfo  {journal} {Phys. Rev. Lett.}\ }\textbf {\bibinfo {volume} {106}},\ \bibinfo {pages} {130506} (\bibinfo {year} {2011})}\BibitemShut {NoStop}%
\bibitem [{\citenamefont {Omran}\ \emph {et~al.}(2019)\citenamefont {Omran}, \citenamefont {Levine}, \citenamefont {Keesling}, \citenamefont {Semeghini}, \citenamefont {Wang}, \citenamefont {Ebadi}, \citenamefont {Bernien}, \citenamefont {Zibrov}, \citenamefont {Pichler}, \citenamefont {Choi} \emph {et~al.}}]{omran2019generation}%
  \BibitemOpen
  \bibfield  {author} {\bibinfo {author} {\bibfnamefont {A.}~\bibnamefont {Omran}}, \bibinfo {author} {\bibfnamefont {H.}~\bibnamefont {Levine}}, \bibinfo {author} {\bibfnamefont {A.}~\bibnamefont {Keesling}}, \bibinfo {author} {\bibfnamefont {G.}~\bibnamefont {Semeghini}}, \bibinfo {author} {\bibfnamefont {T.~T.}\ \bibnamefont {Wang}}, \bibinfo {author} {\bibfnamefont {S.}~\bibnamefont {Ebadi}}, \bibinfo {author} {\bibfnamefont {H.}~\bibnamefont {Bernien}}, \bibinfo {author} {\bibfnamefont {A.~S.}\ \bibnamefont {Zibrov}}, \bibinfo {author} {\bibfnamefont {H.}~\bibnamefont {Pichler}}, \bibinfo {author} {\bibfnamefont {S.}~\bibnamefont {Choi}}, \emph {et~al.},\ }\bibfield  {title} {\bibinfo {title} {Generation and manipulation of {S}chr{\"o}dinger cat states in {R}ydberg atom arrays},\ }\href {https://doi.org/10.1126/science.aax9743} {\bibfield  {journal} {\bibinfo  {journal} {Science}\ }\textbf {\bibinfo {volume} {365}},\ \bibinfo {pages} {570} (\bibinfo {year} {2019})}\BibitemShut {NoStop}%
\bibitem [{\citenamefont {Pogorelov}\ \emph {et~al.}(2021)\citenamefont {Pogorelov}, \citenamefont {Feldker}, \citenamefont {Marciniak}, \citenamefont {Postler}, \citenamefont {Jacob}, \citenamefont {Krieglsteiner}, \citenamefont {Podlesnic}, \citenamefont {Meth}, \citenamefont {Negnevitsky}, \citenamefont {Stadler}, \citenamefont {H\"ofer}, \citenamefont {W\"achter}, \citenamefont {Lakhmanskiy}, \citenamefont {Blatt}, \citenamefont {Schindler},\ and\ \citenamefont {Monz}}]{pogorelov2021compact}%
  \BibitemOpen
  \bibfield  {author} {\bibinfo {author} {\bibfnamefont {I.}~\bibnamefont {Pogorelov}}, \bibinfo {author} {\bibfnamefont {T.}~\bibnamefont {Feldker}}, \bibinfo {author} {\bibfnamefont {C.~D.}\ \bibnamefont {Marciniak}}, \bibinfo {author} {\bibfnamefont {L.}~\bibnamefont {Postler}}, \bibinfo {author} {\bibfnamefont {G.}~\bibnamefont {Jacob}}, \bibinfo {author} {\bibfnamefont {O.}~\bibnamefont {Krieglsteiner}}, \bibinfo {author} {\bibfnamefont {V.}~\bibnamefont {Podlesnic}}, \bibinfo {author} {\bibfnamefont {M.}~\bibnamefont {Meth}}, \bibinfo {author} {\bibfnamefont {V.}~\bibnamefont {Negnevitsky}}, \bibinfo {author} {\bibfnamefont {M.}~\bibnamefont {Stadler}}, \bibinfo {author} {\bibfnamefont {B.}~\bibnamefont {H\"ofer}}, \bibinfo {author} {\bibfnamefont {C.}~\bibnamefont {W\"achter}}, \bibinfo {author} {\bibfnamefont {K.}~\bibnamefont {Lakhmanskiy}}, \bibinfo {author} {\bibfnamefont {R.}~\bibnamefont {Blatt}}, \bibinfo {author} {\bibfnamefont {P.}~\bibnamefont {Schindler}},\ and\ \bibinfo {author}
  {\bibfnamefont {T.}~\bibnamefont {Monz}},\ }\bibfield  {title} {\bibinfo {title} {Compact ion-trap quantum computing demonstrator},\ }\href {https://doi.org/10.1103/PRXQuantum.2.020343} {\bibfield  {journal} {\bibinfo  {journal} {PRX Quantum}\ }\textbf {\bibinfo {volume} {2}},\ \bibinfo {pages} {020343} (\bibinfo {year} {2021})}\BibitemShut {NoStop}%
\bibitem [{\citenamefont {Thomas}\ \emph {et~al.}(2022)\citenamefont {Thomas}, \citenamefont {Ruscio}, \citenamefont {Morin},\ and\ \citenamefont {Rempe}}]{thomas2022efficient}%
  \BibitemOpen
  \bibfield  {author} {\bibinfo {author} {\bibfnamefont {P.}~\bibnamefont {Thomas}}, \bibinfo {author} {\bibfnamefont {L.}~\bibnamefont {Ruscio}}, \bibinfo {author} {\bibfnamefont {O.}~\bibnamefont {Morin}},\ and\ \bibinfo {author} {\bibfnamefont {G.}~\bibnamefont {Rempe}},\ }\bibfield  {title} {\bibinfo {title} {Efficient generation of entangled multiphoton graph states from a single atom},\ }\href {https://doi.org/10.1038/s41586-022-04987-5} {\bibfield  {journal} {\bibinfo  {journal} {Nature}\ }\textbf {\bibinfo {volume} {608}},\ \bibinfo {pages} {677} (\bibinfo {year} {2022})}\BibitemShut {NoStop}%
\bibitem [{\citenamefont {Mooney}\ \emph {et~al.}(2021)\citenamefont {Mooney}, \citenamefont {White}, \citenamefont {Hill},\ and\ \citenamefont {Hollenberg}}]{mooney2021generation}%
  \BibitemOpen
  \bibfield  {author} {\bibinfo {author} {\bibfnamefont {G.~J.}\ \bibnamefont {Mooney}}, \bibinfo {author} {\bibfnamefont {G.~A.}\ \bibnamefont {White}}, \bibinfo {author} {\bibfnamefont {C.~D.}\ \bibnamefont {Hill}},\ and\ \bibinfo {author} {\bibfnamefont {L.~C.}\ \bibnamefont {Hollenberg}},\ }\bibfield  {title} {\bibinfo {title} {Generation and verification of 27-qubit {G}reenberger-{H}orne-{Z}eilinger states in a superconducting quantum computer},\ }\href {https://doi.org/10.1088/2399-6528/ac1df7} {\bibfield  {journal} {\bibinfo  {journal} {J. Phys. Commun.}\ }\textbf {\bibinfo {volume} {5}},\ \bibinfo {pages} {095004} (\bibinfo {year} {2021})}\BibitemShut {NoStop}%
\bibitem [{\citenamefont {Wei}\ \emph {et~al.}(2020)\citenamefont {Wei}, \citenamefont {Lauer}, \citenamefont {Srinivasan}, \citenamefont {Sundaresan}, \citenamefont {McClure}, \citenamefont {Toyli}, \citenamefont {McKay}, \citenamefont {Gambetta},\ and\ \citenamefont {Sheldon}}]{wei2020verifying}%
  \BibitemOpen
  \bibfield  {author} {\bibinfo {author} {\bibfnamefont {K.~X.}\ \bibnamefont {Wei}}, \bibinfo {author} {\bibfnamefont {I.}~\bibnamefont {Lauer}}, \bibinfo {author} {\bibfnamefont {S.}~\bibnamefont {Srinivasan}}, \bibinfo {author} {\bibfnamefont {N.}~\bibnamefont {Sundaresan}}, \bibinfo {author} {\bibfnamefont {D.~T.}\ \bibnamefont {McClure}}, \bibinfo {author} {\bibfnamefont {D.}~\bibnamefont {Toyli}}, \bibinfo {author} {\bibfnamefont {D.~C.}\ \bibnamefont {McKay}}, \bibinfo {author} {\bibfnamefont {J.~M.}\ \bibnamefont {Gambetta}},\ and\ \bibinfo {author} {\bibfnamefont {S.}~\bibnamefont {Sheldon}},\ }\bibfield  {title} {\bibinfo {title} {Verifying multipartite entangled {G}reenberger-{H}orne-{Z}eilinger states via multiple quantum coherences},\ }\href {https://doi.org/10.1103/PhysRevA.101.032343} {\bibfield  {journal} {\bibinfo  {journal} {Phys. Rev. A}\ }\textbf {\bibinfo {volume} {101}},\ \bibinfo {pages} {032343} (\bibinfo {year} {2020})}\BibitemShut {NoStop}%
\end{thebibliography}%

\clearpage
\pagebreak

\appendix

\clearpage 
\pagebreak

% Add S prefix to supplementary figure and table numbering.
\makeatletter
\renewcommand\thetable{S\@arabic\c@table}
\renewcommand\thefigure{S\@arabic\c@figure}
\makeatother

% Reset figure and table counters. 
\setcounter{figure}{0}
\setcounter{table}{0}

\begin{widetext}

\section{Preliminaries}
\label{app-sec:preliminaries}

\subsection{Notation and basic facts}
\label{app-sec:preliminaries/notation}

Throughout our work some symbols such as $s, t, f$ may represent a bitstring or bitvector when so defined. We do not distinguish between bitstrings and bitvectors. We use the following notation:
\begin{itemize}[itemsep=-0.2\parsep]
    \item The length of a bitstring $s$, that is, the number of bits in $s$, is denoted $\abs{s}$.
    \item The Hamming weight of a bitstring $s$, that is, the number of nonzero bits in $s$, is denoted $\wt(s)$.
    \item The $j^{\text{th}}$-bit of a bitstring $s$ is denoted $s_j \in \{0, 1\}$.
    \item The bit-wise sum, equivalently difference, of bitstrings $s$ and $t$ of equal length is $s \oplus t$, which is a bitstring.
    \item The dot product of bitstrings $s$ and $t$ of equal length is $\smash{s \cdot t = \sum_{j = 1}^{\abs{s}} s_j t_j} \,\, (\text{mod} \, 2)$, which is a bit. 
    \item Where unambiguous by context, the concatenation of two bitstrings $s^{(1)}$ and $s^{(2)}$ is written simply $s^{(1)} s^{(2)}$.
    \item The all-zero bitstring is denoted $\vb{0}$, which has all entries zero. The all-one counterpart is denoted $\vb{1}$.
    \item Bitstrings can be used as indices. The basis for the vector, matrix or tensor is simply regarded as being enumerated (labelled) by the bitstrings. Alternatively one can implicitly convert the bitstrings to their decimal values for conventional integer enumeration.
\end{itemize}

We write $[n] = \{1, 2, \ldots, n\}$ for $n \in \mathbb{N}$. We use the following notation for vectors, matrices and tensors:
\begin{itemize}[itemsep=-0.2\parsep]
    \item Vectors are written with boldface, for example $\vb{v}$. Matrices and tensors are not bolded, for example $A$.
    \item Entries are indexed with subscripts. For example $v_j$ is the $\smash{j^\text{th}}$ entry of a vector $\vb{v}$, and $A_{ij}$ is the $\smash{(i, j)^\text{th}}$ entry of a matrix $A$.
    \item $\smash{\norm{\vb{v}}}_p$ is the $p$-norm of a vector $\vb{v}$. 
    \item $\smash{\norm{A}}_p$ for a matrix $A$ denotes the matrix norm induced by the vector $p$-norm. For example $\smash{\norm{A}}_2$ is the spectral norm of $A$ and is the largest singular value of $A$.
    \item $\abs{\vb{v}}$ and $\abs{A}$ are the entry-wise absolute values of a vector $\vb{v}$ and a matrix $A$, themselves a vector and a matrix respectively, not to be confused with the norms of $\vb{v}$ and $A$.
    \item $\vb{v} > \vb{v}'$ means that $\vb{v}$ is entry-wise larger than $\vb{v}'$, that is, $v_j > v'_j$ for every index $j$. Likewise for $\geq, \leq, <$.
    \item The binary convolution between two vectors $\vb{u}$ and $\vb{v}$ of equal length is a vector with entries
    \begin{equation}\begin{split}
        (\vb{u} \ast \vb{v})_s = \sum_{t \in \mathcal{S}} u_t v_{s \oplus t},
    \end{split}\end{equation}
    for each $s \in \mathcal{S}$, where $\mathcal{S}$ is the set of bitstrings indexing $\vb{u}$ and $\vb{v}$.
\end{itemize}

We use conventional notation for Pauli matrices and the Hadamard and $S$ gates,
\begin{equation}\begin{split}
    X = \mqty[
        0 & 1 \\
        1 & 0
    ], \qquad 
    Y = \mqty[
        0 & -i \\
        i & 0
    ], \qquad 
    Z = \mqty[
        1 & 0 \\
        0 & -1
    ], \qquad 
    H = \frac{1}{\sqrt{2}} \mqty[
        1 & 1 \\ 
        1 & -1
    ], \qquad
    S = \mqty[
        1 & 0 \\
        0 & i
    ],
\end{split}\end{equation}
and the CX gate with the first qubit as control and the second qubit as target is written $\ket{0} \bra{0} \otimes \mathbb{I} + \ket{1} \bra{1} \otimes X$. We denote the identity as $\mathbb{I}$, with dimensions inferred from context. The $d$-fold tensor product of an object $A$, for instance quantum state, vector, or matrix, is denoted $A^{\otimes d}$. More generally we write
\begin{equation}\begin{split}
    A^{(s)} = \bigotimes_{j = 1}^{\abs{s}} A^{s_j},
\end{split}\end{equation}
for a bitstring $s$, which is a $d$-fold tensor product over $\{A, \mathbb{I}\}$ as determined by the bitstring $s$. The diagonal entries of tensor products of $Z$ Pauli matrices (which are diagonal) satisfy
\begin{equation}\begin{split}
    \left[ Z^{(s)} \right]_{k k} = (-1)^{k \cdot s},
    \label{app-eq:preliminaries/notation/Z-tensor-entries}
\end{split}\end{equation}
where $k$ is a bitstring index. The entries of tensor products of $H$ satisfy
\begin{equation}\begin{split}
    \sqrt{2^m} \left[ H^{\otimes m} \right]_{f k} 
    = \left[ Z^{(f)} \right]_{k k}
    = (-1)^{f \cdot k},
    \label{app-eq:preliminaries/notation/H-tensor-entries}
\end{split}\end{equation}
where $f$ and $k$ are bitstring indices.

\subsection{Properties of confusion matrices $M$, $Q$ and vector $\vb{q}$}
\label{app-sec:preliminaries/confusion-matrix-properties}

As written in \cref{sec:framework}, in a setting with $m$ measurements, each entry $M_{t s} \in [0, 1]$ records the probability of the quantum device reporting measurement outcome $t \in \mathcal{S} = \{0, 1\}^m$ when the true outcome is $s \in \mathcal{S}$. Correspondingly $M$ is stochastic, \ie
\begin{equation}\begin{split}
    \sum_{t \in \mathcal{S}} M_{t s} 
    = 1 \quad \forall \quad s \in \mathcal{S},
    \label{app-eq:preliminaries/confusion-matrix-properties/M-stochastic}
\end{split}\end{equation}
and consequently,
\begin{equation}\begin{split}
    \norm{M}_1 
    = \max_{s \in \mathcal{S}} \underbrace{ 
        \sum_{t \in \mathcal{S}} \abs{M_{t s}} }_{1}
    = 1.
    \label{app-eq:preliminaries/confusion-matrix-properties/M-norm}
\end{split}\end{equation}

Each entry $q_s \in [0, 1]$ of the symmetrized confusion vector $\vb{q}$ records the probability of a readout error syndrome $s \in \mathcal{S}$ occurring (with BFA applied)---that is, the probability that any true measurement outcome $a \in \mathcal{S}$ is reported as $a \oplus s$. This probability is uniform across $a \in \mathcal{S}$ due to the symmetrization of the readout channels. Correspondingly $\vb{q}$ is a probability distribution,
\begin{equation}\begin{split}
    \sum_{s \in \mathcal{S}} q_s = 1.
    \label{app-eq:preliminaries/confusion-matrix-properties/q-stochastic}
\end{split}\end{equation}

The symmetrized confusion matrix $Q$, with entries $Q_{sf} = q_{s \oplus f} \in [0, 1]$ for $s \in \mathcal{S}$ and $f \in \mathcal{S}$, is doubly stochastic, \ie
\begin{equation}\begin{split}
    \sum_{s \in \mathcal{S}} Q_{s f} 
    &= \sum_{s \in \mathcal{S}} q_{s \oplus f}
    = \sum_{s' \in \mathcal{S}} q_{s'}
    = 1
    \quad \forall \quad f \in \mathcal{S},
    \\
    \sum_{f \in \mathcal{S}} Q_{s f} 
    &= \sum_{f \in \mathcal{S}} q_{s \oplus f}
    = \sum_{s' \in \mathcal{S}} q_{s'}
    = 1
    \quad \forall \quad s \in \mathcal{S},
    \label{app-eq:preliminaries/confusion-matrix-properties/Q-stochastic}
\end{split}\end{equation}
and consequently,
\begin{equation}\begin{split}
    \norm{Q}_1 
    &= \max_{f \in \mathcal{S}} \underbrace{ 
        \sum_{s \in \mathcal{S}} \abs{Q_{sf}} }_{1}
    = 1,
    \qquad
    \norm{Q}_\infty
    = \max_{s \in \mathcal{S}} \underbrace{ 
        \sum_{f \in \mathcal{S}} \abs{Q_{sf}} }_{1}
    = 1.
    \label{app-eq:preliminaries/confusion-matrix-properties/Q-norm}
\end{split}\end{equation}

For a more concrete visualization, below we give explicit forms of $Q$ for $m = 1$, $m = 2$ and $m = 3$ cases,
\begin{equation}\begin{split}
    Q = \mqty[
        q_{0} & q_{1} \\ 
        q_{1} & q_{0}
    ], \qquad 
    Q = \mqty[
        q_{00} & q_{01} & q_{10} & q_{11} \\ 
        q_{01} & q_{00} & q_{11} & q_{10} \\ 
        q_{10} & q_{11} & q_{00} & q_{01} \\ 
        q_{11} & q_{10} & q_{01} & q_{00} \\ 
    ], \qquad 
    Q = \mqty[
        q_{000} & q_{001} & q_{010} & q_{011} & q_{100} & q_{101} & q_{110} & q_{111} \\
        q_{001} & q_{000} & q_{011} & q_{010} & q_{101} & q_{100} & q_{111} & q_{110} \\
        q_{010} & q_{011} & q_{000} & q_{001} & q_{110} & q_{111} & q_{100} & q_{101} \\
        q_{011} & q_{010} & q_{001} & q_{000} & q_{111} & q_{110} & q_{101} & q_{100} \\
        q_{100} & q_{101} & q_{110} & q_{111} & q_{000} & q_{001} & q_{010} & q_{011} \\
        q_{101} & q_{100} & q_{111} & q_{110} & q_{001} & q_{000} & q_{011} & q_{010} \\
        q_{110} & q_{111} & q_{100} & q_{101} & q_{010} & q_{011} & q_{000} & q_{001} \\
        q_{111} & q_{110} & q_{101} & q_{100} & q_{011} & q_{010} & q_{001} & q_{000} \\
    ].
    \label{app-eq:preliminaries/confusion-matrix-properties/Q-explicit-examples}
\end{split}\end{equation}

In an ideal setting without readout errors, $\smash{M_{t s} = \delta_{t s}}$ or equivalently $M = \mathbb{I}$ is the identity matrix. In the symmetrized picture, in the ideal setting the probability distribution $\vb{q}$ has support only on the trivial syndrome, $\smash{q_s = \delta_{s \vb{0}}}$, and correspondingly $Q = \mathbb{I}$ is the identity matrix.

In general, the following properties about the diagonal entries and trace of $Q$ hold,
\begin{equation}\begin{split}
    Q_{s s} = q_{s \oplus s} = q_{\vb{0}}
    \quad \forall \quad s \in \mathcal{S},
    \qquad \qquad
    \tr Q = 2^m q_{\vb{0}}.
    \label{app-eq:preliminaries/confusion-matrix-properties/Q-property-trace}
\end{split}\end{equation}

The determinant of $Q$ can also be analytically evaluated---see \cref{app-eq:preliminaries/Q-diagonalization/Q-trace-det}. This is most compactly done via the diagonalization of $Q$, which we describe in \cref{app-sec:preliminaries/Q-diagonalization}.

Lastly, given an $M$, one can in fact compute the corresponding $\vb{q}$ and $Q$ that characterizes the symmetrized readout error channels, that would have been measured had BFA been used during calibration, and which would be needed for readout error mitigation in subsequent experiment circuits executed with BFA. In particular,
\begin{equation}\begin{split}
    q_s 
    = \frac{1}{2^m} 
        \sum_{\substack{t, t' \in \mathcal{S} \\ t \oplus t' = s}} 
        M_{t t'}
    = \frac{1}{2^m} \sum_{t \in \mathcal{S}}
        M_{t (t \oplus s)},
    \qquad 
    Q_{sf} 
    = q_{s \oplus f}
    = \frac{1}{2^m} \sum_{t \in \mathcal{S}}
        M_{t (t \oplus s \oplus f)},
    \label{app-eq:preliminaries/confusion-matrix-properties/q-Q-from-M}
\end{split}\end{equation}
that is, $q_s$ is the average of all entries $M_{t t'}$ consistent with the syndrome $t \oplus t' = s$, and the construction of $Q$ follows from $\vb{q}$ by definition. The explicit meaning that $\vb{q}$ and $Q$ describe the readout error channels under symmetrization is thus made clear. We clarify, however, that in practice $\vb{q}$ and $Q$ can be directly obtained by including BFA in the calibration circuits, and there is no need to go through $M$. See \cref{app-sec:methods/ro-terminal} for details on the calibration process as implemented in our experiments. This direct calibration with BFA provides resource and precision advantages~\cite{smith2021qubit, hicks2021readout}.

\subsection{Diagonalization of symmetrized confusion matrix $Q$ and the inverse of $Q$}
\label{app-sec:preliminaries/Q-diagonalization}

As noted in \cref{sec:theory-single/general} of the main text, $Q$ is hugely symmetric. Its entries $Q_{sf}$ are invariant under the index shift $(s, f) \to (s \oplus s', f \oplus s')$ for any $s' \in \mathcal{S}$. In particular, $Q$ is symmetric about both its diagonal and anti-diagonal, and so is every of its recursively halved quadrants---see, for example, \cref{app-eq:preliminaries/confusion-matrix-properties/Q-explicit-examples} for visualization. Recall that $m$-fold tensor products of the Pauli matrices form a basis for $2^m \times 2^m$ complex matrices, and of the Pauli matrices, only $\mathbb{I}$ and $X$ are symmetric about both their diagonal and anti-diagonal. Thus it follows that $Q$ must be of the form
\begin{equation}\begin{split}
    Q = \sum_{s \in \mathcal{S}} A_s X^{(s)},
\end{split}\end{equation}
for coefficients $A_s \in \mathbb{R}$, in order for it to possess the symmetries exhibited. To fix the coefficients, one examines the first column of $Q$, which must match $\vb{q}$ by definition. Note that the first column of $X^{(s)}$ has a unit entry at row $s$ and is zero everywhere else, $\smash{\big[X^{(s)}\big]_{r \vb{0}} = \delta_{r s}}$. Thus
\begin{equation}\begin{split}
    Q_{r \vb{0}} 
    = \sum_{s \in \mathcal{S}} A_s \left[ X^{(s)} \right]_{r \vb{0}} 
    = \sum_{s \in \mathcal{S}} A_s \delta_{r s}
    = A_r
    = q_r.
\end{split}\end{equation}

That is,
\begin{equation}\begin{split}
    Q = \sum_{s \in \mathcal{S}} q_s X^{(s)}.
\end{split}\end{equation}

Now note that $X^{(s)}$ is diagonalized by $H^{\otimes m}$, that is, $H^{\otimes m} X^{(s)} H^{\otimes m} = Z^{(s)}$. Then $Q$, which is a linear combination of $X^{(s)}$, is also diagonalized by $H^{\otimes m}$,
\begin{equation}\begin{split}
    H^{\otimes m} Q H^{\otimes m} = \sum_{s \in \mathcal{S}} q_s Z^{(s)}.
    \label{app-eq:preliminaries/Q-diagonalization/Q-diagonalized}
\end{split}\end{equation}

The eigenvalues of $Q$ are thus given by
\begin{equation}\begin{split}
    \lambda_k 
    = \sum_{s \in \mathcal{S}} q_s \left[Z^{(s)}\right]_{k k}
    = \sum_{s \in \mathcal{S}} (-1)^{k \cdot s} q_s,
    \label{app-eq:preliminaries/Q-diagonalization/lambda-solution-1}
\end{split}\end{equation}
where we have invoked the identity in \cref{app-eq:preliminaries/notation/Z-tensor-entries}. Recalling the normalization $\sum_{s \in \mathcal{S}} q_s = 1$, one can obtain an alternate form
\begin{equation}\begin{split}
    \lambda_k 
    = q_{\vb{0}} 
        + \sum_{\substack{s \in \mathcal{S} \\ s \neq \vb{0}}} (-1)^{k \cdot s} q_s
    = \left( 1 - \sum_{\substack{s \in \mathcal{S} \\ s \neq \vb{0}}} q_s \right)
        + \sum_{\substack{s \in \mathcal{S} \\ s \neq \vb{0}}} (-1)^{k \cdot s} q_s
    &= 1 - \sum_{\substack{s \in \mathcal{S} \\ s \neq \vb{0}}} 
        \left[1 - (-1)^{k \cdot s}\right] q_s \\
    &= 1 - 2 \sum_{\substack{s \in \mathcal{S} \\ k \cdot s \neq 0}} q_s,
    \label{app-eq:preliminaries/Q-diagonalization/lambda-solution-2}
\end{split}\end{equation}
which makes the special case $\lambda_{\vb{0}} = 1$ clear. The solutions for the eigenvalues in $Q$ in \cref{app-eq:preliminaries/Q-diagonalization/lambda-solution-1,app-eq:preliminaries/Q-diagonalization/lambda-solution-2} above are reported in \cref{eq:readout-errors/lambda-solution} of the main text. In general, as $Q$ is doubly stochastic, as in \cref{app-eq:preliminaries/confusion-matrix-properties/Q-stochastic}, we are guaranteed that $\abs{\lambda_k} \leq 1$ for all $k \in \mathcal{S}$. In the absence of readout errors ($\smash{q_s = \delta_{s \vb{0}}}$), we straightforwardly have $\vb*{\lambda} = \vb{1}$, that is, $\lambda_k = 1$ uniformly for all $k \in \mathcal{S}$. Lastly we comment that the trace and determinant of $Q$ yield the properties
\begin{equation}\begin{split}
    \tr Q
    &= \sum_{k \in \mathcal{S}} \lambda_k
    = 2^m q_{\vb{0}}, \\    
    \det Q
    &= \prod_{k \in \mathcal{S}} \lambda_k
    = \prod_{k \in \mathcal{S}} 
        \left[ \sum_{s \in \mathcal{S}} (-1)^{k \cdot s} q_s \right] 
    = \prod_{k \in \mathcal{S}} 
        \left[ 1 - 2 \sum_{\substack{s \in \mathcal{S} \\ k \cdot s \neq 0}} q_s \right],
    \label{app-eq:preliminaries/Q-diagonalization/Q-trace-det}
\end{split}\end{equation}
where in the first line we have used our prior result for the trace of $Q$ in \cref{app-eq:preliminaries/confusion-matrix-properties/Q-property-trace}. As $Q$ is stochastic we have that $\abs{\det Q} \leq 1$, and \cref{app-eq:preliminaries/Q-diagonalization/Q-trace-det} implies $\abs{\det Q} = 1$ only when no readout errors are present.

In anticipation of developments in \cref{app-sec:theory}, we examine $Q^{-1}$. Trivially, we remind that in an ideal setting without readout errors, $Q = \mathbb{I}$ and thus $Q^{-1} = \mathbb{I}$. Now, the diagonalization of $Q$ from \cref{app-eq:preliminaries/Q-diagonalization/Q-diagonalized} gives
\begin{equation}\begin{split}
    Q^{-1} 
    = H^{\otimes m} 
        \mqty[\dmat{\frac{1}{\lambda_1}, \frac{1}{\lambda_2}, \ddots, \frac{1}{\lambda_m}}]
        H^{\otimes m},
\end{split}\end{equation}
and thus entries of $Q^{-1}$ can be written as
\begin{equation}\begin{split}
    \left[ Q^{-1} \right]_{s f}
    &= \sum_{k \in \mathcal{S}} 
        \left[ H^{\otimes m} \right]_{s k}
        \left( \frac{1}{\lambda_k} \right)
        \left[ H^{\otimes m} \right]_{k f} \\
    &= \frac{1}{2^m} \sum_{k \in \mathcal{S}} 
        (-1)^{s \cdot k} (-1)^{k \cdot f}
        \left( \frac{1}{\lambda_k} \right),
    \label{app-eq:preliminaries/Q-diagonalization/Q-inverse-general}
\end{split}\end{equation}
where we have used the identities for entries of tensor products of $H$ matrices in \cref{app-eq:preliminaries/notation/H-tensor-entries}. For future use, we note that the first column of $Q^{-1}$ is given by
\begin{equation}\begin{split}
    \left[ Q^{-1} \right]_{s \vb{0}}
    = \frac{1}{2^m} \sum_{k \in \mathcal{S}} 
        (-1)^{s \cdot k}
        \left( \frac{1}{\lambda_k} \right).
    \label{app-eq:preliminaries/Q-diagonalization/Q-inverse-first-column}
\end{split}\end{equation}

Moreover, that $Q$ is hugely symmetric implies the same for $Q^{-1}$---namely, that the entries $[Q^{-1}]_{sf}$ are invariant under the index shift $(s, f) \to (s \oplus s', f \oplus s')$ for any $s' \in \mathcal{S}$, indeed also explicitly observable in \cref{app-eq:preliminaries/Q-diagonalization/Q-inverse-general}. In particular, we observe $\smash{[Q^{-1}]_{s f} = [Q^{-1}]_{(s \oplus f) \vb{0}} = [Q^{-1}]_{\vb{0} (s \oplus f)}}$, and thus each column and row of $Q^{-1}$ contains the same entries but arranged in different orders. This means
\begin{equation}\begin{split}
    \norm{ Q^{-1} }_1 &= \max_{f \in \mathcal{S}} 
        \sum_{s \in \mathcal{S}} \abs{ [Q^{-1}]_{s f} }
        = \sum_{s \in \mathcal{S}} \abs{ [Q^{-1}]_{s \vb{0}} }, \\
    \norm{ Q^{-1} }_\infty &= \max_{s \in \mathcal{S}} 
        \sum_{f \in \mathcal{S}} \abs{ [Q^{-1}]_{s f} }
        = \sum_{f \in \mathcal{S}} \abs{ [Q^{-1}]_{\vb{0} f} }.
    \label{app-eq:preliminaries/Q-diagonalization/Q-inverse-norms}
\end{split}\end{equation}

Also, from $Q Q^{-1} = Q^{-1} Q = \mathbb{I}$, we observe that row and column sums of $Q^{-1}$ evaluate uniformly to unity,
\begin{equation}\begin{split}
    \sum_{k \in \mathcal{S}} Q_{s k} [Q^{-1}]_{k f} = \delta_{s f}
    &\Longrightarrow 
    \sum_{s \in \mathcal{S}} \sum_{k \in \mathcal{S}} 
        Q_{s k} [Q^{-1}]_{k f}
    = \sum_{k \in \mathcal{S}} 
        \underbrace{ \left( \sum_{s \in \mathcal{S}} Q_{s k} \right) }_{1}
        [Q^{-1}]_{k f}
    = \sum_{k \in \mathcal{S}} [Q^{-1}]_{k f}
    = \sum_{s \in \mathcal{S}} \delta_{s f}
    = 1 
    \quad \forall \quad f \in \mathcal{S}, \\
    \sum_{k \in \mathcal{S}} [Q^{-1}]_{s k} Q_{k f}  = \delta_{s f}
    &\Longrightarrow 
    \sum_{f \in \mathcal{S}} \sum_{k \in \mathcal{S}} 
        [Q^{-1}]_{s k} Q_{k f} 
    = \sum_{k \in \mathcal{S}} 
        [Q^{-1}]_{s k}
        \underbrace{ \left( \sum_{f \in \mathcal{S}} Q_{k f} \right) }_{1}
    = \sum_{k \in \mathcal{S}} [Q^{-1}]_{s k}
    = \sum_{f \in \mathcal{S}} \delta_{s f}
    = 1 
    \quad \forall \quad s \in \mathcal{S},
    \label{app-eq:preliminaries/Q-diagonalization/Q-inverse-row-col-sums}
\end{split}\end{equation}
where we have used the fact that $Q$ is doubly stochastic, as in \cref{app-eq:preliminaries/confusion-matrix-properties/Q-stochastic}. Despite this property, we clarify that $Q^{-1}$ is not generally doubly stochastic as its entries are not guaranteed to be non-negative. In fact $Q Q^{-1} = Q^{-1} Q = \mathbb{I}$ implies that each row and column of $Q^{-1}$ must contain at least one negative entry when readout errors are present (when $Q \neq \mathbb{I}$).

\subsection{Tensored readout channels}
\label{app-sec:preliminaries/tensored}

\subsubsection{General setting}
\label{app-sec:preliminaries/tensored/general}

Consider a setting wherein we partition the $m$ measurements into subsets of sizes $\smash{\{m^{(1)}, m^{(2)}, \ldots, m^{(p)}\}}$, where $\smash{m^{(\ell)} \geq 1}$ for each $\ell$ and $\smash{m^{(1)} + m^{(2)} + \cdots + m^{(p)} = m}$, and we make the assumption that there are no correlations in readout errors between measurements in different subsets. That is, readout errors in different subsets are independent; but within a subset correlations in readout error can be present. One then has the decomposition
\begin{equation}\begin{split}
    M = \bigotimes_{\ell = 1}^p M^{(\ell)},
    \label{app-eq:preliminaries/tensored/general/M-factorization}
\end{split}\end{equation}
where $\smash{M^{(\ell)}}$ is a $\smash{2^{m^{(\ell)}} \times 2^{m^{(\ell)}}}$-sized confusion matrix characterizing readout errors affecting the $\ell^\text{th}$ subset of measurements. This is equivalent to a statement of factorization of probabilities
\begin{equation}\begin{split}
    M_{t s} = \prod_{\ell = 1}^p M^{(\ell)}_{t^{(\ell)} s^{(\ell)}},
\end{split}\end{equation}
where $\smash{s = s^{(1)} s^{(2)} \ldots s^{(p)} \in \mathcal{S}}$ and $\smash{t = t^{(1)} t^{(2)} \ldots t^{(p)} \in \mathcal{S}}$ are the corresponding partitionings of the bitstrings over the $p$ subsets, $\smash{s^{(\ell)} \in \{0, 1\}^{m^{(\ell)}}}$ and $\smash{t^{(\ell)} \in \{0, 1\}^{m{(\ell)}}}$ for each $\ell \in [p]$. Likewise,
\begin{equation}\begin{split}
    \vb{q} &= \bigotimes_{\ell = 1}^p \vb{q}^{(\ell)},
    \qquad 
    q_s = \prod_{\ell = 1}^p q^{(\ell)}_{s^{(\ell)}}, \\
    Q &= \bigotimes_{\ell = 1}^p Q^{(\ell)},
    \qquad 
    Q_{sf} = \prod_{\ell = 1}^p Q^{(\ell)}_{s^{(\ell)} f^{(\ell)}},
    \label{app-eq:preliminaries/tensored/general/q-Q-factorization}
\end{split}\end{equation}
with $\smash{s = s^{(1)} s^{(2)} \ldots s^{(p)} \in \mathcal{S}}$ and $\smash{f = f^{(1)} f^{(2)} \ldots f^{(p)} \in \mathcal{S}}$ being partitionings of the bitstrings, and where $\smash{\vb{q}^{(\ell)}}$ is a $\smash{2^{m^{(\ell)}}}$-length probability vector and $\smash{Q^{(\ell)}}$ is the corresponding $\smash{2^{m^{(\ell)}} \times 2^{m^{(\ell)}}}$-sized symmetrized confusion matrix characterizing readout errors affecting the $\ell^\text{th}$ subset of measurements. It follows from this tensor product structure that the eigenvalues of $Q$ factorize,
\begin{equation}\begin{split}
    \vb*{\lambda} = \bigotimes_{\ell = 1}^p \vb*{\lambda}^{(\ell)},
    \qquad
    \lambda_k = \prod_{\ell = 1}^m \lambda^{(\ell)}_{k^{(\ell)}},
    \label{app-eq:preliminaries/tensored/general/lambda-factorization}
\end{split}\end{equation}
where $\smash{\vb*{\lambda}^{(\ell)}}$ are the eigenvalues of $Q^{(\ell)}$ and are given by the same solutions as in \cref{app-eq:preliminaries/Q-diagonalization/lambda-solution-1,app-eq:preliminaries/Q-diagonalization/lambda-solution-2}, but with $\vb{q}^{(\ell)}$ taking the place of $\vb{q}$. Likewise the inverse of $Q$ factorizes,
\begin{equation}\begin{split}
    Q^{-1} = \bigotimes_{\ell = 1}^p Q^{(\ell)^{-1}},
    \qquad 
    \left[ Q^{-1} \right]_{sf} = \prod_{\ell = 1}^p 
        \left[ Q^{(\ell)^{-1}} \right]_{s^{(\ell)} f^{(\ell)}},
    \qquad 
    \left[ Q^{-1} \right]_{s \vb{0}} = \prod_{\ell = 1}^p 
        \left[ Q^{(\ell)^{-1}} \right]_{s^{(\ell)} \vb{0}},
    \label{app-eq:preliminaries/tensored/general/Q-inverse-factorization}
\end{split}\end{equation}
where $\smash{Q^{(\ell)^{-1}}}$ is the inverse of $Q^{(\ell)}$, whose general and first-column entries are given in \cref{app-eq:preliminaries/Q-diagonalization/Q-inverse-general} and \cref{app-eq:preliminaries/Q-diagonalization/Q-inverse-first-column}, but now with $\vb*{\lambda}^{(\ell)}$ taking the place of $\vb{\lambda}$.

\subsubsection{Layer-wise tensored readout channels}
\label{app-sec:preliminaries/tensored/layer-tensored}

In a feedforward circuit with $L > 1$ layers (\ie~the setting of \cref{sec:theory-multiple} of the main text), the discussion above straightforwardly applies when an assumption of layer-wise independence of readout errors is made. That is, readout errors on measurements in different layers of the circuit are taken to be independent, but we accommodate arbitrary correlations between errors in the same layer. Then in the notation of \cref{app-sec:preliminaries/tensored/general} above, one takes $p = L$, and the subsets correspond to the measurements in each layer, $\smash{\{m^{[1]}, m^{[2]}, \ldots, m^{[L]}\}}$. The factorization of $\vb{q}$, $Q$, $Q^{-1}$ and eigenvalues $\vb*{\lambda}$ then follow. In this setting we use square brackets for superscripts, for example $\vb{q}^{[l]}$ instead of $\vb{q}^{(l)}$, to emphasize the physical meaning that the indices enumerate the layers of the feedforward circuit, as is consistent with our writing in the main text.

\subsubsection{Fully tensored readout channels}
\label{app-sec:preliminaries/tensored/fully-tensored}

The setup in \cref{app-sec:preliminaries/tensored/general} likewise applies in a setting where every measurement is assumed to be independent---then $p = m$ and the subsets $\smash{\{m^{(1)}, m^{(2)}, \ldots, m^{(p)}\}}$ are single measurements. To be explicit, in the factorization of $M$ in \cref{app-eq:preliminaries/tensored/general/M-factorization}, we have in this setting
\begin{equation}\begin{split}
    M^{(\ell)} = \mqty[
        1 - M^{(\ell)}_{10} & M^{(\ell)}_{01} \\
        M^{(\ell)}_{10} & 1 - M^{(\ell)}_{01}
    ],
\end{split}\end{equation}
which are $2 \times 2$ confusion matrices each characterizing the readout error affecting the $\ell$-th measurement, for $\ell \in [m]$. The diagonal elements record the probabilities of $0$ and $1$ outcomes being correctly reported, whereas off-diagonals record probabilities of bit-flip errors. With symmetrization, one has
\begin{equation}\begin{split}
    \vb{q}^{(\ell)} = \mqty[ 1 - r_\ell \\ r_\ell ],
\end{split}\end{equation}
where $r_\ell$ is the probability of a readout error occurring on the $\ell$-th measurement. Concretely, the symmetrization means that $\smash{r_\ell = (M^{(\ell)}_{01} + M^{(\ell)}_{10}) / 2}$, which is expressed more generally in \cref{app-eq:preliminaries/confusion-matrix-properties/q-Q-from-M}. Likewise, the symmetrized confusion matrices
\begin{equation}\begin{split}
    Q^{(\ell)} = \mqty[
        q^{(\ell)}_{0} & q^{(\ell)}_{1} \\
        q^{(\ell)}_{1} & q^{(\ell)}_{0}
    ]
    = \mqty[
        1 - r_\ell & r_\ell \\ 
        r_\ell & 1 - r_\ell
    ],
\end{split}\end{equation}
whose diagonal and off-diagonals elements record probabilities of the measurement outcome being correctly and incorrectly reported respectively. The eigenvalues of $\smash{Q^{(\ell)}}$ are easily found to be
\begin{equation}\begin{split}
    \vb*{\lambda}^{(\ell)} = \mqty[1 \\ 1 - 2 r_\ell],
    \label{app-eq:preliminaries/tensored/fully-tensored/lambda}
\end{split}\end{equation}
which coincides with our general solution for $\lambda_k$ in \cref{app-eq:preliminaries/Q-diagonalization/lambda-solution-1,app-eq:preliminaries/Q-diagonalization/lambda-solution-2} or \cref{eq:readout-errors/lambda-solution} of the main text, setting $m = 1$ for a single independent qubit. The inverse of $\smash{Q^{(\ell)}}$ is also easily found,
\begin{equation}\begin{split}
    Q^{(\ell)^{-1}} = \frac{1}{1 - 2 r_\ell} \mqty[
        1 - r_\ell & -r_\ell \\
        -r_\ell & 1 - r_\ell
    ],
    \label{app-eq:preliminaries/tensored/fully-tensored/Q-inverse}
\end{split}\end{equation}
which is consistent with the general solution for entries of $Q^{-1}$ in \cref{app-eq:preliminaries/Q-diagonalization/Q-inverse-general} upon substitution of the present setting.

\clearpage 
\pagebreak

\section{Further Technical Details on Probabilistic Readout Error Mitigation}
\label{app-sec:theory}

\subsection{Theory for a single feedforward layer}
\label{app-sec:theory/single}

\subsubsection{General solution of $\vb*{\alpha}$ for arbitrarily correlated readout channels}
\label{app-sec:theory/single/solution-general}

Here we provide an elaboration of the derivation in \cref{sec:theory-single/general} of the main text, leading to the general solution for $\vb*{\alpha}$ in \cref{eq:theory-single/general/alpha-solution}. First, in the absence of readout errors, the quantum state after the mid-circuit measurements and feedforward is straightforwardly
\begin{equation}\begin{split}
    \overline{\sigma} = \sum_{s \in \mathcal{S}} V_s \Pi_s \rho \Pi_s V_s^\dag,
    \label{app-eq:theory/single/solution-general/state-ideal-1}
\end{split}\end{equation}
where $\Pi_s$ is the measurement projector associated with the outcome $s \in \mathcal{S}$. Equivalently, in the picture of quantum trajectories,
\begin{equation}\begin{split}
    \overline{\sigma} = \sum_{s \in \mathcal{S}} p_s V_s \rho_s V_s^\dag,
    \qquad
    \rho_s = \frac{\Pi_s \rho \Pi_s}{p_s},
    \qquad 
    p_s = \tr \left(\Pi_s \rho\right),
    \label{app-eq:theory/single/solution-general/state-ideal-2}
\end{split}\end{equation}
which was presented in \cref{eq:theory-single/general/state-ideal} of the main text. Technically, here $\rho_s$ is well-defined so long as $p_s \neq 0$; the $p_s = 0$ trajectories are immaterial to $\smash{\overline{\sigma}}$ and can be disregarded. In the presence of readout errors characterized by $\vb{q}$, and imposing a bitmask $f \in \mathcal{S}$ in feedforward, the quantum state after the mid-circuit measurements and feedforward is
\begin{equation}\begin{split}
    \sigma^{(f)}
    = \sum_{s \in \mathcal{S}} \sum_{s' \in \mathcal{S}} 
        q_{s \oplus s'}
        V_{s' \oplus f} \Pi_s \rho \Pi_s V_{s' \oplus f}^\dag
    = \sum_{s \in \mathcal{S}} \sum_{s' \in \mathcal{S}} 
        q_{s \oplus s'} p_s 
        V_{s' \oplus f} \rho_s V_{s' \oplus f}^\dag,
    \label{app-eq:theory/single/solution-general/state-noisy}
\end{split}\end{equation}
where the second equality was presented in \cref{eq:theory-single/general/state-noisy} of the main text. The definition of the tensor $T$ follows, 
\begin{equation}\begin{split}
    T_{b s s'} 
    = \tr \left( O_b V_{s'} \Pi_s \rho \Pi_s V_{s'}^\dag \right)
    = p_s \tr \left( O_b V_{s'} \rho_s V_{s'}^\dag \right),
    \label{app-eq:theory/single/solution-general/T-definition}
\end{split}\end{equation}
as presented in \cref{eq:theory-single/general/T-definition} of the main text. Then from \cref{eq:theory-single/general/expval-O-ideal}, one has
\begin{equation}\begin{split}
    \expval{\overline{O}_b} 
    = \tr T_b
    = \sum_{s \in \mathcal{S}} T_{bss}
    = \sum_{s \in \mathcal{S}} \sum_{s' \in \mathcal{S}} \delta_{s s'} T_{bss}.
    \label{app-eq:theory/single/solution-general/expval-O-ideal}
\end{split}\end{equation}

At the same time, from \cref{eq:theory-single/general/expval-O-f},
\begin{equation}\begin{split}
    \expval{O_b^{(f)}}
    = \sum_{s \in \mathcal{S}} \sum_{s' \in \mathcal{S}} 
        q_{s \oplus s'} T_{b s (s' \oplus f)}
    = \sum_{s \in \mathcal{S}} \sum_{s' \in \mathcal{S}} 
        q_{s \oplus s' \oplus f} T_{b s s'}.
    \label{app-eq:theory/single/solution-general/expval-O-f}
\end{split}\end{equation}

Thus, by the definition of $\widehat{\vb{O}}$ in \cref{eq:theory-single/protocol/estimator-definition},
\begin{equation}\begin{split}
    \expval*{\widehat{O}_b}
    = \sum_{f \in \mathcal{S}} \alpha_f \expval{O_b^{(f)}}
    = \sum_{f \in \mathcal{S}} 
        \sum_{s \in \mathcal{S}} \sum_{s' \in \mathcal{S}} 
        \alpha_f q_{s \oplus s' \oplus f} T_{b s s'}.
    \label{app-eq:theory/single/solution-general/expval-O-mitigated}
\end{split}\end{equation}

We desire $\smash{\expval*{\widehat{\vb{O}}} = \expval*{\overline{\vb{O}}}}$ independent of the tensor $T$, which is unknown to us. This guarantees that the mitigation protocol works for arbitrary quantum circuits of the structure considered, with no additional problem-specific information required. The equality necessitates that the coefficients of every $T_{bss'}$ match on both sides, in particular, between \cref{app-eq:theory/single/solution-general/expval-O-ideal} and \cref{app-eq:theory/single/solution-general/expval-O-mitigated} for every $b \in [B]$, yielding
\begin{equation}\begin{split}
    \sum_{f \in \mathcal{S}}
    \alpha_f q_{s \oplus s' \oplus f} = \delta_{s s'}
    \quad \forall \quad s, s' \in \mathcal{S}.
    \label{app-eq:theory/single/solution-general/linear-system-1}
\end{split}\end{equation}

Noting that $\delta_{s s'} = \delta_{(s \oplus s') \vb{0}}$, the above reduces to
\begin{equation}\begin{split}
    \sum_{f \in \mathcal{S}}
    \alpha_f q_{s \oplus f} = \delta_{s \vb{0}}
    \quad \forall \quad s \in \mathcal{S}.
    \label{app-eq:theory/single/solution-general/linear-system-2}
\end{split}\end{equation}

This linear system can be written as
\begin{equation}\begin{split}
    Q \vb*{\alpha} = \mqty[1 \\ 0 \\ \vdots \\0] 
    \Longleftrightarrow 
    \vb*{\alpha} = Q^{-1} \mqty[1 \\ 0 \\ \vdots \\0].
    \label{app-eq:theory/single/solution-general/linear-system-3}
\end{split}\end{equation}

That is, $\vb*{\alpha}$ is the first column of $Q^{-1}$. Then by \cref{app-eq:preliminaries/Q-diagonalization/Q-inverse-first-column}, we have
\begin{equation}\begin{split}
    \alpha_f
    = \left[ Q^{-1} \right]_{f \vb{0}}
    = \frac{1}{2^m} \sum_{k \in \mathcal{S}} 
        (-1)^{f \cdot k} \left(\frac{1}{\lambda_k}\right),
    \label{app-eq:theory/single/solution-general/alpha-solution-1}
\end{split}\end{equation}
which is the solution reported in \cref{eq:theory-single/general/alpha-solution} of the main text. This exactly coincides with and can be equivalently written as $\vb*{\alpha} = \mathcal{W}(\vb{1} / \vb*{\lambda}) / 2^m$ for $\mathcal{W}$ the unnormalized Walsh-Hadamard transform~\cite{ahmed1975walsh, beer1981walsh}, as also reported in the main text. Noting that $\lambda_{\vb{0}} = 1$, as discussed in \cref{app-sec:preliminaries/Q-diagonalization}, we obtain also the alternate expression
\begin{equation}\begin{split}
    \alpha_f
    = \frac{1}{2^m} \left[ \frac{1}{\lambda_{\vb{0}}} 
        + \sum_{\substack{k \in \mathcal{S} \\ k \neq \vb{0}}} 
        (-1)^{f \cdot k} \left(\frac{1}{\lambda_k}\right) 
        \right]
    = \frac{1}{2^m} \left[ 1
        + \sum_{\substack{k \in \mathcal{S} \\ k \neq \vb{0}}} 
        (-1)^{f \cdot k} \left(\frac{1}{\lambda_k}\right) 
        \right].
    \label{app-eq:theory/single/solution-general/alpha-solution-2}
\end{split}\end{equation}

For illustration, in \cref{app-tab:alpha-solution-examples} we provide explicitly written solutions for $\vb*{\alpha}$ at small $m$. For brevity we have used the form of $\vb*{\alpha}$ in \cref{app-eq:theory/single/solution-general/alpha-solution-2}, and the form of $\vb*{\lambda}$ in \cref{app-eq:preliminaries/Q-diagonalization/lambda-solution-2}.

\begin{table}[!h]
    \centering
    \begin{tabular}{p{1.5cm} p{13cm}}
         \toprule 
         $m$ & Solution \\
         \midrule 
         $1$ & 
             $\begin{aligned}
                \alpha_0 &= \frac{1}{2} \left( 1 + \frac{1}{1 - 2 q_1} \right) &\\
                \alpha_1 &= \frac{1}{2} \left( 1 - \frac{1}{1 - 2 q_1} \right)
            \end{aligned}$ \\
        \midrule
        $2$ & 
            $\begin{aligned}
                \alpha_{00} &= \frac{1}{4} \left( 1
                    + \frac{1}{1 - 2 q_{01} - 2 q_{10}} 
                    + \frac{1}{1 - 2 q_{01} - 2 q_{11}} 
                    + \frac{1}{1 - 2 q_{10} - 2 q_{11}} 
                    \right) &\\
                \alpha_{01} &= \frac{1}{4} \left( 1
                    - \frac{1}{1 - 2 q_{01} - 2 q_{10}} 
                    - \frac{1}{1 - 2 q_{01} - 2 q_{11}} 
                    + \frac{1}{1 - 2 q_{10} - 2 q_{11}} 
                    \right) &\\
                \alpha_{10} &= \frac{1}{4} \left( 1
                    - \frac{1}{1 - 2 q_{01} - 2 q_{10}} 
                    + \frac{1}{1 - 2 q_{01} - 2 q_{11}} 
                    - \frac{1}{1 - 2 q_{10} - 2 q_{11}} 
                    \right) &\\
                \alpha_{11} &= \frac{1}{4} \left( 1
                    + \frac{1}{1 - 2 q_{01} - 2 q_{10}} 
                    - \frac{1}{1 - 2 q_{01} - 2 q_{11}} 
                    - \frac{1}{1 - 2 q_{10} - 2 q_{11}} 
                    \right) &\\
            \end{aligned}$ \\
        \midrule
        $3$ & 
            $\begin{aligned}
                \alpha_{000} = \frac{1}{8} \bigg( 1
                    & + \frac{1}{1 - 2 q_{010} - 2 q_{011} - 2 q_{100} - 2 q_{101}} 
                    + \frac{1}{1 - 2 q_{001} - 2 q_{011} - 2 q_{100} - 2 q_{110}} \\
                    &+ \frac{1}{1 - 2 q_{001} - 2 q_{010} - 2 q_{101} - 2 q_{110}}
                    + \frac{1}{1 - 2 q_{001} - 2 q_{010} - 2 q_{100} - 2 q_{111}} \\
                    &+ \frac{1}{1 - 2 q_{001} - 2 q_{011} - 2 q_{101} - 2 q_{111}}
                    + \frac{1}{1 - 2 q_{010} - 2 q_{011} - 2 q_{110} - 2 q_{111}} \\
                    &+ \frac{1}{1 - 2 q_{100} - 2 q_{101} - 2 q_{110} - 2 q_{111}}
                    \bigg) &\\
                \alpha_{001} = \frac{1}{8} \bigg( 1
                    & - \ldots \bigg) &\\
            \end{aligned}$ \\
        \bottomrule
    \end{tabular}
    \caption{Explicitly written general solutions for $\vb*{\alpha}$ at small $m$ for illustrative purposes. In consideration of space and readability, $\alpha_k$ entries for $m = 3$ and $k \neq \vb{0}$ are left out; they follow the same form as $\alpha_{\vb{0}}$ but with sign flips on individual terms.}
    \label{app-tab:alpha-solution-examples}
\end{table}

\textit{Resource complexity}. In this general setting, a naïve procedure of obtaining $\vb*{\alpha}$ is first to compute $\vb*{\lambda}$ from $\vb{q}$ using \cref{app-eq:preliminaries/Q-diagonalization/lambda-solution-1} or \cref{app-eq:preliminaries/Q-diagonalization/lambda-solution-2}, which comprises $2^m$ entries each involving a sum over extensively many entries of $\vb{q}$. Then $\vb*{\alpha}$ is computed from $\vb*{\lambda}$ using \cref{app-eq:theory/single/solution-general/alpha-solution-1} or \cref{app-eq:theory/single/solution-general/alpha-solution-2}. Each of these steps require $\smash{\order{2^{2m}}}$ time and $\smash{\order{2^m}}$ space in classical resources, for the same overall complexity. But noting that $\vb{q}$, $\vb*{\lambda}$ and $\vb*{\alpha}$ are related by Walsh-Hadamard transforms, as noted in the main text and in the discussion above, one can in fact utilize the fast Walsh–Hadamard transform (FWHT) algorithm~\cite{fino1976unified, beer1981walsh}, which runs in $\order{N \log N}$ time and $\order{N}$ space for an $N$-length input vector, for a more efficient solution computation procedure. This reduces the time complexity of solution computation to $\order{m \cdot 2^m}$ and maintains space complexity at $\order{2^m}$. Explicitly, the procedure goes as follows:
\begin{enumerate}
    \item Perform unnormalized FWHT to compute $\vb*{\lambda} = \mathcal{W}(\vb{q})$, requiring $\order{m \cdot 2^m}$ time and $\order{2^m}$ space.
    \item Compute in-place element-wise division $\vb{1} / \vb*{\lambda}$, requiring $\order{2^m}$ time.
    \item Perform unnormalized FWHT to compute $\vb*{\alpha} = \mathcal{W}(\vb{1} / \vb*{\lambda}) / 2^m$, requiring $\order{m \cdot 2^m}$ time and $\order{2^m}$ space.
\end{enumerate}

Thereafter, computing $\xi$ and $\abs{\vb*{\alpha}} / \xi$ take sub-dominant $\order{2^m}$ time, the latter doable in-place. Lastly, standard algorithms to sample from an explicit probability distribution, such as alias and table methods~\cite{walker1974new, walker1977efficient, vose1991linear, marsaglia2004fast}, allow the drawing of bitmasks $f \in \mathcal{S}$ from the distribution $\abs{\vb*{\alpha}} / \xi$ in $\order{1}$ time per shot, with an initial pre-processing time linear in the number of entries in the distribution, a sub-dominant $\order{2^m}$ in the context here. The overall resource requirements of $\smash{\order{m \cdot 2^m}}$ time and $\smash{\order{2^m}}$ space for solution computation and $\order{1}$ time per shot are reported in \cref{tab:summary} of the main text.

\subsubsection{Simplified solution of $\vb*{\alpha}$ for tensored readout channels}
\label{app-sec:theory/single/solution-tensored}

As outlined in \cref{sec:theory-single/simplified}, making an assumption that readout errors on each of the $m$ measurements are independent, equivalently that the readout error channel acting on the $m$ measurements factorizes into a tensor product of channels acting on individual measurements, allows a simplification of the $\vb*{\alpha}$ solution. In particular, under such a setting, $\vb{q}$, $Q$ and $Q^{-1}$ each factorizes into a tensor product of single-measurement counterparts, as worked through in \cref{app-sec:preliminaries/tensored/fully-tensored}. Then from \cref{app-eq:preliminaries/tensored/general/Q-inverse-factorization,app-eq:preliminaries/tensored/fully-tensored/Q-inverse}, we have
\begin{equation}\begin{split}
    \vb*{\alpha} = \bigotimes_{\ell = 1}^m \vb*{\alpha}^{(\ell)},
    \qquad 
    \vb*{\alpha}^{(\ell)} = \frac{1}{1 - 2 r_\ell} 
        \mqty[ 1 - r_\ell \\ -r_\ell ],
    \label{app-sec:theory/single/solution-tensored/alpha-tensor}
\end{split}\end{equation}
where $\smash{\vb*{\alpha}^{(\ell)}}$ are coefficients for the $\ell$-measurement. This is presented as \cref{eq:theory-single/simplified/tensored-alpha-solution} in the main text. Equivalently,
\begin{equation}\begin{split}
    \alpha_f = \prod_{\ell = 1}^m \alpha^{(\ell)}_{f_\ell},
    \qquad 
    \alpha^{(\ell)}_{b} = \frac{r_\ell^b (r_\ell - 1)^{1 - b}}{2 r_\ell - 1},
    \label{app-sec:theory/single/solution-tensored/alpha-product}
\end{split}\end{equation}
where $b \in \{0, 1\}$. The overhead factor associated with mitigating the $\ell$-measurement is easily computed,
\begin{equation}\begin{split}
    \xi^{(\ell)} 
    = \norm{ \vb*{\alpha}^{(\ell)} }_1
    = \frac{1}{1 - 2 r_\ell},
\end{split}\end{equation}
and \cref{app-sec:theory/single/solution-tensored/alpha-tensor} implies that
\begin{equation}\begin{split}
    \xi 
    = \norm{ \vb*{\alpha} }_1
    = \prod_{\ell = 1}^m \norm{ \vb*{\alpha}^{(\ell)} }_1
    = \prod_{\ell = 1}^m \xi^{(\ell)}
    = \prod_{\ell = 1}^m \frac{1}{1 - 2 r_\ell},
\end{split}\end{equation}
which is presented as \cref{eq:theory-single/simplified/tensored-overhead-factor} of the main text.

\textit{Resource complexity}. The product structure of $\vb*{\alpha}$ and $\xi$ above imply that $\abs{\vb*{\alpha}} / \xi$ is a product over the individual $\abs{\vb*{\alpha}^{(\ell)}} / \xi^{(\ell)}$ probability distributions,
\begin{equation}\begin{split}
    \frac{\abs{\vb*{\alpha}}}{\xi}
    = \bigotimes_{\ell = 1}^m \frac{\abs{\vb*{\alpha}^{(\ell)}}}{\xi^{(\ell)}}
    \Longleftrightarrow 
    \frac{\abs{\alpha_f}}{\xi}
    = \prod_{\ell = 1}^m \frac{\abs{\alpha^{(\ell)}_{f_\ell}}}{\xi^{(\ell)}}.
\end{split}\end{equation}

The probability distributions $\smash{\abs{\vb*{\alpha}^{(\ell)}} / \xi^{(\ell)}}$ and overall rescaling factor $\xi$ to be applied at the end of the mitigation protocol can be computed in $\order{m}$ time and space, as there are $2m$ entries across all $\smash{\vb*{\alpha}^{(\ell)}}$ and $m$ overhead factors $\smash{\xi^{(\ell)}}$ in total, as explicitly written above. Then, for each experiment shot, a round of sampling can be used to draw $f_\ell$ from the distribution $\smash{\abs{\vb*{\alpha}^{(\ell)}} / \xi^{(\ell)}}$ for each $\ell \in [m]$, thus producing a bitmask $f \in \mathcal{S}$ from the product distribution $\smash{\abs{\vb*{\alpha}} / \xi}$ in $\order{m}$ time. These complexities are reflected in \cref{tab:summary} of the main text.

In the simpler setting of uniform readout channels, $r_\ell = r$ for all $\ell \in [m]$ and all the $\smash{\abs{\vb*{\alpha}^{(\ell)}} / \xi^{(\ell)}}$ distributions are identical. Thus, one needs to only compute a single $\smash{\abs{\vb*{\alpha}^{(\ell)}} / \xi^{(\ell)}}$, which requires $\order{1}$ space and time. Then, for each experiment shot, $m$ rounds of sampling can be used to draw $f_\ell$ from the $\smash{\abs{\vb*{\alpha}^{(\ell)}} / \xi^{(\ell)}}$ distribution repeatedly, thus building up the bitmask $f \in \mathcal{S}$ for the shot in $\order{m}$ time. This is also reflected in \cref{tab:summary} of the main text.

\subsection{Multiple feedforward layers}
\label{app-sec:theory/multiple}

\subsubsection{General solution of $\vb*{\alpha}$ for arbitrarily correlated readout channels}
\label{app-sec:theory/multiple/solution-general}

Similar to \cref{app-sec:theory/single/solution-general}, here we provide an elaboration of the discussion in \cref{sec:theory-multiple/general} of the main text. To start, we write the quantum state after the $L$ layers of mid-circuit measurements and feedforward in the absence of readout errors,
\begin{equation}\begin{split}
    \overline{\sigma} 
    = \sum_{s \in \mathcal{S}} 
        W_{ss} \rho W_{ss}^\dag,
    \qquad 
    W_{ss'} = \prod_{l = L}^1 V^{[l]}_{s'^{[l]}} \Pi^{[l]}_{s^{[l]}},
    \label{app-eq:theory/multiple/solution-general/state-ideal}
\end{split}\end{equation}
where $W$ is the sequence of mid-circuit measurements and feedforward unitaries applied, as also defined in \cref{eq:theory-multiple/general/T-definition} of the main text. Here we remind that $\smash{\mathcal{S} = \{0, 1\}^m}$ is the set of possible measurement outcomes over the $m$ total mid-circuit measurements in the quantum circuit, and $\smash{s = s^{[1]} s^{[2]} \ldots s^{[L]} \in \mathcal{S}}$ is the partitioning of bitstring $s$ over the $L$ layers, $\smash{s^{[l]} \in \mathcal{S}^{[l]} = \{0, 1\}^{m^{[l]}}}$. As the computational-basis ($Z$ basis) measurement projectors $\smash{\Pi^{[l]}_{s^{[l]}}}$ are Hermitian,
\begin{equation}\begin{split}
    W^\dag_{ss'} = \prod_{l = 1}^L \Pi^{[l]}_{s^{[l]}} V^{[l]\dag}_{s'^{[l]}}.
\end{split}\end{equation}

Taking $L = 1$ and dropping the layer superscripts that are unneeded, \cref{app-eq:theory/multiple/solution-general/state-ideal} recovers \cref{app-eq:theory/single/solution-general/state-ideal-1} for a single feedforward layer. It is straightforward also to rewrite \cref{app-eq:theory/multiple/solution-general/state-ideal} in a picture of quantum trajectories,
\begin{equation}\begin{split}
    \overline{\sigma} 
    = \sum_{s \in \mathcal{S}} 
        p_s \rho_s,
    \qquad 
    \rho_s
    = \frac{W_{ss} \rho W_{ss}^\dag}{p_s},
    \qquad 
    p_s
    = \tr[ W_{ss} \rho W_{ss}^\dag ],
\end{split}\end{equation}
where $\rho_s$ is the normalized quantum state at the end of the quantum circuit for quantum trajectories defined by the measurement outcome $s \in \mathcal{S}$, and $p_s$ is the probability of those trajectories occurring. Taking $L = 1$ and dropping layer superscripts, this recovers \cref{eq:theory-single/general/state-ideal} of the main text for a single feedforward layer. Reproducing \cref{eq:theory-multiple/general/T-definition}, we define
\begin{equation}\begin{split}
    T_{b s s'}
    &= \tr \left[ O_b 
        W_{s s'} \rho W_{s s'}^\dag
    \right],
\end{split}\end{equation}
which is the expectation value of $O_b$ measured on quantum trajectories that had true measurement outcomes $s$ but feedforward unitaries applied according to $s'$, weighted by the probability of those trajectories occurring. This is directly analogous to \cref{eq:theory-single/general/T-definition} in the single-layer context. Then, using \cref{app-eq:theory/multiple/solution-general/state-ideal} and the definition of $\smash{T_{b s s'}}$, we have
\begin{equation}\begin{split}
    \expval{\overline{O}_b} 
    = \tr \left(O_b \overline{\sigma}\right)
    = \sum_{s \in \mathcal{S}} T_{bss}
    = \tr T_b.
    \label{app-eq:theory/multiple/solution-general/expval-O-ideal}
\end{split}\end{equation}

Next we consider the presence of readout errors. We write the quantum state after the $L$ layers of mid-circuit measurements and feedforward in a case where readout errors have corrupted measurement outcomes $s$ into $s'$, and a bitmask $f \in \mathcal{S}$ has been imposed on the feedforward,
\begin{equation}\begin{split}
    \sigma^{(f)}
    = \sum_{s \in \mathcal{S}} \sum_{s' \in \mathcal{S}} 
        q_{s \oplus s'}
        W_{s (s' \oplus f)}
        \rho 
        W_{s (s' \oplus f)}^\dag.
\end{split}\end{equation}

Then
\begin{equation}\begin{split}
    \expval{O_b^{(f)}}
    = \tr \left( O_b \sigma^{(f)} \right)
    &= \sum_{s \in \mathcal{S}} \sum_{s' \in \mathcal{S}} 
        q_{s \oplus s'} T_{b s (s' \oplus f)}.
    \label{app-eq:theory/multiple/solution-general/expval-O-f}
\end{split}\end{equation}

Note that \cref{app-eq:theory/multiple/solution-general/expval-O-ideal} and \cref{app-eq:theory/multiple/solution-general/expval-O-f} are identical to \cref{app-eq:theory/single/solution-general/expval-O-ideal} and \cref{app-eq:theory/single/solution-general/expval-O-f} in \cref{app-sec:theory/single/solution-general}. Moreover, as discussed in \cref{sec:theory-multiple/protocol}, our definition of the mitigated estimator $\smash{\widehat{\vb{O}}}$ is identical for both the single-layer and multiple-layer contexts. Thus, the remainder of the derivation in \cref{app-sec:theory/single/solution-general} follows through, leading to the solution for $\vb*{\alpha}$ in \cref{app-eq:theory/single/solution-general/alpha-solution-1,app-eq:theory/single/solution-general/alpha-solution-2}, the former presented in \cref{eq:theory-single/general/alpha-solution} of the main text and referenced in \cref{sec:theory-multiple/general} accordingly. The only differences to keep in mind, which are only interpretational and do not change the form of the equations and solutions, are the generalized definitions of $\mathcal{S}$ being the set of possible measurement outcomes across all layers, and $T$ likewise being expectation values on trajectories across all layers, as written above. The classical resource requirements analyzed in \cref{app-sec:theory/single/solution-general} likewise straightforwardly apply.

\subsubsection{Simplified solution of $\vb*{\alpha}$ for layer-wise tensored readout channels}
\label{app-sec:theory/multiple/solution-layer-tensored}

As discussed in \cref{sec:theory-multiple/simplified}, one may examine a setting where readout errors on measurements in different layers are assumed to be independent. This setting falls exactly into that considered in \cref{app-sec:preliminaries/tensored/layer-tensored}. Picking up the discussion there, from \cref{app-eq:preliminaries/tensored/general/q-Q-factorization,app-eq:preliminaries/tensored/general/lambda-factorization,app-eq:preliminaries/tensored/general/Q-inverse-factorization}, we have
\begin{equation}\begin{split}
    \vb{q} = \bigotimes_{l = 1}^L \vb{q}^{[l]},
    \qquad
    \vb*{\lambda} = \bigotimes_{l = 1}^L \vb*{\lambda}^{[l]}
    \Longleftrightarrow 
    \lambda_k = \prod_{l = 1}^L \lambda^{[l]}_{k^{[l]}},
    \qquad 
    \vb*{\alpha} = \bigotimes_{l = 1}^L \vb*{\alpha}^{[l]}
    \Longleftrightarrow 
    \alpha_f = \prod_{l = 1}^L \alpha^{[l]}_{f^{[l]}},
\end{split}\end{equation}
where $\smash{k = k^{[1]} k^{[2]} \ldots k^{[L]}} \in \mathcal{S}$ and $\smash{f = f^{[1]} f^{[2]} \ldots f^{[L]}} \in \mathcal{S}$ are partitionings of the bitstrings over the layers, with $\smash{k^{[l]} \in \mathcal{S}^{[l]}}$ and $\smash{f^{[l]} \in \mathcal{S}^{[l]}}$, and $\smash{\vb{q}^{[l]}}$, $\smash{\vb*{\lambda}^{[l]}}$ and $\smash{\vb*{\alpha}^{[l]}}$ are respectively the confusion probability vector, eigenvalues and mitigator coefficients for layer $l \in [L]$. Part of the above is presented in \cref{eq:theory-multiple/simplified/alpha-factorization} of the main text. The tensor product structure of $\vb*{\alpha}$ also implies that
\begin{equation}\begin{split}
    \xi 
    = \norm{ \vb*{\alpha} }_1
    = \prod_{l = 1}^L \norm{ \vb*{\alpha}^{[l]} }_1
    = \prod_{l = 1}^L \xi^{[l]},
\end{split}\end{equation}
which is presented as \cref{eq:theory-multiple/simplified/xi-factorization} of the main text. To be explicit, $\vb*{\lambda}^{[l]}$ and $\vb*{\alpha}^{[l]}$ are given by the same solutions as in \cref{app-eq:preliminaries/Q-diagonalization/lambda-solution-1,app-eq:preliminaries/Q-diagonalization/lambda-solution-2,app-eq:theory/single/solution-general/alpha-solution-1,app-eq:theory/single/solution-general/alpha-solution-2} or equivalently \cref{eq:readout-errors/lambda-solution,eq:theory-single/general/alpha-solution} of the main text, but with quantities replaced by their layer-wise counterparts---that is, $\smash{(\mathcal{S}, m, \vb{q}, k, f)}$ replaced by $\smash{(\mathcal{S}^{[l]}, m^{[l]}, \vb{q}^{[l]}, k^{[l]}, f^{[l]})}$.

\textit{Resource complexity}. The product structure of $\vb*{\alpha}$ and $\xi$ above imply that $\abs{\vb*{\alpha}} / \xi$ is a product over the layer-wise $\abs{\vb*{\alpha}^{[l]}} / \xi^{[l]}$ probability distributions,
\begin{equation}\begin{split}
    \frac{\abs{\vb*{\alpha}}}{\xi}
    = \bigotimes_{l = 1}^L \frac{\abs{\vb*{\alpha}^{[l]}}}{\xi^{[l]}}
    \Longleftrightarrow 
    \frac{\abs{\alpha_f}}{\xi}
    = \prod_{\ell = 1}^m \frac{\abs{\alpha^{[l]}_{f^{[l]}}}}{\xi^{[l]}}.
\end{split}\end{equation}

Thus, using the solution computation method in \cref{app-sec:theory/single/solution-general}, one can compute $\abs{\vb*{\alpha}^{[l]}} / \xi^{[l]}$ for every $l \in [L]$ in total $\smash{\order*{\sum_{l = 1}^L m^{[l]} \cdot 2^{m^{[l]}}}}$ time and $\smash{\order*{\sum_{l = 1}^L 2^{m^{[l]}}}}$ space. After a sub-dominant pre-processing time of $\smash{\order*{\sum_{l = 1}^L 2^{m^{[l]}}}}$, for each experiment shot, a round of constant-time sampling can be run to draw $\smash{f^{[l]}}$ from the $\smash{\abs{\vb*{\alpha}^{[l]}} / \xi^{[l]}}$ distribution for each $l \in [L]$, thus building up the bitmask $\smash{f = f^{[1]} f^{[2]} \ldots f^{[L]}} \in \mathcal{S}$ for the shot in $\order{L}$ time. Thus the overall complexities are $\smash{\order*{\sum_{l = 1}^L m^{[l]} \cdot 2^{m^{[l]}}}}$ time and $\smash{\order*{\sum_{l = 1}^L 2^{m^{[l]}}}}$ space for solution computation, and $\order{L}$ sampling time per shot. 

Noting that $\smash{\order*{\sum_{l = 1}^L m^{[l]} \cdot 2^{m^{[l]}}} \leq \order*{\sum_{l = 1}^L m^{[l]} \cdot 2^{\overline{m}}}} \leq \order*{m \cdot 2^{\overline{m}}}$ and $\smash{\order*{\sum_{l = 1}^L 2^{m^{[l]}}}} \leq \smash{\order*{L \cdot 2^{\overline{m}}}}$, where $\smash{\overline{m} = \max_{l \in [L]} m^{[l]}}$ is the maximum number of measurements in a layer, we can more succinctly bound the time and space complexities as $\smash{\order*{m \cdot 2^{\overline{m}}}}$ and $\smash{\order*{L \cdot 2^{\overline{m}}}}$ respectively, which are reported in \cref{tab:summary} of the main text.

\subsubsection{Simplified solution of $\vb*{\alpha}$ for fully tensored readout channels}
\label{app-sec:theory/multiple/solution-fully-tensored}

This setting is exactly equivalent to that considered in \cref{app-sec:theory/single/solution-tensored}; thus, we direct readers to the discussion provided therein.

\clearpage
\pagebreak

\subsection{Sampling overhead of mitigated estimator $\widehat{\vb{O}}$}
\label{app-sec:theory/sampling-overhead}

We reproduce the definition of our estimator,
\begin{equation}\begin{split}
    \widehat{\vb{O}} 
    \coloneqq \sum_{f \in \mathcal{S}} \alpha_f \vb{O}^{(f)}
    = \xi \underbrace{ \sum_{f \in \mathcal{S}} 
        \frac{\abs{\alpha_f}}{\xi} \sgn(\alpha_f) \vb{O}^{(f)} }_{\widetilde{\vb{O}}},
\end{split}\end{equation}
as in \cref{eq:theory-single/protocol/estimator-definition} of the main text. We have labelled the estimator before the final rescaling by $\xi = \norm{\vb*{\alpha}}_1$ as $\smash{\widetilde{\vb{O}}}$. Here we generically regard multiple feedforward layers, of which a single feedforward layer is a special case, and thus the results obtained apply to both \cref{sec:theory-single,sec:theory-multiple} of the main text as referenced therein.

Recall the general \texttt{PROM} protocol, as described in \cref{alg:prom-single-layer,alg:prom-multiple-layer}, which is to draw a bitstring $f$ from the $\abs{\vb*{\alpha}} / \xi$ probability distribution in each shot of the experiment circuit and measure $\smash{\sgn(\alpha_f) \vb{O}^{(f)}}$ on that experiment shot. The observables measured in each shot are $\vb{O} = \mqty[O_1 & O_2 & \ldots & O_B]$. Thus, the shot yields a measurement result that comprises an eigenvalue of each $O_b$ measured up to sign, which is bounded between $\smash{[-\norm{O_b}_2, +\norm{O_b}_2]}$, where the spectral norm $\norm{O_b}_2$ reflects the largest singular value of the observable $O_b$. Then their convex linear combination $\smash{\widetilde{O}_b}$ are all bounded within $\smash{[-\norm{O_b}_2, +\norm{O_b}_2]}$. In a hypothetical ideal scenario without readout errors, not employing \texttt{PROM}, in each shot the measurement of $\vb{O}$ likewise gives a result with entries bounded within $\smash{[-\norm{O_b}_2, +\norm{O_b}_2]}$.

We follow a standard line of analysis, similar to, for example, Ref.~\onlinecite{yuan2021universal} albeit in a different context. We compare the mitigated estimator $\smash{\widehat{\vb{O}}}$ and the ideal estimator $\smash{\overline{\vb{O}}}$, the latter accessible only in a hypothetical noiseless setting, averaged over $n$ independent experiment shots,
\begin{equation}\begin{split}
    \widehat{\vb{O}}^{\{n\}} = \frac{1}{n} \sum_{j = 1}^n \widehat{\vb{O}},
    \qquad 
    \overline{\vb{O}}^{\{n\}} = \frac{1}{n} \sum_{j = 1}^n \overline{\vb{O}},
\end{split}\end{equation}
and we likewise define
\begin{equation}\begin{split}
    \widetilde{\vb{O}}^{\{n\}} 
    = \frac{1}{n} \sum_{j = 1}^n \widetilde{\vb{O}}.
\end{split}\end{equation}

We demand precision $\epsilon$ and confidence $1 - \delta$ that holds over all observables, explicitly
\begin{equation}\begin{split}
    \Pr\left[ \abs{ \widehat{O}^{\{n\}}_b - \expval{\overline{O}_b} } \geq \epsilon \right] \leq \delta
    \qquad \forall \quad b \in [B], \\
    \Pr\left[ \abs{ \overline{O}^{\{n\}}_b - \expval{\overline{O}_b} } \geq \epsilon \right] \leq \delta
    \qquad \forall \quad b \in [B],
\end{split}\end{equation}
and we want to compute the scaling of the number of shots $n$ required with these parameters. Standard application of Hoeffding's inequality gives a tail bound
\begin{equation}\begin{split}
    \Pr\left[ \abs{ \widetilde{O}^{\{n\}}_b - \expval{\widetilde{O}_b} } \geq \chi \right]
    &= \Pr\left[ \abs{ \xi \widetilde{O}^{\{n\}}_b - \xi \expval{\widetilde{O}_b} } \geq \xi \chi \right] \\
    &= \Pr\left[ \abs{ \widehat{O}^{\{n\}}_b - \expval{\widehat{O}_b} } \geq \xi \chi \right]
    \leq 2 \exp( -\frac{n \chi^2}{2 \norm{O_b}_2^2} ),
\end{split}\end{equation}
and setting $\xi \chi = \epsilon$, we have
\begin{equation}\begin{split}
    \Pr\left[ \abs{ \widehat{O}^{\{n\}}_b - \expval{\widehat{O}_b} } \geq \epsilon \right]
    \leq 2 \exp( -\frac{n \epsilon^2}{2 \xi^2 \norm{O_b}_2^2} )
    \leq \delta
    \Longrightarrow 
    n \propto \frac{2 \xi^2 \norm{O_b}_2^2}{\epsilon^2} \log(\frac{2}{\delta}).
    \label{eq-sec:theory/sampling-overhead/mitigated-n}
\end{split}\end{equation}

On the other hand, Hoeffding's inequality applied to the ideal estimator gives
\begin{equation}\begin{split}
    \Pr\left[ \abs{ \overline{O}^{\{n\}}_b - \expval{\overline{O}_b} } \geq \epsilon \right]
    \leq 2 \exp( -\frac{n \epsilon^2}{2 \norm{O_b}_2^2} )
    \leq \delta
    \Longrightarrow 
    n \propto \frac{2 \norm{O_b}_2^2}{\epsilon^2} \log(\frac{2}{\delta}).
    \label{eq-sec:theory/sampling-overhead/ideal-n}
\end{split}\end{equation}

Comparing \cref{eq-sec:theory/sampling-overhead/mitigated-n,eq-sec:theory/sampling-overhead/ideal-n}, we observe that to reach the same precision at the same confidence as in a setting of noiseless circuit execution, the number of shots of the noisy circuit required is larger by a multiplicative factor of $\xi^2$. We thus conclude that the sampling overhead of the mitigation is quantified by $\xi^2$, as stated in the main text.

\subsection{Sensitivity of error mitigation to differences in inputs}
\label{app-sec:theory/sensitivity-analysis}

Consider $\vb{q}$ and $Q$ inputs with their corresponding $\vb*{\alpha}$ solution for the \texttt{PROM} protocol, and a different $\vb{q}'$ and $Q'$ with their $\vb*{\alpha}'$ solution. We denote the overhead factors of the two solutions $\xi = \norm{\vb*{\alpha}}_1$ and $\xi' = \norm{\vb*{\alpha}'}_1$. We analyze the relation between the difference in $\vb{q}$ and $\vb{q}'$ and that in $\vb*{\alpha}$ and $\vb*{\alpha}'$, and thus the mitigated estimators produced by the protocol. Recall that $\vb*{\alpha}$ is the solution of a linear system specified by $Q$---\ie~\cref{app-eq:theory/single/solution-general/linear-system-3}. Thus, we start from a standard result from the theory of sensitivity of linear systems~\cite{deif1986sensitivity, rohn1989new},
\begin{equation}\begin{split}
    \frac{ \norm{\vb*{\alpha}' - \vb*{\alpha}}_p }{ \norm{\vb*{\alpha}}_p }
    \leq 
    \frac{ \kappa_p(Q) r(Q)}{1 - \kappa_p(Q) r(Q)},
    \qquad 
    \kappa_p(Q) = \norm{ Q^{-1} }_p \norm{ Q }_p,
    \qquad 
    r(Q) = \frac{\norm{ Q' - Q }_p}{\norm{ Q }_p},
    \label{app-eq:theory/sensitivity-analysis/condition-number-definition}
\end{split}\end{equation}
valid provided $Q$ is non-singular and $\kappa_p(Q) r(Q) < 1$, for any vector $p$-norm and the associated induced matrix norm, and where $\kappa_p(Q)$ is the norm-wise condition number of $Q$ under that norm. Here for our purposes we examine the $1$-norm. Note from \cref{app-eq:preliminaries/confusion-matrix-properties/Q-norm} that $\norm{Q}_1 = 1$. On the other hand
\begin{equation}\begin{split}
    \norm{Q' - Q}_1
    = \max_{f \in \mathcal{S}} \sum_{s \in \mathcal{S}} \abs{ Q'_{sf} - Q_{sf} }
    = \max_{f \in \mathcal{S}} \underbrace{ \sum_{s \in \mathcal{S}} 
            \abs{ q'_{s \oplus f} - q_{s \oplus f} }
        }_{2 d(\vb{q}', \vb{q})}
    = 2 d(\vb{q}', \vb{q}),
\end{split}\end{equation}
where $0 \leq d(\vb{q}', \vb{q}) \leq 1$ is the total variation distance between the probability distributions $\vb{q}'$ and $\vb{q}$. Moreover, recall that $\vb*{\alpha}$ is the first column of $Q^{-1}$ as derived in \cref{app-eq:theory/single/solution-general/linear-system-3,app-eq:theory/single/solution-general/alpha-solution-1}, and that $\norm*{ Q^{-1} }_1$ reduces to the 1-norm of its first column (or indeed any of its columns), as shown in \cref{app-eq:preliminaries/Q-diagonalization/Q-inverse-norms}. Thus,
\begin{equation}\begin{split}
    \norm{ Q^{-1} }_1 = \norm{ \vb*{\alpha} }_1 = \xi,
    \qquad 
    \kappa_1(Q) = \norm{ Q^{-1} }_1 \norm{ Q }_1 = \xi.
\end{split}\end{equation}

Putting the above together,
\begin{equation}\begin{split}
    \frac{\norm{\vb*{\alpha}' - \vb*{\alpha}}_1}{\xi}
    \leq 
    \frac{2 \xi d(\vb{q}', \vb{q})}
        {1 - 2 \xi d(\vb{q}', \vb{q})},
\end{split}\end{equation}
valid provided $2 \xi d(\vb{q}', \vb{q}) < 1$. Then, using the reverse triangle inequality, we observe the following sensitivity bound on the overhead factor $\xi'$ given differing $\vb{q}$ and $\vb{q}'$ inputs,
\begin{equation}\begin{split}
    \frac{\abs{ \xi' - \xi }}{\xi}
    = \frac{\abs{ \norm{ \vb*{\alpha}' }_1 - \norm{ \vb*{\alpha} }_1 }}{\xi}
    \leq \frac{\norm{ \vb*{\alpha}' - \vb*{\alpha} }_1}{\xi}
    = \frac{2 \xi d(\vb{q}', \vb{q})}
        {1 - 2 \xi d(\vb{q}', \vb{q})}.
    \label{app-eq:theory/sensitivity-analysis/sensitivity-bound-overhead-factor}
\end{split}\end{equation}

Next we examine the mitigated estimators
\begin{equation}\begin{split}
    \widehat{\vb{O}} = \sum_{f \in \mathcal{S}} \alpha_f \vb{O}^{(f)}, \qquad
    \widehat{\vb{O}}' = \sum_{f \in \mathcal{S}} \alpha_f' \vb{O}^{(f)}.
\end{split}\end{equation}

We observe the following sensitivity bound on the mitigated expectation values $\smash{\expval*{\widehat{O}_b'}}$ given differing $\vb{q}$ and $\vb{q}'$ inputs,
\begin{equation}\begin{split}
    \abs{ \expval{\widehat{O}'_b} - \expval{\widehat{O}_b} }
    = \abs{ \sum_{f \in \mathcal{S}} 
        \left( \alpha_f' - \alpha_f \right) \expval{O_b^{(f)}}
    }
    &\leq \sum_{f \in \mathcal{S}} 
        \abs{ \alpha_f' - \alpha_f } \abs{ \expval{O_b^{(f)}} } \\
    &\leq \sum_{f \in \mathcal{S}} 
        \abs{ \alpha_f' - \alpha_f } \cdot \norm{O_b}_2 \\
    &= \norm{ \vb*{\alpha}' - \vb*{\alpha} }_1 \cdot \norm{O_b}_2 \\
    &\leq \frac{2 \xi^2 d(\vb{q}', \vb{q})}
        {1 - 2 \xi d(\vb{q}', \vb{q})} \cdot \norm{O_b}_2,
    \label{app-eq:theory/sensitivity-analysis/sensitivity-bound-mitigated-expval}
\end{split}\end{equation}
for each $b \in [B]$, where spectral norm $\norm{O_b}_2$ corresponds to the largest singular value of the observable $O_b$.

\subsubsection{Corollary---general bound on sampling overheads}
\label{app-sec:theory/sensitivity-analysis/corollary-sampling-overhead}

Consider the sensitivity bound in \cref{app-eq:theory/sensitivity-analysis/sensitivity-bound-overhead-factor} but with primed and unprimed quantities swapped, and take $\vb{q}'$ to correspond to the ideal scenario without readout errors---such that $\smash{q'_s = \delta_{s \vb{0}}}$ and thus $\smash{\alpha'_f = \delta_{f \vb{0}}}$ and $\xi' = 1$. We leave $\vb{q}$, with corresponding solution $\vb*{\alpha}$ and sampling overhead $\xi$, to be completely arbitrary. That is, $\vb{q}$ accommodates arbitrary correlations in readout errors. Suppose the total readout error probability is at most $\eta$, that is, $\smash{q_{\vb{0}} \geq 1 - \eta}$. Then
\begin{equation}\begin{split}
    d(\vb{q}, \vb{q}')
    = d(\vb{q}', \vb{q})
    = \frac{1}{2} \sum_{s \in \mathcal{S}} \abs{q_s - q'_s}
    = \frac{1}{2} \bigg( 
       \underbrace{ \abs{q_{\vb{0}} - q'_{\vb{0}}} }_{\leq 1 - (1 - \eta)}
       + \underbrace{ \sum_{\substack{s \in \mathcal{S} \\ s \neq \vb{0}}} \abs{q_s - q'_s} }_{\leq \eta}
       \bigg) 
    \leq \eta,
    \label{app-eq:theory/sensitivity-analysis/corollary-sampling-overhead/distance-q-qp}
\end{split}\end{equation}
and by the sensitivity bound,
\begin{equation}\begin{split}
    \frac{\abs{ \xi - \xi' }}{\xi'}
    \leq \frac{2 \xi' d(\vb{q}, \vb{q}')}
        {1 - 2 \xi' d(\vb{q}, \vb{q}')}
    \Longrightarrow 
    \abs{ \xi - 1 }
    \leq \frac{2 \eta}
        {1 - 2 \eta}
    \Longrightarrow
    \xi
    \leq 
    \frac{1}{1 - 2 \eta},
\end{split}\end{equation}
valid for $\eta \in [0, 1/2)$.

\subsubsection{Corollary---general bound for errors in observables due to readout errors in feedforward}
\label{app-sec:theory/sensitivity-analysis/corollary-error}

Now we consider the sensitivity bound in \cref{app-eq:theory/sensitivity-analysis/sensitivity-bound-mitigated-expval}. We take $\vb{q}'$ to be the exact characterization of readout errors on the device, and $\vb{q}$ to correspond to a scenario without readout errors ($\smash{q_s = \delta_{s \vb{0}}}$), such that $\smash{\alpha_f = \delta_{f \vb{0}}}$ and $\xi = 1$. That is, employing the \texttt{PROM} protocol with $\vb{q}$ is tantamount to performing no readout error mitigation at all---since only $f = \vb{0}$ is ever sampled. Thus
\begin{equation}\begin{split}
    \expval{\widehat{O}_b} = \expval{O_b^{\mathrm{raw}}}, \qquad 
    \expval{\widehat{O}'_b} = \expval{\overline{O}_b},
\end{split}\end{equation}
where $\smash{\expval{O_b^{\mathrm{raw}}}}$ is the expectation value of observable $O_b$ measured in a setting with readout errors described by $\vb{q}'$ but with no readout error mitigation applied, and $\smash{\expval{\overline{O}_b}}$ is the ideal expectation value as would be measured in a noiseless setting, the latter as written with the same notation in \cref{sec:theory-single,sec:theory-multiple} of the main text.

Suppose the total readout error probability is at most $\eta$, that is, $\smash{q'_{\vb{0}} \geq 1 - \eta}$. Then \cref{app-eq:theory/sensitivity-analysis/corollary-sampling-overhead/distance-q-qp} identically applies. The error in the expectation value of observable $O_b$ measured in a setting with readout errors (characterized by $\vb{q}'$) that impact the mid-circuit measurements and feedforward, but with no readout error mitigation applied, with respect to an ideal setting without readout errors, is then bounded by
\begin{equation}\begin{split}
    \abs{ \expval{O_b^{\mathrm{raw}}} - \expval{\overline{O}_b} }
    = \abs{ \expval{\widehat{O}'_b} - \expval{\widehat{O}_b} }
    \leq
    \frac{2 \xi^2 d(\vb{q}', \vb{q})}
        {1 - 2 \xi d(\vb{q}', \vb{q})} \cdot \norm{O_b}_2
    \leq 
    \frac{2 \eta}{1 - 2 \eta} \cdot \norm{O_b}_2,
\end{split}\end{equation}
valid for $\eta \in [0, 1/2]$. 

\clearpage 
\pagebreak

\section{Technical Details on Hardware Demonstration \& Benchmarking}
\label{app-sec:methods}

Here we provide implementation and technical details on our experiments on quantum hardware, as presented in \cref{sec:expts} of the main text.

\subsection{Quantum hardware}
\label{app-sec:methods/hardware}

We utilized IBM transmon-based superconducting quantum devices in our experiments. These included 27-qubit devices \{\textit{ibm\_algiers}, \textit{ibmq\_kolkata}, \textit{ibm\_mumbai}\} hosting Falcon-family processors, and 127-qubit devices \{\textit{ibm\_osaka}, \textit{ibm\_kyoto}\} hosting Eagle-family processors. The basis gate sets of all devices comprise 1-qubit gates $\smash{\{X, \sqrt{X}, \mathrm{RZ}\}}$, with the $\sigma^z$-rotation ($\mathrm{RZ}$) gate implemented virtually in hardware via framechanges at zero error and duration~\cite{mckay2017efficient}, and 2-qubit gate CX for Falcon processors and echoed cross-resonance (ECR) for Eagle processors~\cite{sundaresan2020reducing}. The CX and ECR gates are equivalent up to 1-qubit rotations. We constructed all experiment circuits using arbitrary 1-qubit and CX gates, as shown in \cref{fig:expts/reset/circuit-diagram,fig:expts/ghz/circuit-diagram,fig:expts/teleport/circuit-diagram} of the main text and \cref{app-fig:methods/circuits}, and transpiled to the native gate set of each device before execution. 

Typical performance metrics of the devices, such as gate errors, gate times, readout errors, readout times, relaxation $T_1$ and dephasing $T_2$ times, are provided in \cref{app-tab:quantum-device-characteristics}.

\subsection{Readout error mitigation for terminal measurements}
\label{app-sec:methods/ro-terminal}

\subsubsection{Background}
\label{app-sec:methods/ro-terminal/background}

In addition to implementing the proposed probabilistic readout error mitigation (\texttt{PROM}) in our experiments, we utilized also a standard readout error mitigation scheme for terminal measurements---\ie~measurements for the observables of interest at the end of the quantum circuits---to correct for readout errors on those measurements. This improves accuracy of results on the measured observables, which would otherwise be distorted by readout noise, as is appropriate for our purpose of investigating the effectiveness of our \texttt{PROM} protocol. 

As described in \cref{sec:theory-single,sec:theory-multiple} of the main text, \texttt{PROM} targets readout errors for mid-circuit measurements and feedforward, and is not precisely intended for terminal measurements. The protocol nonetheless works when applied, for terminal measurements are a special case of mid-circuit ones with no further quantum operations, but is not as resource-efficient as more restrictive schemes specialized for terminal measurements. In practice, it is best to employ \texttt{PROM} in conjunction with a separate readout error mitigation scheme for terminal measurements, as was done here.

To set the stage and for completeness, we first describe a general procedure in \cref{app-sec:methods/ro-terminal/background/general} without symmetrization. Then we describe in particular the procedure when readout channels are symmetrized, as in our present setting, and specific implementation details in our experiments in \cref{app-sec:methods/ro-terminal/background/symmetrized}.

\subsubsection{Standard procedure without symmetrization of readout error channels}
\label{app-sec:methods/ro-terminal/background/general}

We first summarize a standard prescription for terminal readout error mitigation~\cite{kandala2017hardware, kandala2019error, jurcevic2021demonstration}. We consider the confusion matrix $M$, as described in \cref{sec:framework} of the main text and \cref{app-sec:preliminaries/confusion-matrix-properties}, over $m$ qubits. Recall that each entry $M_{t s}$ records the probability of the quantum device reporting measurement outcome $t \in \mathcal{S} = \{0, 1\}^m$ when the true outcome is $s \in \mathcal{S}$. Prior or alongside the experiment, calibration circuits can be run to estimate $M$. In particular, in a calibration quantum circuit, the register of $m$ qubits is prepared in computational basis state $\ket{s}$, which requires simply the application of $X$ gates on the initial state $\smash{\ket{0}^{\otimes m}}$ on appropriate qubits, and then measured in the computational basis. The probability of observing measurement bitstring $t$ gives $M_{t s}$ by definition. Accordingly $\abs{S} = 2^m$ distinct quantum circuits need to be run, to enumerate over all $s \in \mathcal{S}$.

Given an $M$, linear inversion can be performed to mitigate measured experiment counts. In particular, suppose an experiment yielded counts $\vb{c}$, where entry $c_s$ records the number of shots of the quantum circuit that produced terminal measurement outcome $s \in \mathcal{S}$. Then $\vb{c}$ and the true counts $\vb{c}'$ that would hypothetically be observed without readout errors are related by
\begin{equation}\begin{split}
    c_t = \sum_{s \in \mathcal{S}} M_{t s} c'_s
    \Longleftrightarrow 
    \vb{c} = M \vb{c}',
\end{split}\end{equation}
as mentioned in \cref{sec:framework} of the main text, and accordingly the true counts can be approximately recovered as
\begin{equation}\begin{split}
    \vb{c}' = M^{-1} \vb{c},
\end{split}\end{equation}
where $M^{-1}$ is numerically computed. Such a linear inversion does not guarantee that the recovered $\vb{c}'$ is non-negative, as would be expected for a physical counts vector. In cases where non-negativity is to be enforced, an additional mapping step~\cite{smolin2012efficient} can be performed to find the closest non-negative counts distribution to the recovered $\vb{c}'$, or alternatively a least-squares fit with non-negative constraints (and possibly other physicality constraints) can be used in place of a direct numerical inverse~\cite{geller2020rigorous, maciejewski2020mitigation, nachman2020unfolding, jurcevic2021demonstration}.

The need to execute $\abs{S} = 2^m$ quantum circuits to calibrate $M$, which is of size $\abs{S} \times \abs{S} = 2^m \times 2^m$, and the subsequent classical computation of its inverse, becomes problematic in resource requirements at large $m$, say exceeding ${\sim} 14$ qubits. For this reason some studies (\eg~Refs.~\onlinecite{geller2020rigorous, koh2022simulation, koh2022stabilizing, koh2023measurement, koh2023observation}) employ a tensored readout error mitigation scheme, wherein the full register of $m$ qubits is partitioned into $p$ sub-registers comprising $\smash{m^{(1)}, m^{(2)}, \ldots, m^{(p)}}$ qubits each, and one assumes that there are no correlations in readout errors between measurements in different sub-registers. This is precisely the setting described in \cref{app-sec:preliminaries/tensored/general}. In particular, in such a setting the confusion matrix factorizes, $\smash{M = M^{(1)} \otimes M^{(2)} \otimes \ldots \otimes M^{(p)}}$, and $\smash{M^{(\ell)}}$ for each sub-register can be calibrated separately, each requiring $2^{m_\ell}$ calibration quantum circuits. As the qubits in the calibration circuits for different $\ell \in [p]$ are disjoint, the circuits can in fact be merged breadth-wise, thus requiring $2^{\max{(m_1, m_2, \ldots, m_p)}}$ calibration circuits in total instead of $2^{m_1} + 2^{m_2} + \ldots + 2^{m_p}$. The inverses of $\smash{M^{(\ell)}}$, each a $2^{m_\ell} \times 2^{m_\ell}$ matrix, can also be computed separately, since $\smash{M^{-1}} = \smash{[M^{(1)}]^{-1}} \otimes \smash{[M^{(2)}]^{-1}} \otimes \ldots \otimes \smash{[M^{(p)}]^{-1}}$. In practice, it is not necessary to explicit construct $M^{-1}$, which is expensive. Rather one simply applies each $\smash{[M^{(\ell)}]^{-1}}$ to the relevant sectors of $\vb{c}$. 

In certain scenarios, one may have calibrated an $M$ covering measurements on a larger set of qubits than that to be mitigated. The larger than needed $M$ can be used after marginalizing out the unused measurements. We label the calibrated qubits in $M$ by $\mathcal{X} = [m]$, and we retain the notation $\mathcal{S} = \{0, 1\}^m$ for the set of possible measurement outcomes on the $m$ measurements. Let the terminally measured qubits present on the quantum circuit be $\mathcal{X}' \subseteq \mathcal{X}$. The circuit contains $m' = \abs{\mathcal{X}'}$ measurements. Then the confusion matrix characterizing readout errors on those $m'$ measurements, to be used for terminal readout error mitigation on the circuit, is
\begin{equation}\begin{split}
    M'_{t' s'} 
    = \frac{1}{2^{m - m'}}
        \sum_{\substack{t \in \mathcal{S} \\ t_{\mathcal{X}'} = t'}}
        \sum_{\substack{s \in \mathcal{S} \\ s_{\mathcal{X}'} = s'}}
        M_{t s},
    \label{app-eq:methods/ro-terminal/background/general/marginalization}
\end{split}\end{equation}
where $\smash{s_{\mathcal{X}'}}$ denotes the substring of $s$ formed by entries at indices $\mathcal{X}'$. Thus, in principle, in an experiment with multiple quantum circuits containing terminal measurements on different sets of qubits, \cref{app-eq:methods/ro-terminal/background/general/marginalization} allows one to calibrate a larger $M$ over the union of measured qubits across all circuits; then for each circuit the marginalized $M'$ is used for error mitigation. In practice, error characteristics of a quantum device drifts over time, thus re-calibrations are typically required for long experiments.

\subsubsection{Procedure with symmetrization of readout error channels}
\label{app-sec:methods/ro-terminal/background/symmetrized}

Applying BFA~\cite{smith2021qubit, hicks2021readout} symmetrizes the readout error channels, which are then characterized by the confusion vector $\vb{q}$ and associated symmetrized confusion matrix $Q$, as described in the main text (\cref{sec:framework,sec:theory-single/general}) and in \cref{app-sec:preliminaries/confusion-matrix-properties}. Recall that each entry $q_s$ records the probability of a readout error syndrome $s \in \mathcal{S}$ on the quantum device, that is, the probability of reporting outcome $a \oplus t \in \mathcal{S}$ when the true outcome is some $a \in \mathcal{S}$. By definition, $Q_{sf} = q_{s \oplus f}$ for $s \in \mathcal{S}$ and $f \in \mathcal{S}$. Then in an experiment, the observed counts $\vb{c}$ on terminal measurements and the true counts $\vb{c}'$ that would hypothetically be observed without readout errors are related by
\begin{equation}\begin{split}
    c_t = \sum_{a \in \mathcal{S}} q_{a \oplus t} c'_a
    \Longleftrightarrow 
    \vb{c} = \vb{c}' \ast \vb{q} 
\end{split}\end{equation}
where $\ast$ denotes convolution (see \cref{app-sec:preliminaries/notation} for definition), or equivalently
\begin{equation}\begin{split}
    c_t = \sum_{a \in \mathcal{S}} Q_{a t} c'_a
    \Longleftrightarrow 
    \vb{c} = Q \vb{c}'.
\end{split}\end{equation}

Accordingly the true counts is recovered as
\begin{equation}\begin{split}
    \vb{c}' = Q^{-1} \vb{c}.
\end{split}\end{equation}

In our implementation, we additionally enforce non-negativity of $\vb{c}'$ by mapping to the  closest non-negative counts distribution~\cite{smolin2012efficient} in Euclidean norm, a method that was also outlined in \cref{app-sec:methods/ro-terminal/background/general} above. 

To calibrate $\vb{q}$, a single calibration circuit preparing the $\smash{\ket{0}^{\otimes m}}$ computational basis state and then measuring all qubits (with BFA twirling applied) suffices~\cite{smith2021qubit, hicks2021readout}. The probability of observing a measurement outcome $s \in \mathcal{S}$ gives $q_s$ by definition. This is in contrast to \cref{app-sec:methods/ro-terminal/background/general}, which was without symmetrization and required $2^m$ calibration circuits preparing all $2^m$ computational basis states. The prior discussion on tensored calibration and terminal readout error mitigation schemes in \cref{app-sec:methods/ro-terminal/background/general} directly carries over but with $Q$ replacing $M$.

In our experiments, we calibrated $\vb{q}$ and thus $Q$ on the measured qubits present on the quantum circuits. When $\vb{q}$ and $Q$ on subsets of qubits were required, for example when the mid-circuit measurements supporting feedforward or terminal measurements occurred on a subset of qubits, or when layer-wise or fully tensored assumptions were to be imposed on our $\texttt{PROM}$ protocol (as in \cref{sec:expts} of the main text), we performed marginalization on the larger than needed $\vb{q}$ and $Q$, identical to \cref{app-eq:methods/ro-terminal/background/general/marginalization} but with $Q$ replacing $M$. As the error characteristics of quantum devices drift over time, we interleaved multiple rounds of calibration circuits with experiment circuits throughout our experiments.

\subsection{Dynamical decoupling}
\label{app-sec:methods/dynamical-decoupling}

Inevitably, in quantum circuits of practical interest there exist periods of time during which certain qubits are idle as they wait for operations on other qubits to be completed. For example, sequences of gates may need to be executed on other qubits before a qubit is itself involved in $2$-qubit operations with those qubits, or the qubit could be waiting for mid-circuit measurements for feedforward operations to follow. These idle periods can be observed in the circuit diagrams of \cref{fig:expts/reset/circuit-diagram,fig:expts/ghz/circuit-diagram,fig:expts/teleport/circuit-diagram}. 

During idle periods, qubits are still subject to decoherence (\eg~thermal relaxation and dephasing) noise. Dynamical decoupling sequences~\cite{viola1999dynamical}, which run though an alternating schedule of basis changes that overall multiply to the identity, can be implemented to suppress the effect of such decoherence. At a basic level, dynamical decoupling refocuses the dephasing noise experienced by the qubits, such that the accumulated contributions cancel to a certain order at the end of the sequence.

In our experiments, we employ a combination of commonly used XY4 and XY8 dynamical decoupling sequences~\cite{kim2023scalable, jurcevic2021demonstration, pokharel2018demonstration}, the former inserted in idle periods approximately up to the duration of a CX gate on the device (${\sim} \SI{400}{\nano\second}$) and the latter for longer idle durations, such as those induced by waiting for mid-circuit measurements. This selection is not specifically fine-tuned, and we do not expect device performance to differ significantly for minor modifications to the exact sequences. We remark that there exist more sophisticated dynamical decoupling sequences that cancel noise to higher orders or offer other theoretical advantages~\cite{khodjasteh2005fault, uhrig2007keeping}; however, empirical results~\cite{ezzell2023dynamical} indicate that simpler sequences can perform better in practice, owing in part to the presence of errors on the gates introduced by the dynamical decoupling sequence, and the fact that noise backgrounds often differ from theoretical assumptions (\ie~in form and temporal uniformity).

While dynamical decoupling has been broadly employed to great effect on quantum devices (see, in addition to the references above, \eg~Refs.~\onlinecite{pokharel2024better, pokharel2023demonstration, kim2023evidence}), we comment on a couple of limitations that are relevant in the present context. First, in reality perfect cancellation of noise incurred during idle periods does not happen, as the noise background is not uniform over time. This may be especially so when measurements are being performed on the device, which are generally considered non-smooth processes. Second, the superconducting quantum processors we utilized do not yet support pulse operations during the feedforward latency~\cite{baumer2024quantum}, that is, during the short period of time (${\sim} \SI{200}{\nano\second}$) that the mid-circuit measurement outcomes are processed by classical control hardware and feedforward operations decided. Therefore in our experiments, dynamical decoupling sequences are not applied during these latency periods. These factors in general limit the effectiveness of dynamical decoupling on circuits with feedforward. Engineering improved dynamical decoupling methods for such contexts is an active area of research~\cite{baumer2024quantum, tong2024empirical}.

\subsection{Measuring Pauli expectation values}
\label{app-sec:methods/pauli-expval}

As a preliminary to subsequent discussions, we describe explicitly the primitive used to measure Pauli expectation values in our experiments. As with most quantum devices, the superconducting processors we used perform all measurements in the computational basis, that is, the $Z$ basis. Hence to measure in other Pauli bases a change of basis is required. To measure the expectation value of an $n$-qubit Pauli operator $P \in \{\mathbb{I}, X, Y, Z\}^n$, the following operation comprising single-qubit rotations is performed,
\begin{equation}\begin{split}
    B_P = \bigotimes_{k = 1}^n U_{P_k},
    \qquad 
    U_{g} = \begin{dcases}
        \mathbb{I} & g \in \{\mathbb{I}, Z\} \\
        H & g = X \\
        H S^\dag & g = Y,
    \end{dcases}
\end{split}\end{equation}
before measuring all qubits in the support of $P$ (\ie~qubits on which $P$ acts non-trivially) in the computational basis. In an experiment shot, these measurements yield a bitstring $b$. The measurement outcome of $P$ yielded by that shot is $+1$ when $\wt(b)$ is even and $-1$ when $\wt(b)$ is odd.

\subsection{Measuring fidelity of GHZ states via stabilizer expectation values}
\label{app-sec:methods/ghz-fidelity}

In \cref{sec:expts/ghz} of the main text, we are interested in measuring the fidelity $\mathcal{F}$ of the prepared $n$-qubit state at the end of the quantum circuit $\rho_{\mathrm{c}}$ against the ideal GHZ state $\rho_{\textsc{ghz}}$,
\begin{equation}\begin{split}
    \mathcal{F} = \left[ \tr(\sqrt{ 
        \sqrt{\rho_{\textsc{ghz}}} \rho_{\mathrm{c}} \sqrt{\rho_{\textsc{ghz}}} }) \right]^2
    = \bra{\psi_{\textsc{ghz}}} \rho_{\mathrm{c}} \ket{\psi_{\textsc{ghz}}},
    \qquad 
    \rho_{\textsc{ghz}} = \ket{\psi_{\textsc{ghz}}} \bra{\psi_{\textsc{ghz}}},
    \qquad 
    \ket{\psi_{\textsc{ghz}}} = \frac{\ket{0}^{\otimes n} + \ket{1}^{\otimes n}}{\sqrt{2}}.
\end{split}\end{equation}

In general, the fidelity between two quantum states can be measured through the SWAP test~\cite{buhrman2001quantum, barenco1997stabilization} and related variants. But the size of the circuit required, in particular, the need to prepare both quantum states in parallel qubit registers and the hugely expensive controlled-SWAP operation between both registers, makes this direct measurement impractical on present hardware. The need to prepare an ideal GHZ state to SWAP against is also fundamentally problematic in our context. 

In our work, we used a method based on measuring Pauli expectation values to calculate the fidelity $\mathcal{F}$, as introduced in Refs.~\onlinecite{flammia2011direct, silva2011practical} and used, for example, in Refs.~\onlinecite{baumer2023efficient, cao2023generation} which we follow. We refer readers to these references for the broader theory, which in fact applies for more general quantum states, not only the GHZ state. In the present context, the fidelity $\mathcal{F}$ can be decomposed as
\begin{equation}\begin{split}
    \mathcal{F} 
    = \frac{1}{2^n} \sum_{P \in \mathcal{P}} \expval{P}_{\rho_{\mathrm{c}}}
    = \frac{1}{2^n} \sum_{P \in \mathcal{P}} \tr[ P \rho_{\mathrm{c}} ],
\end{split}\end{equation}
where $\mathcal{P}$ are the stabilizers of the GHZ state, comprising exactly $2^n$ elements. That is, each element $P \in \mathcal{P}$ satisfies $\smash{P \ket{\psi_{\textsc{ghz}}} = \ket{\psi_{\textsc{ghz}}}}$ and is a Pauli operator over $n$ qubits. Explicitly, the set of stabilizers can be written as
\begin{equation}\begin{split}
    \mathcal{P} = 
    \left\{ Z^{(x)} : 
        x \in \{0, 1\}^n, \, \wt(x) \,\, \text{even} \right\}
    \cup 
    \left\{ (-1)^{\wt(x) / 2} X^{(\vb{1} \oplus x)} Y^{(x)} :
        x \in \{0, 1\}^n, \, \wt(x) \,\, \text{even} \right\}.
\end{split}\end{equation}

To measure $\expval{P}$ for each $P \in \mathcal{P}$, we use the basic method described in \cref{app-sec:methods/pauli-expval}, accounting for the possible minus sign of $P$ thereafter. We comment that a random sampling of $P \in \mathcal{P}$ suffices to estimate the fidelity $\mathcal{F}$ up to sampling error~\cite{flammia2011direct, silva2011practical, baumer2023efficient, cao2023generation}, but we were able to measure all stabilizers in our experiments, which also eliminated potential considerations about sampling noise in the interpretation of data.

A further reduction in quantum resources, in particular the number of quantum circuits to be executed, can be achieved by simultaneously measuring a maximally sized subset of commuting Pauli operators on each circuit, for instance as employed in Refs.~\onlinecite{gokhale2019minimizing, koh2023measurement} in varying contexts. The basis change circuit components for such measurements can be solved for efficiently through a stabilizer tableau approach. But as these basis rotations are generally entangling and are no longer independent $1$-qubit rotations, they require a number of CX gates to implement (generically scaling with the number of qubits). As our objective is to measure the fidelity of the prepared state against the GHZ state, we chose to avoid this technique to avoid introducing additional error in the measurements.

Lastly we remark that there exist alternative methods to measure GHZ state fidelities. A commonly used method is that of parity oscillations, as used, for example, in Refs.~\onlinecite{sackett2000experimental, leibfried2005creation, monz2011fourteen, omran2019generation, pogorelov2021compact, thomas2022efficient}. Applied towards our present purpose, we highlight that care should be taken to suitably account for the effect of relative phases in the prepared state, that is, that a phase $\phi \neq 0$ may be present in the prepared state $\smash{(\ket{0}^{\otimes n} + e^{i \phi} \ket{1}^{\otimes n}) / \sqrt{2}}$. A third method, multiple quantum coherences~\cite{mooney2021generation, wei2020verifying}, has also been demonstrated in prior literature. We chose the stabilizer measurement approach in our experiment as it is straightforward and robust both theoretically and in implementation on the quantum device.

\clearpage 
\pagebreak

\subsection{Additional circuit diagrams}
\label{app-sec:methods/circuits}

For clarity we provide illustrations of a number of relevant quantum circuit structures and components in \cref{app-fig:methods/circuits}.

\begin{figure}[!h]
    \centering
    \includegraphics[width = 0.95\linewidth]{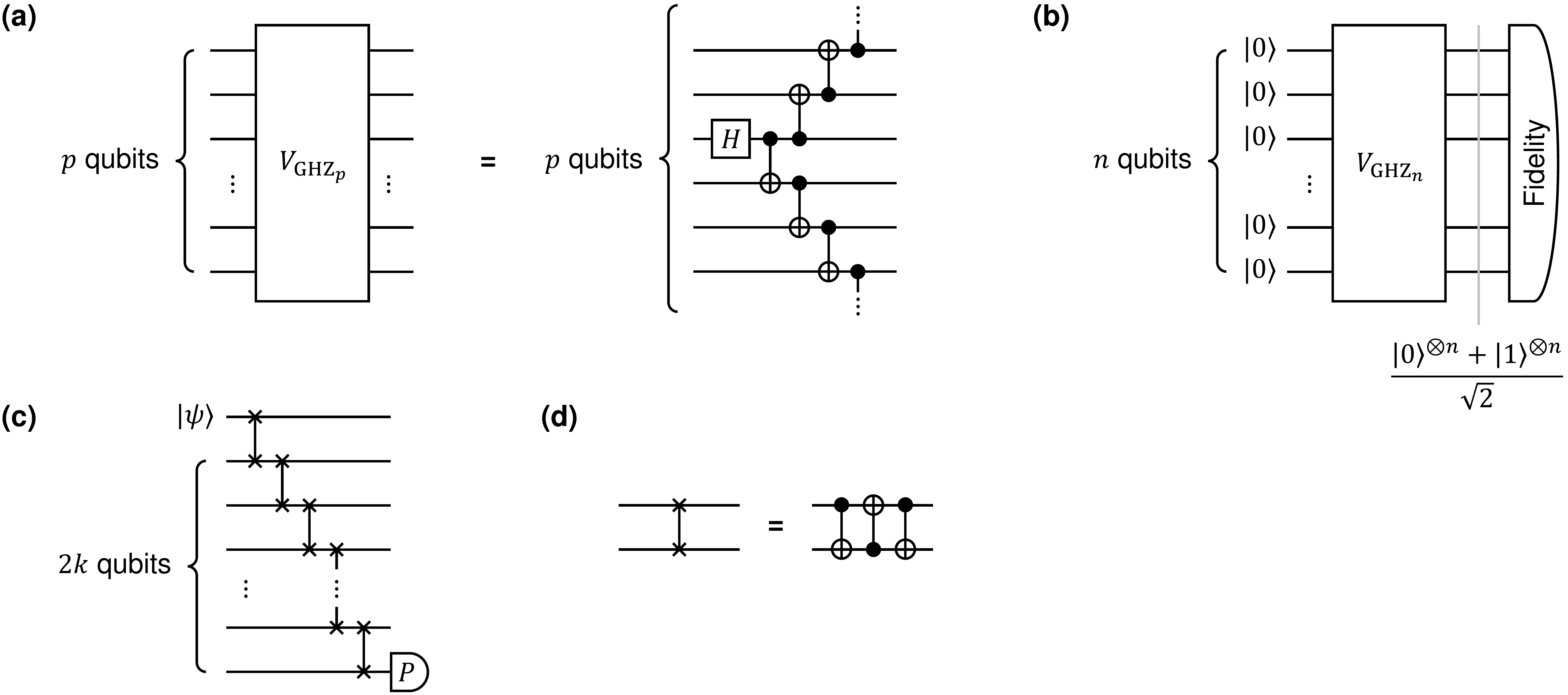}
    \phantomsubfloat{\label{app-fig:methods/circuits/ghz-block}}
    \phantomsubfloat{\label{app-fig:methods/circuits/ghz-unitary}}
    \phantomsubfloat{\label{app-fig:methods/circuits/teleport-unitary}}
    \phantomsubfloat{\label{app-fig:methods/circuits/swap-decomposition}}
    \caption{\textbf{Additional circuit diagrams.} \textbf{(a)} Structure of the $V_{\textsc{ghz}_p}$ block acting on $p$ qubits, which unitarily prepares a $p$-qubit GHZ state on an $\smash{\ket{0}^{\otimes p}}$ input. A Hadamard gate is applied on a qubit, and a fan-out of CX gates spanning all $p$ qubits follow. Thus $p - 1$ CX gates are used, with depth $\ceil{p / 2}$. \textbf{(b)} Unitary GHZ state preparation circuit as used in experiments (\cref{sec:expts/ghz} of the main text) for comparison. \textbf{(c)} Unitary circuit that transports an input qubit state $\ket{\psi}$ a distance of $2k$ qubits, as used in staged quantum state teleportation experiments (\cref{sec:expts/teleport} of the main text) for comparison. $P \in \{X, Y, Z\}$ denotes a Pauli-basis terminal measurement. \textbf{(d)} Standard decomposition of a SWAP gate into $3$ CX gates, which is optimal.}
    \label{app-fig:methods/circuits}
\end{figure}

\begin{turnpage}
    \begin{table}
        \centering
        \begin{tabular}{
                p{3cm} p{1.5cm} p{1.8cm} 
                p{1.5cm} p{1.5cm} p{1.5cm} 
                p{1.5cm} p{1.5cm} p{1.5cm} 
                p{1.5cm} p{1.5cm} p{1.5cm}}
            \toprule 
                Device & Qubits & 2Q Gate &
                \multicolumn{3}{c}{1Q Gate Error ($\times 10^{-4}$)} &
                \multicolumn{3}{c}{2Q Gate Error ($\times 10^{-3}$)} &
                \multicolumn{3}{c}{Readout Error ($\times 10^{-2}$)} \\ 
            \cmidrule(lr){4-6} 
            \cmidrule(lr){7-9} 
            \cmidrule(lr){10-12}
            & & & 
                P10 & P90 & Med. &
                P10 & P90 & Med. &
                P10 & P90 & Med. \\
            \midrule 
            \textit{ibm\_osaka} & 127 & ECR 
                & 1.23 & 12.5 & 2.41 
                & 4.03 & 19.3 & 7.15
                & 0.65 & 8.94 & 1.99 \\
            \textit{ibm\_kyoto} & 127 & ECR
                & 1.46 & 10.5 & 2.68
                & 4.56 & 19.4 & 8.01 
                & 0.61 & 8.15 & 1.51 \\
            \textit{ibm\_algiers} & 27 & CX
                & 1.83 & 8.23 & 2.84
                & 5.88 & 15.3 & 8.90
                & 0.59 & 3.59 & 1.01 \\
            \textit{ibmq\_kolkata} & 27 & CX
                & 1.90 & 4.75 & 2.55
                & 4.96 & 17.2 & 8.62
                & 0.56 & 14.5 & 1.23 \\
            \textit{ibmq\_mumbai} & 27 & CX
                & 1.58 & 3.75 & 2.08
                & 5.66 & 11.5 & 7.52
                & 1.24 & 5.87 & 1.70 \\
            \bottomrule
        \end{tabular} \\[\baselineskip]
        \begin{tabular}{
                p{3cm}
                p{3.2cm}
                p{3.2cm}
                p{1.3cm} p{1.3cm} p{1.3cm}
                p{1.3cm} p{1.3cm} p{1.3cm}
                p{1.3cm} p{1.3cm} p{1.3cm}}
            \toprule 
                Device &
                1Q Gate Time ($\si{\nano\second}$) &
                Readout Time ($\si{\nano\second}$) &
                \multicolumn{3}{c}{2Q Gate Times ($\si{\nano\second}$)} &
                \multicolumn{3}{c}{$T_1$ ($\si{\micro\second}$)} &
                \multicolumn{3}{c}{$T_2$ ($\si{\micro\second}$)} \\ 
            \cmidrule(lr){4-6} 
            \cmidrule(lr){7-9} 
            \cmidrule(lr){10-12}
            & & &
                P10 & P90 & Med. &
                P10 & P90 & Med. &
                P10 & P90 & Med. \\
            \midrule 
            \textit{ibm\_osaka}
                & 60
                & 1400
                & 660 & 660 & 660
                & 127 & 396 & 275 
                & 19.8 & 322 & 151 \\
            \textit{ibm\_kyoto}
                & 60
                & 1400
                & 660 & 660 & 660
                & 124 & 320 & 223 
                & 28.2 & 237 & 115 \\
            \textit{ibm\_algiers}
                & 35.6
                & 910
                & 263 & 622 & 370
                & 93.1 & 184 & 136 
                & 44.3 & 252 & 144 \\
            \textit{ibmq\_kolkata}
                & 35.6
                & 640
                & 391 & 556 & 462
                & 58.5 & 139 & 96.4 
                & 22.8 & 119 & 63.8 \\
            \textit{ibmq\_mumbai}
                & 35.6
                & 3513
                & 277 & 640 & 405
                & 54.6 & 130 & 108
                & 58.4 & 251 & 149 \\
            \bottomrule
        \end{tabular}
        \caption{Performance characteristics of quantum devices used in experiments. For quantities that vary between qubits on the same device, we provide $10^{\text{th}}$ and $90^{\text{th}}$ percentile and median values evaluated over all qubits on the device. Characteristics summarized here are based on routine (${\sim}$daily) calibration data of the machines, pulled over a duration of two weeks in the middle of experiments.}
        \label{app-tab:quantum-device-characteristics}
    \end{table}
\end{turnpage}

\clearpage 
\pagebreak

\section{Additional Experiment Results and Simulations}
\label{app-sec:data}

\subsection{Mid-circuit dynamic qubit reset}
\label{app-sec:data/reset}

\subsubsection{Additional experiment results}
\label{app-sec:data/reset/experiment}

As referenced in \cref{sec:expts/reset} of the main text, in \cref{app-fig:data/reset/experiment/data-n-repm-r3} we present additional experiment results acquired on two other superconducting quantum devices. The results are qualitatively similar to that shown in \cref{fig:expts/reset/fidelity-system,fig:expts/reset/fidelity-spectator}. In \cref{app-fig:data/reset/experiment/data-w-repm-r3} we include repetition-based readout error mitigation protocols \texttt{Rep-ALL} and \texttt{Rep-MAJ} at $r = 3$, which are not at all competitive with \texttt{PROM} or even without readout error mitigation for mid-circuit measurements and feedforward.

\begin{figure}[!h]
    \centering
    \includegraphics[width = 0.82\linewidth]{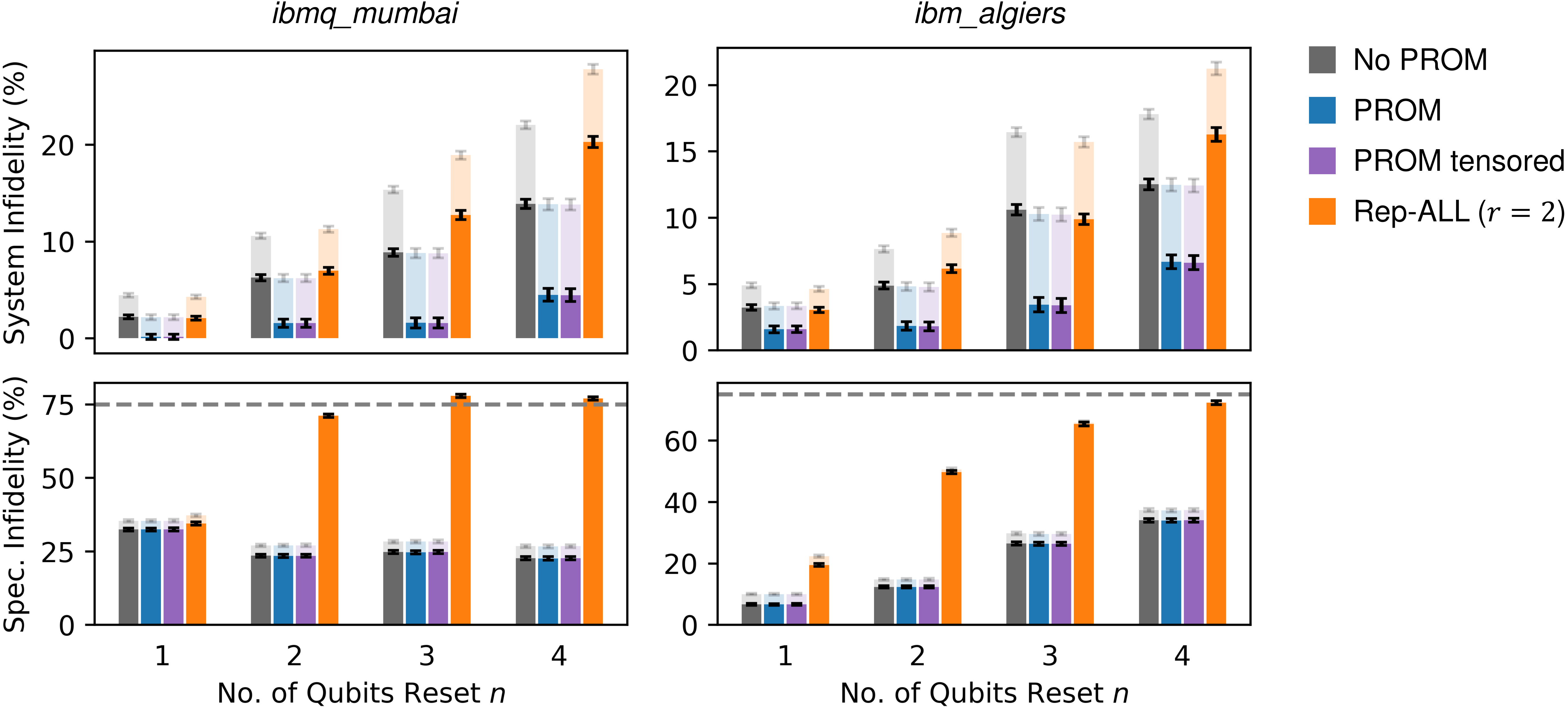}
    \vspace{0.5\baselineskip}
    \caption{\textbf{Additional dynamic qubit reset experiment results.} Infidelities of system qubits after dynamic resets against the ideal $\smash{\ket{0}^{\otimes n}}$ post-reset state, and infidelities of spectator qubits against $\smash{\ket{0}^{\otimes 2}}$ after witnessing reset events conditioned on system qubits being correctly reset, comparing \texttt{PROM}, \texttt{PROM} assuming tensored readout channels, \texttt{Rep-ALL} at $r = 2$, and without readout error mitigation for mid-circuit measurements and feedforward. Dashed line demarcates $75\%$ infidelity expected for a maximally mixed spectator state. In all plots, solid bars are with terminal readout error mitigation; translucent bars without. Experiment performed on superconducting quantum device \textit{ibmq\_mumbai} and \textit{ibm\_algiers} as labeled, complementing results on \textit{ibmq\_kolkata} as presented in \cref{fig:expts/reset/fidelity-system,fig:expts/reset/fidelity-spectator} of the main text. Error bars are standard deviations across $1000$ trials per qubit chain, each with $10000$ shots; data averaged over ${\sim}8$ randomly chosen qubit chains.}
    \label{app-fig:data/reset/experiment/data-n-repm-r3}
\end{figure}

\begin{figure}[!h]
    \centering
    \includegraphics[width = \linewidth]{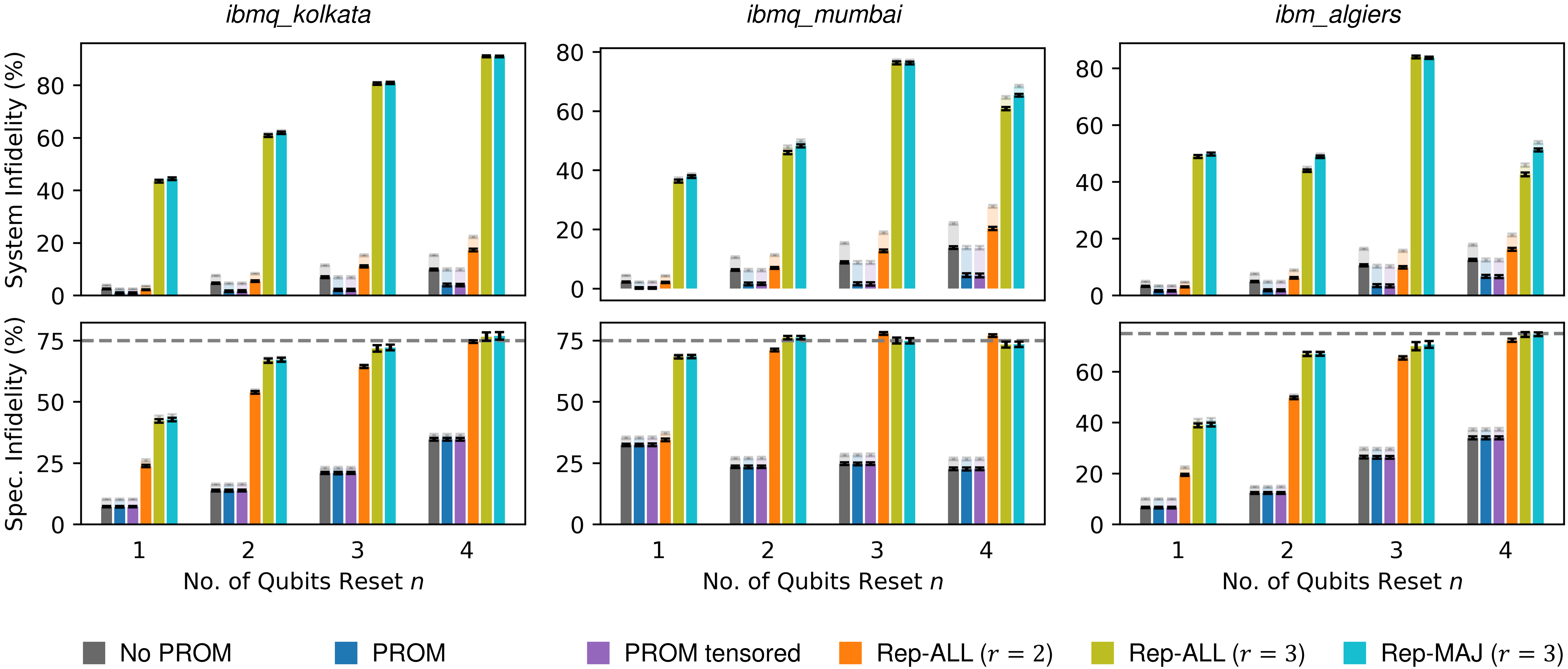}
    \vspace{0.5\baselineskip}
    \caption{\textbf{Additional dynamic qubit reset experiment results.} Zoomed-out version of \cref{fig:expts/reset/fidelity-system,fig:expts/reset/fidelity-spectator} of the main text and \cref{app-fig:data/reset/experiment/data-n-repm-r3} that includes repetition-based readout error mitigation methods \texttt{Rep-ALL} and \texttt{Rep-MAJ} at $r = 3$.}
    \label{app-fig:data/reset/experiment/data-w-repm-r3}
\end{figure}

\clearpage 
\pagebreak

\subsection{Shallow-depth GHZ state preparation}
\label{app-sec:data/ghz}

\subsubsection{Additional experiment results}
\label{app-sec:data/ghz/experiment}

As referenced in \cref{sec:expts/ghz} of the main text, in \cref{app-fig:data/ghz/experiment/data-n04-n06-n-repm-r3} we present additional experiment results acquired on a second superconducting quantum device for GHZ system sizes $n = 4, 6$. The results are qualitatively similar to that shown in \cref{fig:expts/ghz/fidelities-n-4-6}. In \cref{app-fig:data/ghz/experiment/data-n04-n06-w-repm-r3} we include repetition-based readout error mitigation protocols \texttt{Rep-ALL} and \texttt{Rep-MAJ} at $r = 3$, which are not at all competitive with \texttt{PROM} or even without readout error mitigation for mid-circuit measurements and feedforward.

\vspace{\baselineskip}

\begin{figure}[!h]
    \centering
    \includegraphics[width = \linewidth]{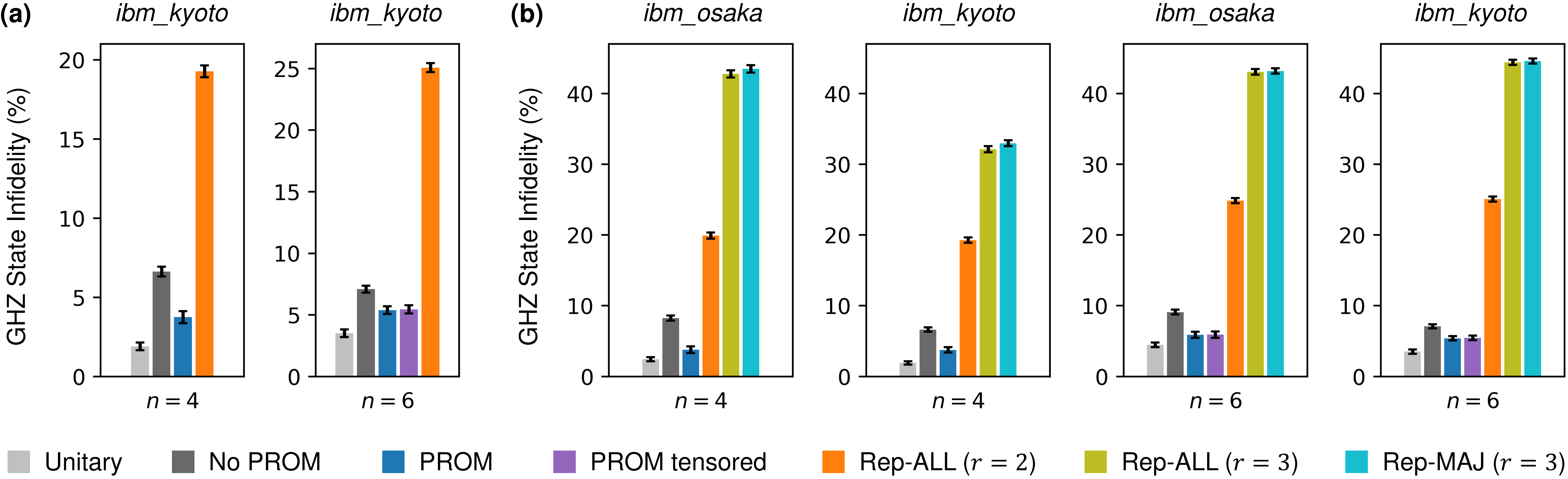}
    \phantomsubfloat{\label{app-fig:data/ghz/experiment/data-n04-n06-n-repm-r3}}
    \phantomsubfloat{\label{app-fig:data/ghz/experiment/data-n04-n06-w-repm-r3}}
    \vspace{-2\baselineskip}
    \vspace{0.5\baselineskip}
    \caption{\textbf{Additional GHZ state preparation experiment results.} \textbf{(a)} Infidelities of the prepared state against the ideal $n$-qubit GHZ state, comparing \texttt{PROM}, \texttt{PROM} assuming tensored readout channels, \texttt{Rep-ALL} at $r = 2$, and without readout error mitigation for mid-circuit measurements and feedforward. Also shown are infidelities on a unitary circuit without feedforward ($\smash{V_{\textsc{ghz}_n}}$) for comparison. Data acquired on the \textit{ibm\_kyoto} superconducting quantum device, complementing results on \textit{ibm\_osaka} as presented in \cref{fig:expts/ghz/fidelities-n-4-6} of the main text. \textbf{(b)} Zoomed-out version of \cref{fig:expts/ghz/fidelities-n-4-6} of the main text and (a) that includes repetition-based readout error mitigation methods \texttt{Rep-ALL} and \texttt{Rep-MAJ} at $r = 3$. Readout error mitigation for terminal measurements was applied for all data. The $n = 4, 6$ experiments employ $(b, p) = (2, 1)$, $(3, 1)$ circuit structures respectively. Error bars are standard deviations across $100$ trials each with $10000$ shots.}
    \label{app-fig:data/ghz/experiment/data-n04-n06}
\end{figure}

\clearpage 
\pagebreak

\subsection{Staged quantum state teleportation}
\label{app-sec:data/teleport}

\subsubsection{Additional experiment results}
\label{app-sec:data/teleport/experiment-data}

As referenced in \cref{sec:expts/teleport} of the main text, in \cref{app-fig:data/reset/teleport/data-n-repm-r2} we present additional experiment results acquired on a second superconducting quantum device. The results are qualitatively similar to that shown in \cref{fig:expts/teleport/expvals}. In \cref{app-fig:data/reset/teleport/data-w-repm-r2} we include repetition-based readout error mitigation protocol \texttt{Rep-ALL} at $r = 2$, which is not at all competitive with \texttt{PROM} or even without readout error mitigation for mid-circuit measurements and feedforward.

\vspace{\baselineskip}

\begin{figure}[!h]
    \centering
    \includegraphics[width = 0.74\linewidth]{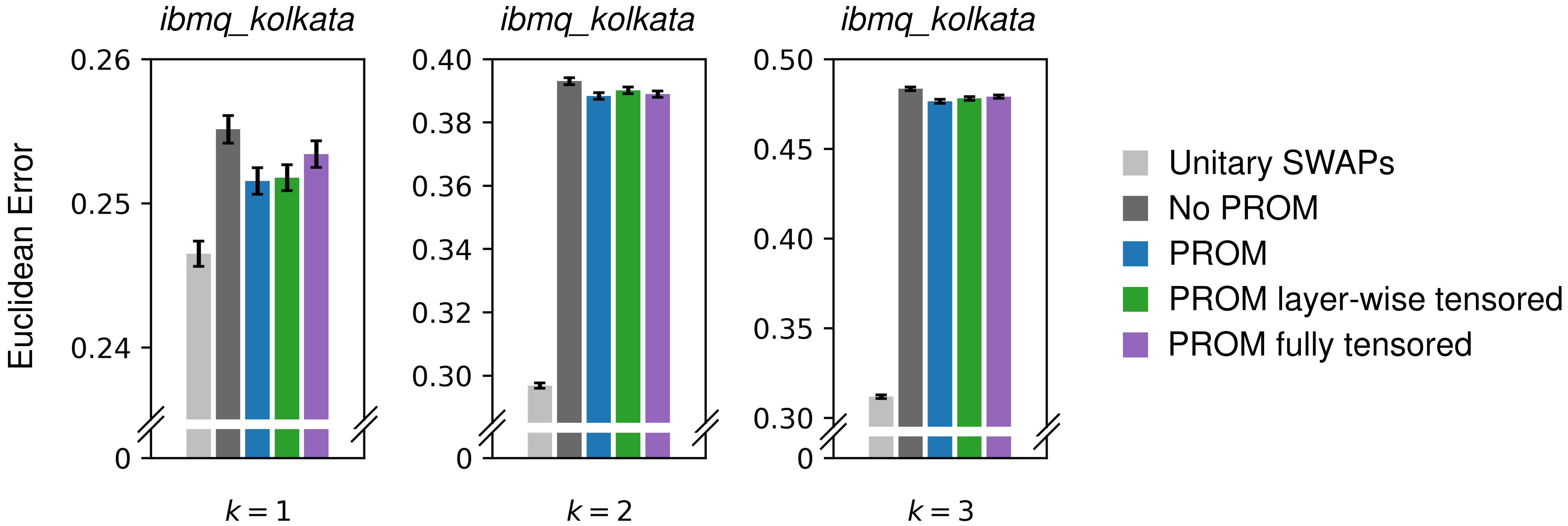}
    \caption{\textbf{Additional quantum state teleportation experiment results.} Euclidean errors $\smash{\Delta = (\delta X^2 + \delta Y^2 + \delta Z^2)^{1/2}}$ after $k \in \{1, 2, 3\}$ teleportations, comparing \texttt{PROM}, \texttt{PROM} assuming layer-wise and fully tensored readout channels, and without readout error mitigation for mid-circuit measurements and feedforward. Also shown are errors on a unitary circuit with SWAP gates that transport the input state $\ket{\psi}$ across the same qubits. Readout error mitigation for terminal measurements was applied for all data. Experiments performed on the superconducting quantum device \textit{ibmq\_kolkata}, complementing results on \textit{ibm\_osaka} as presented in \cref{fig:expts/teleport/expvals} of the main text. Error bars are standard deviations across $100$ trials each with $1000000$ shots.}
    \label{app-fig:data/reset/teleport/data-n-repm-r2}
\end{figure}

\begin{figure}[!h]
    \centering
    \includegraphics[width = \linewidth]{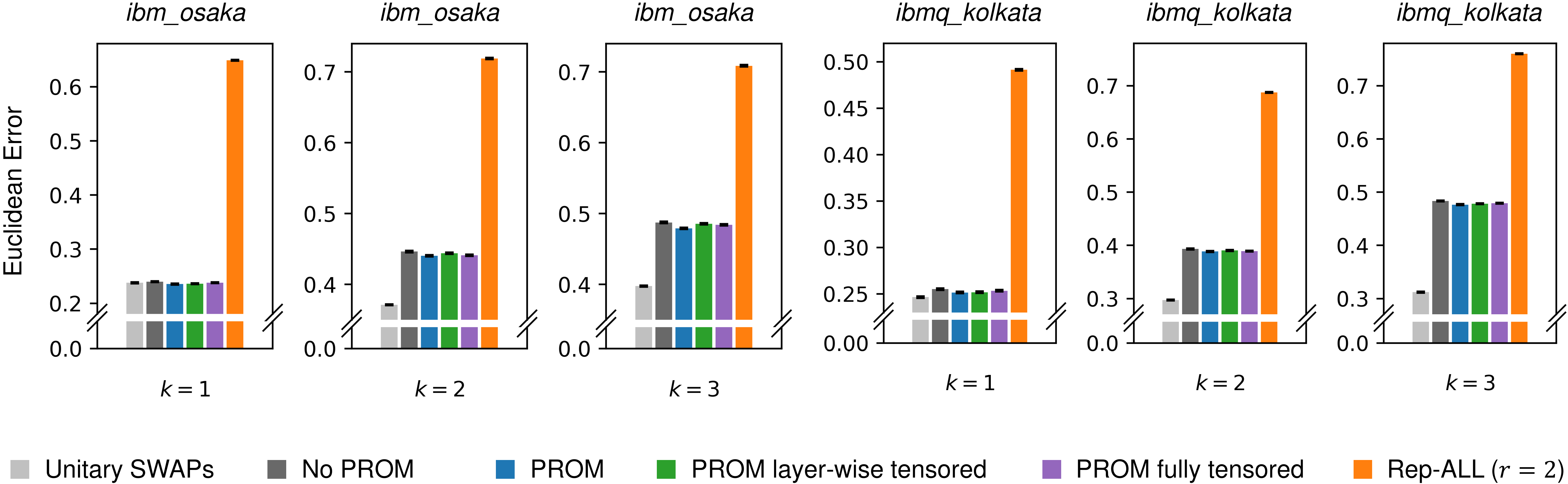}
    \caption{\textbf{Additional quantum state teleportation experiment results.} Zoomed-out version of \cref{fig:expts/teleport/expvals} of the main text and \cref{app-fig:data/reset/teleport/data-n-repm-r2} that includes repetition-based readout error mitigation method \texttt{Rep-ALL} at $r = 2$.}
    \label{app-fig:data/reset/teleport/data-w-repm-r2}
\end{figure}

\clearpage 
\pagebreak

\end{widetext}

\end{document}